\documentclass[useAMS,usenatbib,usegraphicx]{mn2e}
\usepackage{times}
\usepackage{amssymb}
\usepackage{xcolor}
\usepackage[normalem]{ulem}

\title[CFHTLenS: Relation between galaxy DM haloes and baryons]{CFHTLenS: The relation between galaxy dark matter haloes and baryons from weak gravitational lensing}
\author[CFHTLenS]{
\parbox{\textwidth}{
\vspace{-0.5cm}
\raggedright
\mbox{Malin~Velander\unskip$^{1,2}$\thanks{E-mail: mbmvelander@gmail.com},}
\mbox{Edo~van~Uitert\unskip$^{1,3}$,}
\mbox{Henk~Hoekstra\unskip$^{1,4}$,}
\mbox{Jean~Coupon\unskip$^{5}$,}
\mbox{Thomas~Erben\unskip$^{3}$,}
\mbox{Catherine~Heymans\unskip$^{6}$,}
\mbox{Hendrik~Hildebrandt\unskip$^{7,3}$,}
\mbox{Thomas~D.~Kitching\unskip$^{6}$,}
\mbox{Yannick~Mellier\unskip$^{8,9}$,}
\mbox{Lance~Miller\unskip$^{2}$,}
\mbox{Ludovic~Van~Waerbeke\unskip$^{7}$,}
\mbox{Christopher Bonnett\unskip$^{10}$,}
\mbox{Liping~Fu\unskip$^{11}$,}
\mbox{Stefania~Giodini\unskip$^{1}$,}
\mbox{Michael~J.~Hudson\unskip$^{12,13}$,}
\mbox{Konrad~Kuijken\unskip$^{1}$,}
\mbox{Barnaby~Rowe\unskip$^{14,15}$,}
\mbox{Tim~Schrabback\unskip$^{1,16,3}$,}
\mbox{Elisabetta~Semboloni\unskip$^{1}$}
}
\vspace{0.4cm}\\
\parbox{\textwidth}{
$^{1}$Leiden Observatory, Leiden University, Niels Bohrweg 2, 2333 CA Leiden, The Netherlands \\
$^{2}$Department of Physics, Oxford University, Keble Road, Oxford OX1 3RH, UK \\
$^{3}$Argelander Institute for Astronomy, University of Bonn, Auf dem H{\"u}gel 71, 53121 Bonn, Germany \\
$^{4}$Department of Physics and Astronomy, University of Victoria, Victoria, BC V8P 5C2, Canada \\
$^{5}$Institute of Astronomy and Astrophysics, Academia Sinica, P.O. Box 23-141, Taipei 10617, Taiwan \\
$^{6}$Scottish Universities Physics Alliance, Institute for Astronomy, University of Edinburgh, Royal Observatory, Blackford Hill, Edinburgh, EH9 3HJ, UK \\
$^{7}$University of British Columbia, Department of Physics and Astronomy, 6224 Agricultural Road, Vancouver, B.C. V6T 1Z1, Canada \\
$^{8}$Institut d'Astrophysique de Paris, Universit\'{e} Pierre et Marie Curie - Paris 6, 98 bis Boulevard Arago, F-75014 Paris, France \\
$^{9}$Institut d'Astrophysique de Paris, CNRS, UMR 7095, 98 bis Boulevard Arago, F-75014 Paris, France \\
$^{10}$Institut de Ciencies de l'Espai, CSIC/IEEC, F. de Ciencies, Torre C5 par-2, Barcelona 08193, Spain \\
$^{11}$Key Lab for Astrophysics, Shanghai Normal University, 100 Guilin Road, 200234, Shanghai, China \\
$^{12}$Dept. of Physics and Astronomy, University of Waterloo, Waterloo, ON, N2L 3G1, Canada \\
$^{13}$Perimeter Institute for Theoretical Physics, 31 Caroline Street N, Waterloo, ON, N2L 1Y5, Canada \\
$^{14}$Department of Physics and Astronomy, University College London, Gower Street, London WC1E 6BT, U.K \\
$^{15}$California Institute of Technology, 1200 E California Boulevard, Pasadena CA 91125, USA \\
$^{16}$Kavli Institute for Particle Astrophysics and Cosmology, Stanford University, 382 Via Pueblo Mall, Stanford, CA 94305-4060, USA
\vspace{-0.5cm}}
}

\begin{document}
\date{
\vspace{-0.5cm}
}
\pagerange{\pageref{firstpage}--\pageref{lastpage}} \pubyear{2013}
\maketitle

\label{firstpage}

\defcitealias{vhv11}{VU11}

\begin{abstract}
We present a study of the relation between dark matter halo mass and the baryonic content of their host galaxies, quantified through galaxy luminosity and stellar mass. Our investigation uses $154\,\mathrm{deg}^2$ of Canada-France-Hawaii Telescope Lensing Survey (CFHTLenS) lensing and photometric data, obtained from the CFHT Legacy Survey. To interpret the weak lensing signal around our galaxies we employ a galaxy-galaxy lensing halo model which allows us to constrain the halo mass and the satellite fraction. Our analysis is limited to lenses at redshifts between 0.2 and 0.4, split into a red and a blue sample. We express the relationship between dark matter halo mass and baryonic observable as a power law with pivot points of \mbox{$10^{11}\,h_{70}^{-2}\,L_{\odot}$} and \mbox{$2\times10^{11}\,h_{70}^{-2}\,M_{\odot}$} for luminosity and stellar mass respectively. For the luminosity-halo mass relation we find a slope of 
\mbox{$1.32\pm0.06$}
and a normalisation of
\mbox{$1.19^{+0.06}_{-0.07}\times10^{13}\,h_{70}^{-1}\,M_{\odot}$}
for red galaxies, while for blue galaxies the best-fit slope is 
\mbox{$1.09^{+0.20}_{-0.13}$}
and the normalisation is
\mbox{$0.18^{+0.04}_{-0.05}\times10^{13}\,h_{70}^{-1}\,M_{\odot}$}.
Similarly, we find a best-fit slope of 
\mbox{$1.36^{+0.06}_{-0.07}$}
and a normalisation of
\mbox{$1.43^{+0.11}_{-0.08}\times10^{13}\,h_{70}^{-1}\,M_{\odot}$}
for the stellar mass-halo mass relation of red galaxies, while for blue galaxies the corresponding values are
\mbox{$0.98^{+0.08}_{-0.07}$}
and
\mbox{$0.84^{+0.20}_{-0.16}\times10^{13}\,h_{70}^{-1}\,M_{\odot}$}.
All numbers convey the 68\% confidence limit. For red lenses, the fraction which are satellites inside a larger halo tends to decrease with luminosity and stellar mass, with the sample being nearly all satellites for a stellar mass of \mbox{$2\times10^{9}\,h_{70}^{-2}\,M_{\odot}$}. The satellite fractions are generally close to zero for blue lenses, irrespective of luminosity or stellar mass. This, together with the shallower relation between halo mass and baryonic tracer, is a direct confirmation from galaxy-galaxy lensing that blue galaxies reside in less clustered environments than red galaxies. We also find that the halo model, while matching the lensing signal around red lenses well, is prone to over-predicting the large-scale signal for faint and less massive blue lenses. This could be a further indication that these galaxies tend to be more isolated than assumed.
\end{abstract}

\begin{keywords}
cosmology: observations -- gravitational lensing: weak -- galaxies: haloes -- dark matter \vspace{-1.0cm}
\end{keywords}

\clearpage
\section{Introduction}
In order to fully understand the mechanisms behind galaxy formation, the connection between galaxies and the extensive dark matter haloes in which they are enveloped must be studied in exhaustive detail. In pursuit of this precision, reliable mass estimates of both the baryonic and the dark matter content of galaxies are required. The visible component may be evaluated using galaxy properties such as the luminosity or the stellar mass, properties which can be derived via stellar synthesis models \citep{khw03,gcb05,bel01,src07}. The dark matter, on the other hand, cannot be observed directly but must be examined through its gravitational influence on the surroundings. At the largest scales reached by haloes, optical tracers such as satellite galaxies are scarce. Furthermore, estimates of halo mass from satellite galaxy kinematics \citep[see, for example,][]{mbc11} do not only require spectroscopic measurements of a very large number of objects, which are unfeasible with current instrumentation, but they also require the application of the virial theorem with all its associated assumptions. To study any and all galaxies it is therefore desirable to use probes independent of these tracers, and independent of the physical state of the halo, but with the power to explore a large range of scales. These requirements are all satisfied by weak gravitational lensing.

Gravitational lensing is a fundamental consequence of gravity. As light from distant objects travels through the Universe it is deflected by intervening matter. This deflection causes the distant objects, or sources, to appear distorted (and magnified). In the weak regime the distortion is minute, and only by studying the shapes of a large number of sources can information about the foreground gravitational field be extracted. By examining the average lensing distortion as a function of distance from foreground galaxies, or lenses, the density profiles of their dark matter haloes may be directly investigated; this technique is known as galaxy-galaxy lensing. First detected by \citet{bbs96}, the field of galaxy-galaxy lensing has been growing rapidly, with increasing precision as survey area grows. Our understanding of the underlying physics also increases as the interpretation of the signal becomes more sophisticated. Simulations predict that dark matter haloes are well approximated by Navarro-Frenk-White profiles \citep*[NFW;][]{nfw96} and comparing such a profile to the observed signal around isolated lenses results in halo mass estimates. Galaxies and their haloes are not generally isolated, however, but reside in clustered environments. The ramification is that the interpretation of the observed galaxy-galaxy lensing signal around foreground lenses becomes more complicated since the signal from neighbouring haloes also influences the result. To address this problem a number of approaches have been employed. Early studies modelled the lensing signal by associating all matter with galaxies and comparing the resulting shear field to the observations in a maximum-likelihood approach \citep{sch97,hgd98,hyg04}. In this case the clustering of galaxies was accounted for through the observed positions and \citet{hgd98} explicitly attempted to correct for the offset signal seen by satellite galaxies in larger haloes. It was, however, an approximate description. Alternatively the issue can be circumvented by selecting only isolated lenses \citep[see][]{hhy05}. This inevitably leads to a large reduction in the number of lenses, and the sample is no longer representative as it does not probe the full range of environments.

Over the past decade a new approach has gained traction: the weak lensing halo model \citep{coo02,guz02,mts05,vhv11,ltb11}. Within the halo model framework, all haloes are represented as distinct entities, each with a galaxy at the centre. Enclosed in each main halo are satellite galaxies surrounded by subhaloes. In this work we seek to employ the halo model to gain a more accurate picture of galaxy-size dark matter haloes, allowing for a more precise analysis of the link between galaxies and the dark matter haloes they reside in. For this purpose we use image data from the completed Canada-France-Hawaii Telescope Legacy Survey (CFHTLS), and weak lensing and photometric redshift catalogues produced by the Canada-France-Hawaii Telescope Lensing Survey \citep[CFHTLenS\footnote{\tt www.cfhtlens.org};][]{hvm12,mhk13,hek12}. This work improves on the preliminary galaxy-galaxy lensing analysis carried out using a small subset of the CFHTLS and a single-halo model fit to the inner regions only \citep{phh07}. Furthermore, unlike \citet{ckm12} who studied the clustering signal of galaxies for the full CFHTLS-Wide to constrain the evolution in redshift of the stellar-to-halo mass relation, our analysis is based on galaxy-galaxy lensing, which can directly constrain the average halo mass of galaxies on small scales.

Three recent studies to use the galaxy-galaxy lensing halo model to constrain these relations are \citet{msk06}, \citet{vhv11} (hereafter~\citetalias{vhv11}) and \citet{ltb12}. \citet{msk06} studied the halo masses of lenses from the full area of the fourth data release of the Sloan Digital Sky Survey \citep[SDSS DR4;][]{aaa06} using a galaxy-galaxy lensing halo model. The SDSS is very wide, but also very shallow which means that for low luminosity galaxies it is highly powerful while it lacks the depth to constrain the halo masses of higher-luminosity galaxies which are at higher redshifts on average. A similar study was performed by \citetalias{vhv11} using an earlier implementation of the halo model software used for this paper. That study exploited a 300~deg$^2$ overlap between the SDSS DR7 and the intermediate-depth second Red-sequence Cluster Survey \citep[RCS2;][]{ggy11}. The SDSS data were used to identify the lenses, but the lensing analysis was performed on the RCS2, improving greatly at the high mass end on the previous analysis based on the shallow SDSS alone. However, while the \citetalias{vhv11} lenses had accurate spectroscopic redshift estimates, there were no redshift estimates available for the sources at the time. Thus the work presented here has, aside from the increased depth down to $i'_{\mathrm{AB}}=24.7$, a further advantage over the \citetalias{vhv11} analysis owing to the high-precision photometric redshifts available for all objects used in our analysis \citep[see][]{hek12}. This makes it possible to cleanly separate lenses from sources and therefore minimises the contamination by satellites. It also allows for optimal weighting of the lensing signal.

\citet{ltb12} combined several techniques to constrain the relation between halo mass and stellar mass using data from the deep space-based Cosmic Evolution Survey \citep[COSMOS;][]{saa07}. They did not, however, refine their results by splitting their lens sample according to galaxy type. In a follow-up paper, \citet{tgl12} did split the COSMOS sample into star forming and passive galaxies to study the redshift evolution of the same relation, but limited their study to massive galaxies located centrally in a group-sized halo. Thanks to the large area and depth of the CFHTLenS, we are in this paper able to investigate the relation for blue and red galaxies separately without limiting our sample in that way. We provide here a detailed comparison between our results and those quoted in \citet{msk06}, \citetalias{vhv11} and \citet{ltb12}, but leave \citet{tgl12} due to the large difference in sample selection between our analysis and theirs.

This paper is organised as follows: we introduce the data in Section~\ref{cfhtls:sec:data}, and in Section~\ref{cfhtls:sec:method} we review our halo model and the formalism behind it. We investigate the lensing signal as a function of luminosity in Section~\ref{cfhtls:sec:luminosity} and as a function of stellar mass in Section~\ref{cfhtls:sec:stellarmass}. In Section~\ref{cfhtls:sec:comparison} we compare our results to the three previous studies introduced above and we conclude in Section~\ref{cfhtls:sec:conclusion}. The following cosmology is assumed throughout \citep[WMAP7;][]{ksd11}: $(\Omega_{M},\Omega_{\Lambda},h,\sigma_{8},w) = (0.27,0.73,0.70,0.81,-1)$. All numbers reported throughout the paper have been obtained with this cosmology, and all $h$ are factored into the numbers. The $h$-scaling of each quantity is made explicit via the use of $h_{70} = H_0/70 = 1$ where appropriate.

\section{Data}\label{cfhtls:sec:data}
In this paper we present a galaxy-galaxy weak lensing analysis of the entire Wide part of the Canada-France-Hawaii Telescope Legacy Survey (CFHTLS-Wide). The unique combination of area and depth makes this survey ideal for weak lensing analyses. The CFHTLS was a joint project between Canada and France which commenced in 2003 and which is now completed. The survey area was imaged using the Megaprime wide field imager mounted at the prime focus of the Canada-France-Hawaii Telescope (CFHT) and equipped with the MegaCam camera. MegaCam comprises an array of $9\times4$ CCDs and has a field of view of 1~deg$^2$. The wide synoptic survey covers an effective area of about 154~deg$^2$ in five bands: $u^*$, $g'$, $r'$, $i'$ and $z'$. This area is composed of four independent fields, W1--4, each with an area of 23-64~deg$^2$ and with a full multi-colour depth of $i'_{\mathrm{AB}}=24.7$ (source in the CFHTLenS catalogue). The images have been independently reduced within the CFHTLenS collaboration, and for details on this data reduction process, we refer to \citet{ehl09,ehm12}. 

CFHTLenS has measured accurate shapes and photometric redshifts for $8.7\times10^6$ galaxies \citep[][]{hvm12,mhk13,hek12}. The shear estimates for the sources used in this work have been obtained using {\em lens}fit as detailed in \citet{mhk13}, and thoroughly tested for systematics within the CFHTLenS collaboration \citep[see][]{hvm12}. All sources also have multi-band photometric redshift estimates as detailed in \citet{hek12}. The catalogues we use in this work are discussed in \citet{hvm12}, \citet{mhk13} and \citet{hek12}, with the exception of the stellar mass estimates. These estimates were obtained and tested for this paper and we therefore describe them in detail below.

\subsection{Stellar masses}\label{cfhtls:sec:stellarmassestimates}

\begin{figure}
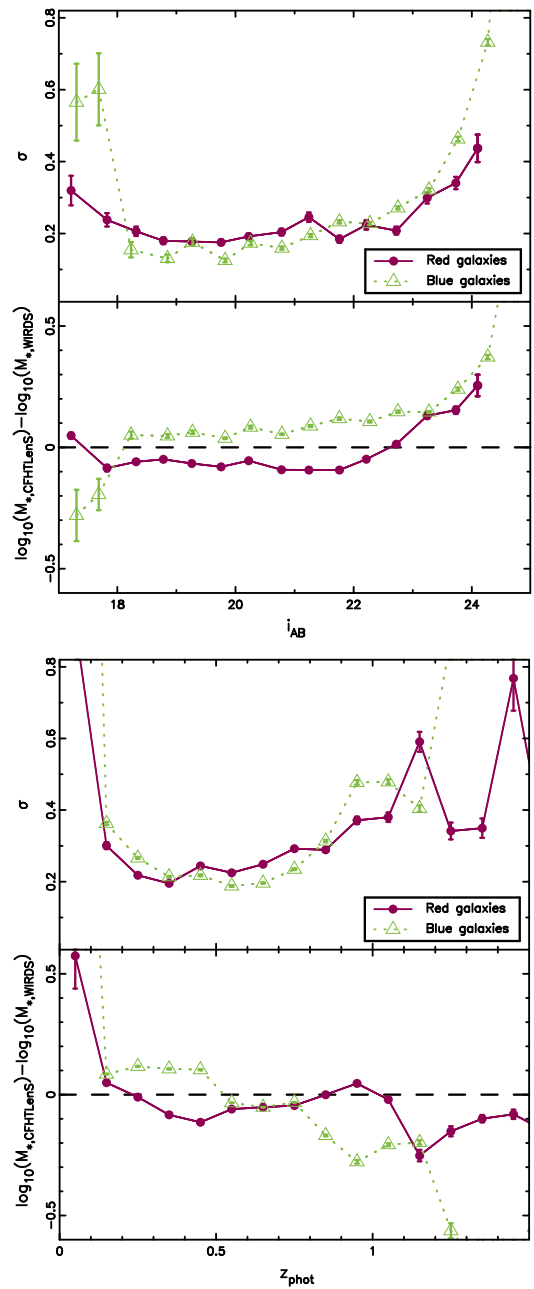

\begin{center}
\includegraphics[width=84mm,angle=270]{wirds.mag.ps}\\
\vspace{0.2cm}
\includegraphics[width=84mm,angle=270]{wirds.z.ps}
\end{center}
\caption{CFHTLenS stellar masses compared to those from the CFHT WIRCam Deep Survey (WIRDS) as a function of $i'_{\mathrm{AB}}$-magnitude (top) and redshift (bottom) for red (dark purple solid dots) and blue (light green open triangles) galaxies. The upper panels in each plot show the dispersion in log stellar mass and the lower panels show the bias of the CFHTLenS stellar mass estimates relative to the WIRDS stellar mass estimates.}
\label{cfhtls:fig:stellarmasscomp}
\end{figure}

Our primary photometry analysis uses the Bayesian photometric redshift software {\sc bpz} \citep{ben00,cbs06} to estimate photometric redshifts after performing an extinction correction on the multi-colour magnitudes. Using {\sc bpz} with a simple set of six modified \citet{cww80} templates is our preferred method to estimate redshifts when using only five optical bands \citep[see][]{hek12}, and we note that it has been shown that the {\sc bpz} software is as accurate for photometric redshift estimates as the alternative Bayesian {\sc LePhare\footnote{www.cfht.hawaii.edu/$\sim$arnouts/lephare.html}} \citep{acm99,iam06} software \citep[see][for a comparison]{hak10}. For physical parameters such as stellar mass estimates, however, our preferred method is to use a more complex set of galaxy templates. Using {\sc LePhare} and \citet{bru03} templates has been proven to be a robust method to estimate physical parameters \citep[see][]{isl10} and so we choose to use {\sc LePhare} to estimate stellar masses. For a consistent analysis we also compute rest-frame luminosities from the same spectral template as used for the stellar mass estimates.

We derive our stellar mass estimates by fitting synthetic spectral energy distribution (SED) templates while keeping the redshift fixed at the {\sc bpz} maximum likelihood estimate. The SED templates are based on the stellar population synthesis (SPS) package developed by \citet{bru03} assuming a \cite{cha03} initial mass function (IMF). Following \citet{isl10}, our initial set of templates includes 18 models using two different metallicities ($\mathrm{Z}_1=0.008\,\mathrm{Z}_\odot$ and $\mathrm{Z}_2=0.02\,\mathrm{Z}_\odot$) and nine exponentially decreasing star formation rates $\propto e^{-t/\tau}$, where $t$ is time and $\tau$ takes the values $\tau=0.1, 0.3, 1, 2, 3, 5, 10, 15, 30\,\mathrm{Gyr}$. The final template set is then generated over 57 starburst ages ranging from 0.01 to 13.5~Gyr, and seven extinction values ranging from 0.05 to 0.3 using a \cite{cab00} extinction law. \citet{isl10} investigated the possible sources of uncertainty and bias by comparing stellar mass estimates between methods. The expected difference between our estimates and those based on a Salpeter IMF \citep{awl07}, a ``diet'' Salpeter IMF \citep{bel08}, or a Kroupa IMF \citep{bmb06} is $-0.24$~dex, $-0.09$~dex, or 0~dex respectively \citep[see][]{isl10}. In their Section 4.2, \citet{isl10} further argue that the choice of extinction law may lead to a systematic difference of $0.14$, and the choice of SPS model to a median difference of 0.13--0.15~dex, with differences reaching 0.24~dex for massive galaxies with a high star formation rate.

We determine the errors on our stellar mass estimates via the 68\% confidence limits of the SED fit, using the full probability distribution function. However, since we fix the redshift these errors tell us only how good the model fit is, and do not account for uncertainties in the photometric redshift estimates \citep[see Section 5.2 of][]{hek12}. To assess the stellar mass uncertainty due to photometric redshift errors we therefore compare our mass estimates to those of the CFHT WIRCam Deep Survey \citep[WIRDS;][]{bhm12}. The WIRDS stellar masses were derived from the CFHTLS Deep fields with additional broad-band near-infrared data using the same method as described here. We are thus comparing our CFHTLenS stellar mass estimates to other estimates which are also based on photometric data, but which have deeper photometry leading to a more robust stellar mass estimate. The additional near-infrared data allows us to rely on these estimates up to a redshift of 1.5 \citep{pbl07}. For our comparison we use a total of 134,290 galaxies in the overlap between the CFHTLenS and WIRDS data, splitting our sample into red and blue galaxies using their photometric type $T_{\rm BPZ}$. $T_{\rm BPZ}$ is a number in the range of $[1.0,6.0]$ representing the best-fit SED and we define our red and blue samples as galaxies with $T_{\rm BPZ}<1.5$ and $2.0<T_{\rm BPZ}<4.0$ respectively, where the latter captures most spiral galaxies. A colour-colour comparison confirms that these samples are well defined. In Figure~\ref{cfhtls:fig:stellarmasscomp} we show the comparison between our stellar mass estimates and those from WIRDS as a function of magnitude (top, with galaxies in the redshift range $[0.2,0.4]$) and redshift (bottom, with galaxies in the magnitude range $[17.0,23.5]$).

For the range of lens redshifts used in this paper, \mbox{$0.2\leq z_{\mathrm{lens}}\leq 0.4$}, the total dispersion compared to WIRDS is then \mbox{$\sim0.2$~dex} for both red and blue galaxies. The lower panel in the bottom plot of Figure~\ref{cfhtls:fig:stellarmasscomp} shows that for red galaxies our stellar masses are in general slightly lower than the WIRDS estimates, with the opposite being true for blue galaxies. For galaxies brighter than \mbox{$i'_{\mathrm{AB}}\sim18$}, both the dispersion and the bias increase due to biases in the redshift estimates \citep[see][]{hek12}. The bias and dispersion also increase rapidly at magnitudes fainter than \mbox{$i'_{\mathrm{AB}}\sim23$}, again due to redshift errors.

We emphasise that this comparison with WIRDS quantifies only the statistical stellar mass uncertainty due to errors in the photometric redshifts and due to our particular template choice. Since the mass estimates from both datasets have been derived using identical method and template set, the systematic errors affecting stellar mass estimates are not taken into account above. The uncertainties arising from the choice of models and dust extinction law adds 0.15~dex and 0.14~dex respectively to the error budget, as mentioned above, resulting in a total uncertainty of $\sim0.3$~dex.

\subsection{Lens and source sample}\label{cfhtls:sec:cats}
\begin{figure}
\includegraphics[height=84mm,angle=270]{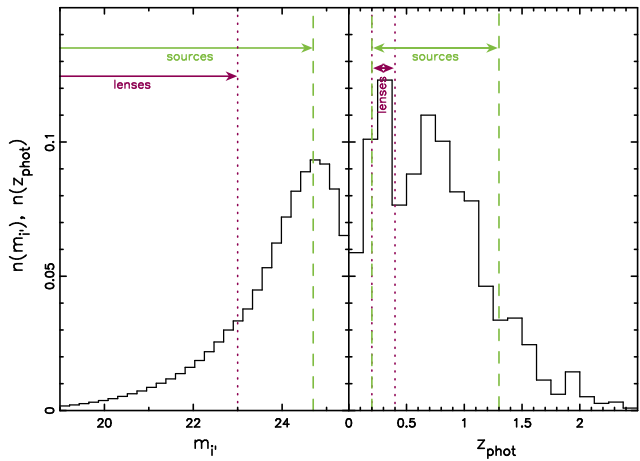}
\caption{Magnitude (left panel) and photometric redshift (right panel) distributions of galaxies in the CFHTLenS catalogue. For the left panel we show all galaxies in the CFHTLenS, while for the right panel we limit our sample to magnitudes brighter than $i'_{\mathrm{AB}}=24.7$. The upper limit of lens (source) magnitude used is shown with a dark purple dotted (light green dashed) line in the left panel, while our lens (source) redshift selection is marked with dark purple dotted (light green dashed) lines in the right panel. Though the lens and source selections appear to overlap in redshift, sources are always selected such that they are well separated from lenses in redshift (see Section~\ref{cfhtls:sec:cats}). Furthermore, close pairs are down-weighted as described in Section~\ref{cfhtls:sec:wgglensing}.}
\label{cfhtls:fig:redshifts}
\end{figure}

The depth of the CFHTLS enables us to investigate lenses with a large range of lens properties and redshifts, which in turn grants us the opportunity to thoroughly study the evolution of galaxy-scale dark matter haloes. As discussed by \citet{hek12}, the use of photometric redshifts inevitably entails some bias in redshift estimates, and also in derived quantities such as luminosity and stellar mass. Our analysis is sensitive even to a small bias since our lenses are selected to reside at relatively low redshifts of \mbox{$0.2\leq z_{\mathrm{lens}}\leq 0.4$}, where $z$ is understood to be the peak of the photometric redshift probability density function, unless explicitly stated otherwise (see Figure~\ref{cfhtls:fig:redshifts}). Because our lensing signal is detected with high precision, we empirically correct for this bias using the overlap with a spectroscopic sample as described in Appendix~\ref{cfhtls:app:photozbias}. Throughout this paper, we then use the corrected redshifts, luminosities and stellar masses for our lenses. For the full survey area we achieve a lens count of $N_{\mathrm{lens}}=1.1\times10^6$.

We then split our lens sample in luminosity or stellar mass bins as described in Sections~\ref{cfhtls:sec:luminosity} and \ref{cfhtls:sec:stellarmass} to investigate the halo mass trends as a function of lens properties. Since we have access to multi-colour data, we are also able to further divide our lenses in each bin into a red and a blue sample using photometric type as described in Section~\ref{cfhtls:sec:stellarmassestimates}. We also ensure that our lenses are brighter than $i'_{\mathrm{AB}}<23$ which corresponds to an 80\% completeness of the spectroscopic redshift sample we use to quantify the redshift bias discussed above. The high completeness ensures that the spectroscopic sample is a good representation of our total galaxy sample. The galaxy sample is dominated by blue late-type galaxies for $i'_{\mathrm{AB}}>22$, however, and we are thus unable to perform a reliable redshift bias correction for red lenses at fainter magnitudes due to a lack of objects. We therefore exclude red lenses with $i'_{\mathrm{AB}}>22$ while allowing blue lenses to magnitudes as faint as $i'_{\mathrm{AB}}=23$. This selection is also illustrated in Figure~\ref{cfhtls:fig:redshifts}.

To minimise any dilution of our lensing signal due to photometric redshift uncertainties, we use an approach similar to that of \citet{ltb12} and use only sources for which the redshift 95\% confidence limit does not overlap with the lens redshift. We further ensure that the lens and source are separated by at least 0.1 in redshift space. To verify the effectiveness of this separation, we compare the source number counts around our lenses to that around random points \citep[as suggested by][Section~4.1]{sjf04}. This test shows no significant evidence of contamination. The source magnitude is only limited by the maximum CFHTLenS analysis depth of \mbox{$i'_{\mathrm{AB}}\sim24.7$} \citep[see][]{hvm12,mhk13,hek12}. Note that we do not apply a redshift bias correction to source redshifts as there is no existing spectroscopic redshift survey at these faint limits. While it is important to correct our lenses for such a bias since the derived baryonic observables such as luminosity and stellar mass depend strongly on redshift, it is less important for the sources as the lensing signal scales with the ratio $D_{ls}/D_{s}$, where $D_{s}$ and $D_{ls}$ are the angular diameter distances to the source, and between the lens and source respectively. This ratio is insensitive to small biases in the source redshifts. Our source count for the full survey (excluding masked areas) is then $N_{\mathrm{source}}=5.6\times10^6$, corresponding to an effective source density of $10.6\,\mathrm{arcmin}^{-2}$ where we use the source density definition from \citet[][Equation~1]{hvm12}.

The high quality of the CFHTLenS shear measurements has been verified via a series of systematics tests presented in \citet{hvm12} and \citet{mhk13}. To further illustrate the robustness of the shears we perform two separate analyses specifically designed to test the galaxy-galaxy lensing signal. First, we use a sample of magnitude-selected lenses across the entire survey and compare the resulting weak lensing signal to that found by \citet{phh07} for a 22~deg$^2$ subset of the CFHTLS data, and to that found by \citetalias{vhv11} for RCS2. Both previous analyses use shear measurement software based on the class of methods first introduced by \citet*{ksb95} and known as KSB. The details of the comparison may be found in Appendix~\ref{cfhtls:app:parkerAndRCS2}, and we find that the signal we measure agrees well with these earlier shear measurements. The second test, as described in Appendix~\ref{cfhtls:app:seeing}, uses the seeing of the images to test for any potential multiplicative bias still remaining. We find that this bias is consistent with zero.

\section{Method}\label{cfhtls:sec:method}
To analyse the dark matter haloes surrounding our lenses we use a method known as galaxy-galaxy lensing, and compare the measured signal with a halo model. In this section we will introduce the basic formalism and give an overview of our halo model.

\subsection{Galaxy-galaxy lensing}\label{cfhtls:sec:wgglensing}
The first-order lensing distortion, shear, is a stretch tangentially about a lens, induced by the foreground structure on the shape of a background source galaxy. Assuming that sources are randomly oriented intrinsically, the net alignment caused by lensing can be measured statistically from large source samples. In a galaxy-galaxy lensing analysis, source galaxy distortions are averaged in concentric rings centred on lens galaxies. We measure the tangential shear, $\gamma_t$, as a function of radial distance from the lens this way, and also the cross shear, $\gamma_{\times}$, which is a 45$^\circ$ rotated signal. When averaged azimuthally, the cross shear can never be induced by a single lens which means that it may be used as a systematics check. The amplitude of the tangential shear is directly related to the differential surface density $\Delta\Sigma(r)=\overline{\Sigma}(<r)-\Sigma(r)$, i.e.~the difference between the mean projected surface mass density enclosed by $r$ and the surface density at $r$, via
\begin{equation}
\Delta\Sigma(r) = \Sigma_{\rm crit}\langle\gamma_t(r)\rangle
\label{cfhtls:eq:surfaceDensity}
\end{equation}
with $\Sigma_{\rm crit}$ the critical surface density
\begin{equation}
\Sigma_{\rm crit} = \frac{c^2}{4\pi G}\frac{D_s}{D_lD_{ls}}
\label{cfhtls:eq:criticalSurfaceDensity}
\end{equation}
where $D_l$ is the angular diameter distance to the lens, and $D_s$ and $D_{ls}$ are defined as before. Here, $c$ is the speed of light and $G$ is the gravitational constant. By comparing differential surface densities rather than tangential shears, the geometric factor is neutralised and the amplitudes of the signals can be directly contrasted between different samples. The only caveat is that the properties of lenses depend on the lens redshift so this difference still has to be taken into account.

We calculate the weighted average shear in each distance bin from the lens by combining the shear measurement weight $w$ with the geometric lensing efficiency $\eta=(D_lD_{ls})/D_s$ as described in \citet[][Appendix B.4]{vks11}. By using $\eta$ we down-weight close pairs and can minimise any influence of redshift inaccuracies on the measured signal that way. We quantify any remaining redshift systematics by calculating a correction factor for each mass estimate based on the redshift error distribution; see Appendix~\ref{cfhtls:app:photozs} for details on how this is done. The average shear, scaled to a reference redshift, is then given by
\begin{equation}
\langle\gamma_t(r)\rangle = \frac{\sum w_i(\gamma_{t,i}\,\eta^{-1}_{f,i})\,\eta^2_{f,i}}{\sum w_i\eta^2_{f,i}}
\end{equation}
where $\eta_f=\eta/\eta_{\mathrm{ref}}$ is the lensing efficiency weight factor with $\eta_{\mathrm{ref}}$ a reference lensing efficiency value. The lensing weight $w_i$ is defined in Equation 8 of \citet{mhk13}, and accounts both for the ellipticity measurement error and for the intrinsic shape noise. Finally, we convert the average shear to $\Delta \Sigma(r)$ using the $\Sigma_{\rm crit}$ computed for the reference lens and source redshifts.

The CFHTLenS shears are affected by a small but non-negligible multiplicative bias. \citet{mhk13} have modelled this bias using a set of simulations specifically created as a `clone' of the CFHTLenS, obtaining a calibration factor $m(\nu_{\mathrm{SN}},r_{\mathrm{gal}})$ as a function of the signal-to-noise ratio, $\nu_{\mathrm{SN}}$, and size of the source galaxy, $r_{\mathrm{gal}}$. Rather than dividing each galaxy shear by a factor $(1+m)$, which would lead to a biased calibration as discussed in \citet{mhk13}, we apply it to our average shear measurement in each distance bin using the correction
\begin{equation}
1+K(r) = \frac{\sum w_i\eta_{f,i}[1+m(\nu_{\mathrm{SN},i},r_{\mathrm{gal},i})]}{\sum w_i\eta_{f,i}}\;.
\end{equation}
The lensing signal is then calibrated as follows:
\begin{equation}
\langle\gamma^{\mathrm{cal}}(r)\rangle = \frac{\langle\gamma(r)\rangle}{1+K(r)}\;.
\label{cfhtls:eq:corrfac}
\end{equation}
The effect of this correction term on our galaxy-galaxy analysis is to increase the average lensing signal amplitude by at most 6\%. Though there will be some uncertainty associated with this term, \citet{kfh13} find that it has a negligible effect on their shear covariance matrix. The calibration factor $m$ enters linearly in our Equation~\ref{cfhtls:eq:corrfac}, while it is squared in the \citet{kfh13} correlation function correction factor, thus amplifying its effect. The conclusion we draw is therefore that the impact of the calibration factor uncertainty will be insignificant in this work. We also apply the additive $c$-term correction discussed in \citet{hvm12} but find that it does not change our results either.

The circular averaging over lens-source pairs makes this type of analysis robust against small-scale systematics introduced by for example PSF residuals in the shape measurement catalogues. Because the galaxy-galaxy lensing signal is more resilient to systematics than cosmic shear, we choose to maximise our signal-to-noise by using the full CFHTLenS area (except for masked areas) rather than removing the fields that have not passed the cosmic shear systematics test described in \citet{hvm12}. However, there could be spurious large-scale signal present owing to areas being masked, or from lenses close to an edge, such that the circular average does not cover all azimuthal angles. We correct for such spurious signal using a catalogue of random lens positions situated outside any masked areas; the number of random lenses used is 50,000 per square-degree field, which amounts to more than ten times as many as real lenses. The stacked lensing signal measured around these random lenses is evidence of incomplete circular averages and will be present in the observed stacked lensing signal as well. Because of our high sampling of this random points signal, we can correct the observed signal measured in each field by subtracting the signal around the random lenses. This random points test is discussed in more detail in \citet{mhs05}. The test shows that for this data, individual fields do indeed display a signal around random lenses which is to be expected, even in the absence of any shape measurement error, due to cosmic shear and shot noise, and due to the masking effect mentioned above. Averaged over the entire CFHTLenS area the random lens signal is insignificant relative to the signal around true lenses ranging from $\sim0.5\%$ to $\sim5\%$ over the angular range used in this analysis. Additionally, to ascertain whether including the fields that fail the cosmic shear systematics test biases our results, we compare the tangential shear around all galaxies with~$19.0<i'_{\mathrm{AB}}<22.0$ in the fields that respectively pass and fail this test, and find no significant differences between the signals.

\subsection{The halo model}\label{cfhtls:sec:halomodel}
\begin{figure}
\includegraphics[height=84mm,angle=270]{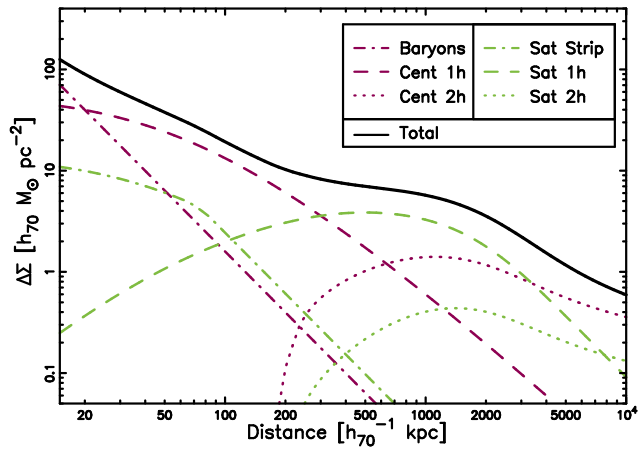}
\caption{Illustration of the halo model used in this paper. Here we have used a halo mass of $M_{200}=10^{12}\,h_{70}^{-1}\,M_{\odot}$, a stellar mass of $M_*=5\times10^{10}\,h_{70}^{-2}\,M_{\odot}$ and a satellite fraction of $\alpha=0.2$. The lens redshift is $z_{\mathrm{lens}} = 0.5$. Dark purple lines represent quantities tied to galaxies which are centrally located in their haloes while light green lines correspond to satellite quantities. The dark purple dash-dotted line is the baryonic component, the light green dash-dotted line is the stripped satellite halo, dashed lines are the 1-halo components induced by the main dark matter halo and dotted lines are the 2-halo components originating from nearby haloes.}
\label{cfhtls:fig:haloModelExample}
\end{figure}
To accurately model the weak lensing signal observed around galaxy-size haloes, we have to account for the fact that galaxies generally reside in clustered environments. In this work we do this by employing the halo model software first introduced in \citetalias{vhv11}. For full details on the exact implementation we refer to \citetalias{vhv11}; here we give a qualitative overview.

Our halo model builds on work presented in \citet{guz02} and \citet{mts05}, where the full lensing signal is modelled by accounting for the central galaxies and their satellites separately. We assume that a fraction $(1-\alpha)$ of our galaxy sample reside at the centre of a dark matter halo, and the remaining objects are satellite galaxies surrounded by subhaloes which in turn reside inside a larger halo. In this context $\alpha$ is the satellite fraction of a given sample.

The lensing signal induced by central galaxies consists of two components: the signal arising from the main dark matter halo (the 1-halo term $\Delta\Sigma^{\rm 1h}$) and the contribution from neighbouring haloes (the 2-halo term $\Delta\Sigma^{\rm 2h}$). The two components simply add to give the lensing signal due to central galaxies:
\begin{equation}
\Delta\Sigma_{\rm cent} = \Delta\Sigma_{\rm cent}^{\rm 1h} + \Delta\Sigma_{\rm cent}^{\rm 2h}\;.
\label{cfhtls:eq:deltaSigmaCent}
\end{equation}
In our model we assume that all main dark matter haloes are well represented by an NFW density profile \citep*{nfw96} with a mass-concentration relationship as given by \citet{dsk08}. The halo model parameters resulting from an analysis such as ours (see, for example, Section~\ref{cfhtls:sec:luminosity}) are not very sensitive to the exact halo concentration, however, as discussed in \citetalias{vhv11} and in Appendix~\ref{cfhtls:app:assumptions}. To compute the 2-halo term, we use the non-linear power spectrum from \citet{spj03}. We also assume that the dependence of the galaxy bias on mass follows the prescription from \citet{smt01}, incorporating the adjustments described in \citet{twz05}. Note that this mass-bias relation is empirically calibrated on large numerical simulations, and does not discriminate between different galaxy types. Finally, we note that the central term essentially assumes a delta function in halo mass as a function of a given observable since we do not integrate over the halo mass distribution. For a given luminosity bin, for example, the particular mass distribution within that bin therefore has to be accounted for. We do correct our measured halo mass for this in the following sections, assuming a log-normal distribution, and the correction method is described in Appendices~\ref{cfhtls:app:photozs} and \ref{cfhtls:app:binscatter} for the luminosity and stellar mass analysis respectively.

We model satellite galaxies as residing in subhaloes whose spatial distribution follows the dark matter distribution of the main halo. The number density of satellites in a halo of a given mass is described by the halo occupation distribution (HOD) which is commonly parameterised through a power law of the form $\langle N \rangle = M^{\epsilon}$. Following \citet{mts05}, we set $\epsilon=1$ for masses above a characteristic mass scale, defined to be three times the typical halo mass of a set of lenses. For masses below this threshold, we use $\epsilon=2$. In our model, the subhaloes have  been tidally stripped of dark matter in the outer regions. As \citet{mts05} did, we adopt a truncated NFW profile, choosing a truncation radius of $0.4r_{200}$ beyond which the lensing signal is proportional to $r^{-2}$, where $r$ is the physical distance from the lens. This choice results in about 50\% of the subhalo dark matter being stripped, and we acquire a satellite term which supplies signal on small scales. Thus satellite galaxies add three further components to the total lensing signal: the contribution from the stripped subhalo ($\Delta\Sigma^{\rm strip}$), the satellite 1-halo term which is off-centre since the satellite galaxy is not at the centre of the main halo, and the 2-halo term from nearby haloes. Just as for the central galaxies, the three terms add to give the satellite lensing signal:
\begin{equation}
\Delta\Sigma_{\rm sat} = \Delta\Sigma_{\rm sat}^{\rm strip} + \Delta\Sigma_{\rm sat}^{\rm 1h} + \Delta\Sigma_{\rm sat}^{\rm 2h}\;.
\label{cfhtls:eq:deltaSigmaSat}
\end{equation}

There is an additional contribution to the lensing signal, not yet considered in the above equations. This is the signal induced by the lens baryons ($\Delta\Sigma^{\rm bar}$). This last term is a refinement of the halo model presented in \citetalias{vhv11}, necessary since weak lensing measures the total mass of a system and not just the dark matter mass. Following \citet{ltb11} we model the baryonic component as a point source with a mass equal to the mean stellar mass of the lenses in the sample:
\begin{equation}
\Delta\Sigma^{\rm bar} = \frac{\langle M_*\rangle}{\pi r^2}\;.
\label{cfhtls:eq:deltaSigmaBar}
\end{equation}
This term is fixed by the stellar mass of the lens, and we do not fit it. Note that we choose not to include the baryonic term for neighbouring haloes, but its contribution is negligible.

Finally, to obtain the total lensing signal of a galaxy sample of which a fraction $\alpha$ are satellites we combine the baryon, central and satellite galaxy signals, applying the appropriate proportions:
\begin{equation}
\Delta\Sigma = \Delta\Sigma^{\rm bar} + (1-\alpha)\Delta\Sigma_{\rm cent} + \alpha\Delta\Sigma_{\rm sat}\;.
\label{cfhtls:eq:deltaSigmaTotal}
\end{equation}
All components of our halo model are illustrated in Figure~\ref{cfhtls:fig:haloModelExample}. In this example the dark matter halo mass is $M_{200}=10^{12}\,h_{70}^{-1} M_{\odot}$, the stellar mass is $M_*=5\times10^{10}\,h_{70}^{-2}\,M_{\odot}$, the satellite fraction is $\alpha=0.2$, the lens redshift is $z_{\rm lens}=0.5$ and $D_{ls}/D_{s}=0.5$. On small scales the 1-halo components are prominent, while on large scales the 2-halo components dominate.

We note here that the halo model is necessarily based on a number of assumptions. Some of these assumptions may be overly stringent or inaccurate, and some may differ from assumptions made in other implementations of the galaxy-galaxy halo model. To be able to make useful comparisons with other studies (such as the comparison done in this paper, see Section~\ref{cfhtls:sec:comparison}), particularly considering the statistical power and accuracy afforded by the CFHTLenS, we attempt to provide a quantitative impression of how large a role the assumptions actually play in determining the halo mass and satellite fractions. The full evaluation is recounted in Appendix~\ref{cfhtls:app:assumptions} where we study the effect of the following modelling choices: the inclusion of a baryonic component, the NFW mass-concentration relation as applied to the central halo profile, the truncation radius of the stripped satellite component, the distribution of satellites within a given halo, the HOD and the bias prescription. Our general finding is that, given reasonable spans in the parameters affecting these choices, the best-fit halo mass can change by up to $\sim$15--20\% for each individual assumption tested. The magnitude of the effect depends on the luminosity or stellar mass, and bins with a greater satellite fraction will often be more strongly affected. In essentially all cases the effect is subdominant to observational errors and we therefore do not take them into account in what follows, though we do acknowledge that several effects may conspire to cause a non-negligible change to our results.

\section{Luminosity trend}\label{cfhtls:sec:luminosity}
\begin{figure}
\includegraphics[height=84mm,angle=270]{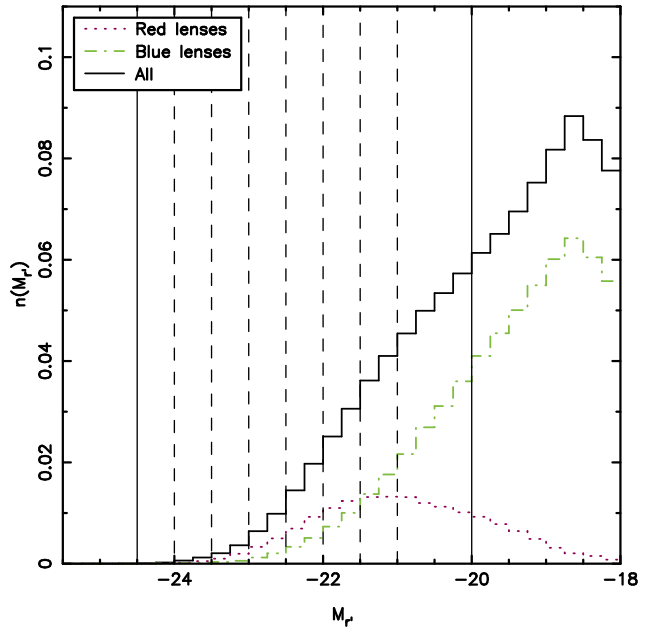}
\caption{$r'$-band absolute magnitude distribution in the CFHTLenS catalogues for lenses with redshifts \mbox{$0.2\leq z_{\mathrm{lens}}\leq0.4$} (black solid histogram). The distribution of red (blue) lenses is shown in dotted dark purple (dot-dashed light green). Our lens bins are marked with vertical lines.}
\label{cfhtls:fig:magSelection}
\end{figure}
\begin{table}
\caption{Details of the luminosity bins. (1) Absolute magnitude range; (2) Number of lenses; (3) Mean redshift; (4) Fraction of lenses that are blue.}
\label{cfhtls:tab:magbins1}
\begin{center}
\begin{tabular}{@{}lcccc}
\hline
Sample 	& $M_{r'}$$^{(1)}$ 	& $n_{\rm lens}$$^{(2)}$ 	& $\langle z\rangle$$^{(3)}$ & $f_{\rm blue}$$^{(4)}$ \\
\hline
L1 & [-21.0,-20.0] & 91224   & 0.32 & 0.70 \\
L2 & [-21.5,-21.0] & 33633   & 0.32 & 0.45 \\
L3 & [-22.0,-21.5] & 23075   & 0.32 & 0.32 \\
L4 & [-22.5,-22.0] & 12603   & 0.32 & 0.20 \\
L5 & [-23.0,-22.5] & 5344    & 0.32 & 0.11 \\
L6 & [-23.5,-23.0] & 1704    & 0.31 & 0.05 \\
L7 & [-24.0,-23.5] & 344     & 0.30 & 0.03 \\
L8 & [-24.5,-24.0] & 76      & 0.30 & 0.09 \\
\hline
\end{tabular}
\end{center}
\end{table}
\begin{figure*}
\includegraphics[height=168mm,angle=270]{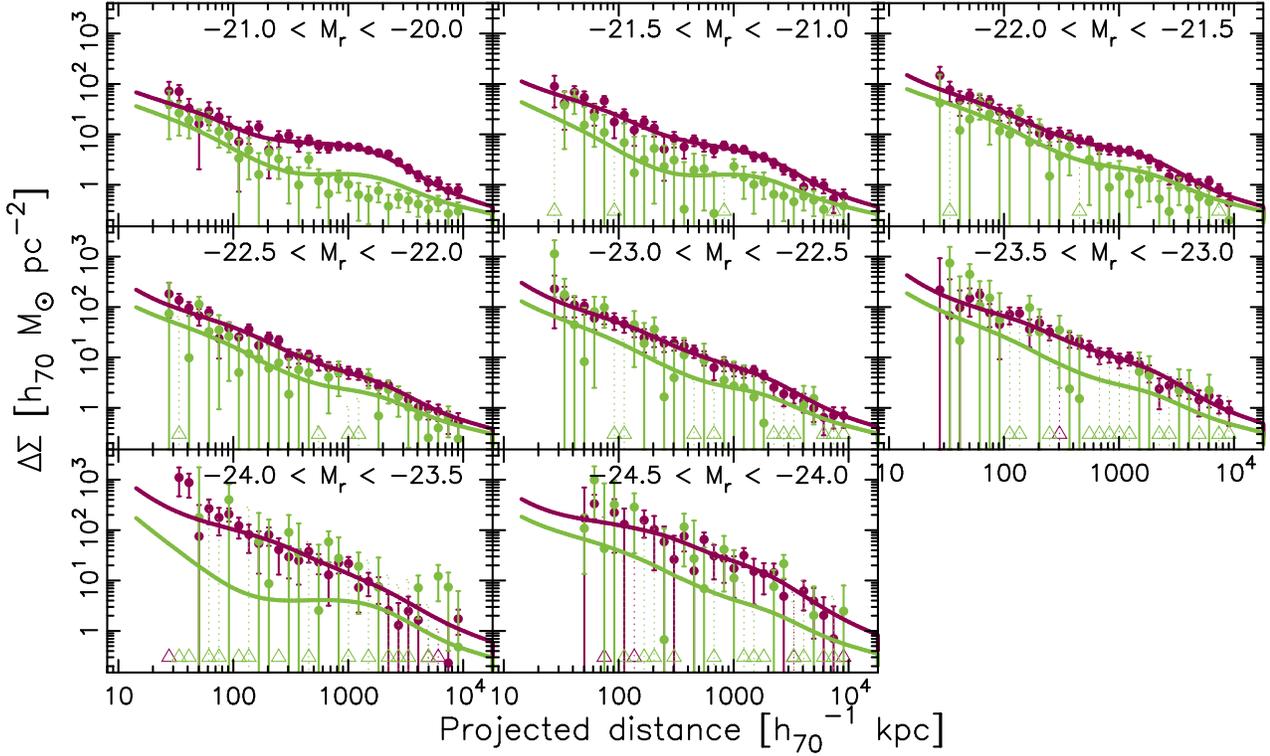}
\caption{Galaxy-galaxy lensing signal around lenses which have been split into luminosity bins according to Table \ref{cfhtls:tab:magbins1}, modelled using the halo model described in Section~\ref{cfhtls:sec:halomodel}. The dark purple (light green) dots represent the measured differential surface density, $\Delta\Sigma$, of the red (blue) lenses, and the solid line is the best-fit halo model. Triangles represent negative points that are included unaltered in the model fitting procedure, but that have here been moved up to positive values as a reference. The dotted error bars are the unaltered error bars belonging to the negative points. The squares represent distance bins containing no objects. For a detailed decomposition into the halo model components, we refer to Appendix~\ref{cfhtls:app:luminositydetails}.}
\label{cfhtls:fig:magbins}
\end{figure*}

The luminosity of a galaxy is an easily obtainable indicator of its baryonic content. To investigate the relation between dark matter halo mass and galaxy mass we therefore split the lenses into 8 bins according to MegaCam absolute $r'$-band magnitudes as detailed in Table~\ref{cfhtls:tab:magbins1} and illustrated in Figure~\ref{cfhtls:fig:magSelection}. The lens property averages quoted in this and forthcoming tables are pure averages and do not include the lensing weights, unless explicitly specified. The choice of bin limits follows the lens selection in \citetalias{vhv11}. This choice will allow us to directly compare our results to the results shown in \citetalias{vhv11} since the RCS2 data have been obtained using the same filters and telescope. We also split each luminosity bin into red and blue subsamples as described in Section~\ref{cfhtls:sec:stellarmassestimates} and proceed to measure the galaxy-galaxy lensing signal for each sample, with errors obtained via bootstrapping $10^4$ times over the full CFHTLenS area, where the number of bootstraps ensure convergence of the mean. We then fit the signal between $50\,h_{70}^{-1}\,\mathrm{kpc}$ and $2\,h_{70}^{-1}\,\mathrm{Mpc}$ with our halo model using a $\chi^2$ analysis. Only the halo mass $M_{200}$ and the satellite fraction $\alpha$ are left as free parameters while we keep all other variables fixed. When fitting, we assume that the covariance matrix of the lensing measurements is diagonal. Off-diagonal elements are generally present due to cosmic variance and shape noise, but \citet{ctm12} find that for a lens sample at a redshift range similar to that of our lenses the covariance matrix is diagonal up to $\sim$1 Mpc, which corresponds well to the largest scale we include in our fits (this is also confirmed via visual inspection of our matrices). Furthermore, Figure 7.2 from the PhD thesis of Jens R\"odiger\footnote{http://hss.ulb.uni-bonn.de/2009/1790/1790.htm} shows that the off-diagonal elements are comparatively small. Hence we do not expect that the off-diagonal elements in the $\chi^2$ fit will have a significant impact on the best-fit parameters. The results are shown in Figure~\ref{cfhtls:fig:magbins} for all luminosity bins and for each red and blue lens sample, with details of the fitted halo model parameters quoted in Table~\ref{cfhtls:tab:magbins2}. The halo masses in this table have been corrected for various contamination effects as detailed in Section~\ref{cfhtls:sec:luminosityScaling} and Appendix~\ref{cfhtls:app:corrections}. Note that the number of blue lenses in the two highest-luminosity bins, L7 and L8, is too low to adequately constrain the halo mass. In the following sections, these two blue bins have therefore been removed from the analysis of blue lenses.

\begin{table*}
\caption{Results from the halo model fit for the luminosity bins. (1) Mean luminosity for red lenses $[10^{10}\,h_{70}^{-2}\,L_{\odot}]$; (2) Mean stellar mass for red lenses $[10^{10}\,h_{70}^{-2}\,M_{\odot}]$; (3) Scatter-corrected best-fit halo mass for red lenses $[10^{11}\,h_{70}^{-1}\,M_{\odot}]$; (4) Best-fit satellite fraction for red lenses; (5) Mean luminosity for blue lenses $[10^{10}\,h_{70}^{-2}\,L_{\odot}]$; (6) Mean stellar mass for blue lenses $[10^{10}\,h_{70}^{-2}\,M_{\odot}]$; (7) Scatter-corrected best-fit halo mass for blue lenses $[10^{11}\,h_{70}^{-1}\,M_{\odot}]$; (8) Best-fit satellite fraction for blue lenses. The fitted parameters are quoted with their $1\sigma$ errors. Note that the blue results from the L7 and L8 bins are not used for fitting the power law relation in Section~\ref{cfhtls:sec:luminosityScaling}.}
\label{cfhtls:tab:magbins2}
\begin{center}
\begin{tabular}{@{}lcccccccc}
\hline
Sample 	& $\langle L_r^{\mathrm{red}}\rangle$$^{(1)}$ & $\langle M_*^{\mathrm{red}}\rangle$$^{(2)}$ & $M_h^{\rm red}$$^{(3)}$ & $\alpha^{\rm red}$$^{(4)}$ & $\langle L_r^{\mathrm{blue}}\rangle$$^{(5)}$ & $\langle M_*^{\mathrm{blue}}\rangle$$^{(6)}$ & $M_h^{\rm blue}$$^{(7)}$ & $\alpha^{\rm blue}$$^{(8)}$ \\
\hline
L1 & 0.91 & 1.83 & $5.64^{+1.62}_{-1.36}$ & $0.25^{+0.03}_{-0.03}$ & 1.08 & 0.50 & $1.73^{+0.55}_{-0.39}$ & $0.00^{+0.01}_{-0.00}$\\
L2 & 1.74 & 3.74 & $13.6^{+2.02}_{-2.29}$ & $0.14^{+0.02}_{-0.02}$ & 2.23 & 1.10 & $1.50^{+1.05}_{-0.86}$ & $0.00^{+0.01}_{-0.00}$\\
L3 & 2.73 & 5.97 & $19.4^{+3.39}_{-2.88}$ & $0.11^{+0.02}_{-0.02}$ & 3.52 & 1.83 & $8.33^{+2.40}_{-2.44}$ & $0.00^{+0.01}_{-0.00}$\\
L4 & 4.28 & 9.35 & $39.3^{+6.88}_{-5.08}$ & $0.05^{+0.03}_{-0.03}$ & 5.51 & 3.00 & $9.68^{+4.97}_{-3.85}$ & $0.00^{+0.02}_{-0.00}$\\
L5 & 6.69 & 14.9 & $60.4^{+8.96}_{-9.01}$ & $0.08^{+0.04}_{-0.04}$ & 8.44 & 4.63 & $12.7^{+10.9}_{-8.18}$ & $0.00^{+0.05}_{-0.00}$\\
L6 & 10.4 & 23.9 & $109^{+22.1}_{-18.4}$ & $0.13^{+0.07}_{-0.07}$ & 13.7 & 7.88 & $21.2^{+33.2}_{-18.9}$ & $0.00^{+0.09}_{-0.00}$\\
L7 & 16.4 & 35.6 & $309^{+54.6}_{-75.1}$ & $0.02^{+0.14}_{-0.02}$ & --- & --- & --- & --- \\
L8 & 25.4 & 20.3 & $690^{+294}_{-183}$ & $0.20^{+0.00}_{-0.20}$ & --- & --- & --- & --- \\
\hline
\end{tabular}
\end{center}
\end{table*}

As expected, the amplitude of the signal increases with luminosity for both red and blue samples indicating an increased halo mass. In general, for identical luminosity selections blue galaxies have less massive haloes than red galaxies do. For the red sample, lower luminosity bins display a slight bump at scales of \mbox{$\sim1\,h_{70}^{-1}\,\mathrm{Mpc}$}. This is due to the satellite 1-halo term becoming important and indicates that a significant fraction of the galaxies in those bins are in fact satellite galaxies inside a larger halo. On the other hand, brighter red galaxies are more likely to be located centrally in a halo. The blue galaxy halo models also display a bump for the lower luminosity bins, but this feature is at larger scales than the satellite 1-halo term. The signal breakdown shown in Figure~\ref{cfhtls:fig:magbinsBlue} (Appendix~\ref{cfhtls:app:luminositydetails}) reveals that this bump is due to the central 2-halo term arising from the contribution of nearby haloes. We note, however, that in these low-luminosity blue bins, the model overestimates the signal at projected separations greater than $\sim$2$h_{70}^{-1}$Mpc. This could be an indicator that our description of the galaxy bias, while accurate for red lenses, results in too high a bias for blue lenses. Alternatively, the discrepancy may suggest that the regime where the 1-halo term transitions into the 2-halo term is not accurately described due to inherent limitations of the halo model, such as non-linear galaxy biasing, halo exclusion representation and inaccuracies in the non-linear matter power spectrum (see Section~\ref{cfhtls:sec:halomodel}). To optimally model the regime in question, the handling of these factors should perhaps be dependent on galaxy type, but that is not done here. The reason is that we do not currently have enough data available to investigate this regime in detail. In the future, however, it should be explored further.

\subsection{Luminosity scaling relations}\label{cfhtls:sec:luminosityScaling}
\begin{figure}
\includegraphics[height=84mm,angle=270]{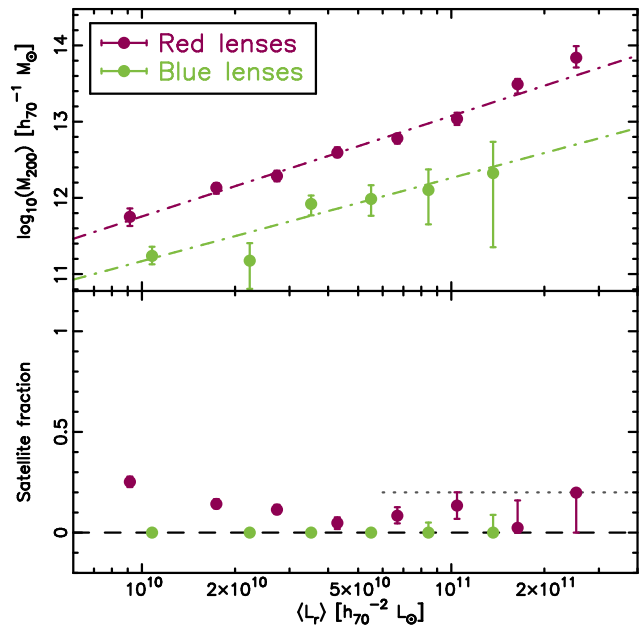}
\caption{Satellite fraction $\alpha$ and bias-corrected halo mass $M_{200}$ as a function of $r'$-band luminosity. Dark purple (light green) dots represent the results for red (blue) lens galaxies, and the dash-dotted lines show the power law scaling relations fit to the Figure~\ref{cfhtls:fig:magbins} galaxy-galaxy lensing signal (rather than to the points shown) as described in the text. The dotted line in the lower panel shows the $\alpha$ prior applied to the highest-luminosity bins.}
\label{cfhtls:fig:magbinsAlpha}
\end{figure}
\begin{figure}
\includegraphics[height=84mm,angle=270]{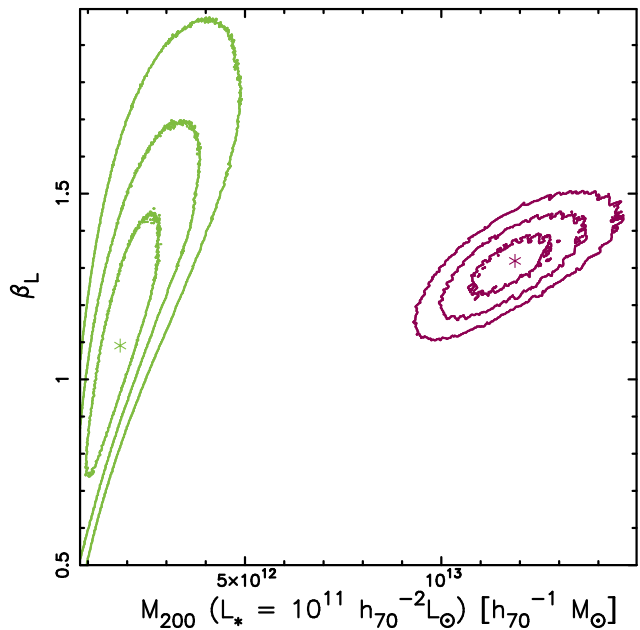}
\caption{Constraints on the power law fits shown in Figure~\ref{cfhtls:fig:magbinsAlpha}. In dark purple (light green) we show the constraints on the fit for red (blue) lenses, with lines representing the 67.8\%, 95.4\% and 99.7\% confidence limits and stars representing the best-fit value.}
\label{cfhtls:fig:lumCon}
\end{figure}

Before determining the relation between halo mass and luminosity we have to correct our raw halo mass estimates for two systematic effects. Firstly, we rely on photometric redshift estimates which do not benefit from the absolute accuracy of spectroscopic redshifts. We can therefore not be certain that a lens which is thought to be at a certain redshift is in fact at that redshift. If the redshift is different, then the derived luminosity will also be different which means that the lens may have been placed in the wrong bin. Though the lenses can scatter randomly according to their individual redshift errors, the net effect will be to scatter lenses from bins with higher abundances to those with lower abundances. The measured halo mass will therefore be biased. To correct for this effect we create mock lens catalogues and allow the objects to scatter according to their redshift error distributions. Secondly, the halo masses in a given luminosity bin will not be evenly distributed, which means that the measured halo mass does not necessarily correspond to the mean halo mass. The derivation of the factor we apply to our halo masses to correct for both these effects is detailed in Appendix~\ref{cfhtls:app:photozs}.

The estimated halo masses for all luminosity bins, corrected for the above scatter effects, are shown as a function of luminosity in the top panel of Figure~\ref{cfhtls:fig:magbinsAlpha}. Red lenses display a slightly steeper relationship between halo mass and luminosity than blue lenses, and the haloes of the blue galaxies are in general less massive for a given luminosity bin. Following \citetalias{vhv11}, we fit a power law of the form
\begin{equation}
M_{200} = M_{0,L}\left(\frac{L}{L_{\mathrm{fid}}}\right)^{\beta_L}
\label{cfhtls:eq:lumpower}
\end{equation}
with $L_{\mathrm{fid}}=10^{11}\,h_{70}^{-2}\,L_{r',\odot}$ a scaling factor chosen to be the $r'$-band luminosity of a fiducial galaxy. Rather than fitting to the final mass estimates we fit this relation directly to the lensing signals themselves (taking the scatter correction into account). We do this because the error bars are asymmetric in the former case, but the difference in results between the two fitting techniques is small.

For our red lenses we find
\mbox{$M_{0,L}=1.19^{+0.06}_{-0.07}\times10^{13}\,h_{70}^{-1}\,M_{\odot}$}
and
\mbox{$\beta_L=1.32\pm0.06$},
while for our blue lenses the corresponding numbers are
\mbox{$M_{0,L}=0.18^{+0.04}_{-0.05}\times10^{13}\,h_{70}^{-1}\,M_{\odot}$}
and
\mbox{$\beta_L=1.09^{+0.20}_{-0.13}$}.
The parameters are quoted with their $1\sigma$ errors and the constraints for these fits are shown in Figure~\ref{cfhtls:fig:lumCon}. Here we again see that the red lenses are better constrained than the blue. This is partly because we have more red lenses in most bins, and partly because red lenses in general are more massive at a given luminosity.

The mass-to-light ratios, \mbox{$M_{200}/\langle L_r\rangle$}, of our red sample range from
\mbox{$62^{+18}_{-15}\,h_{70}\,M_{\odot}\,L_{\odot}^{-1}$},
at the lowest luminosity bin to
\mbox{$90\pm13\,h_{70}\,M_{\odot}\,L_{\odot}^{-1}$}
for L5. For our blue sample the numbers are
\mbox{$16^{+5}_{-4}\,h_{70}\,M_{\odot}\,L_{\odot}^{-1}$}
for L1 and
\mbox{$15\pm2\,h_{70}\,M_{\odot}\,L_{\odot}^{-1}$}
for L5. Beyond L5, the mass-to-light ratio for red lenses continues to increase, reaching 
\mbox{$272^{+116}_{-72}\,h_{70}\,M_{\odot}\,L_{\odot}^{-1}$}
in bin L8. In these highest luminosity bins a significant fraction of the red lenses may be associated with groups or small clusters, as pointed out by \citetalias{vhv11}.

\subsection{Satellite fraction}\label{cfhtls:sec:satfrac}
The lower panel of Figure~\ref{cfhtls:fig:magbinsAlpha} shows the satellite fraction $\alpha$ as a function of luminosity for both the red and the blue sample. At lower luminosities the satellite fraction is \mbox{$\sim25\%$} for red lenses and as the luminosity increases the satellite fraction decreases. This indicates that a fair fraction of faint red lenses are satellites inside a larger dark matter halo, consistent with previous findings \citep[see][]{msk06,vhv11,ckm12}. In the highest luminosity bins the satellite fraction is difficult to constrain due to the shape of the halo model satellite terms (light green lines in Figure~\ref{cfhtls:fig:haloModelExample}) becoming indistinguishable from the central 1-halo term (dark purple dashed), as discussed in Appendix~\ref{cfhtls:app:luminositydetails}. To ensure that our halo masses are not biased low we follow \citetalias{vhv11} and apply a uniform satellite fraction prior to these bins, allowing a maximum $\alpha$ of 20\%. This prior is marked in Figure~\ref{cfhtls:fig:magbinsAlpha}. For blue lenses, the satellite fraction remains low across all luminosities indicating that almost none of our blue galaxies are satellites, again consistent with previous findings. This may be a sign that blue galaxies in our analysis are in general more isolated than red ones for a given luminosity, a theory corroborated by the low signal on large scales for blue galaxies (see Figure~\ref{cfhtls:fig:magbinsBlue} in Appendix~\ref{cfhtls:app:luminositydetails}). Here we have made no distinction between field galaxies and galaxies residing in a denser environment; for a more in-depth study of this distinction see \citet{ghe13}.

\section{Stellar mass trend}\label{cfhtls:sec:stellarmass}
\begin{figure}
\includegraphics[height=84mm,angle=270]{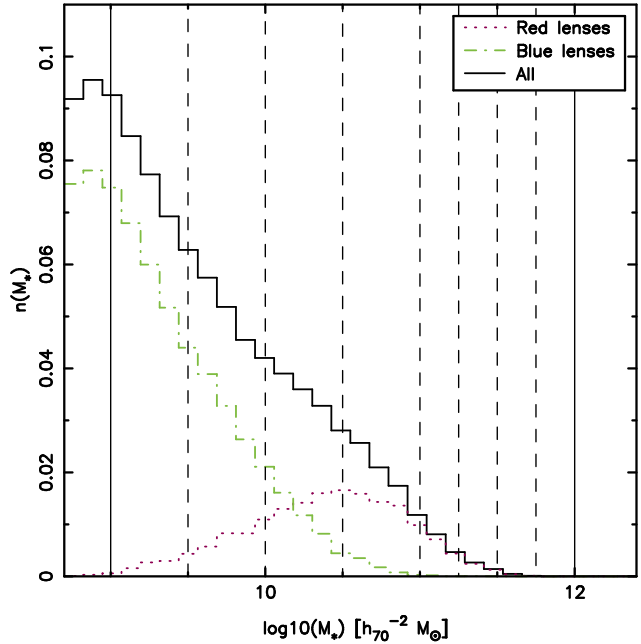}
\caption{Stellar mass distribution in the CFHTLenS catalogues for lenses with redshifts \mbox{$0.2\leq z_{\mathrm{lens}}\leq0.4$} (black solid histogram). The distribution of red (blue) lenses is shown in dotted dark purple (dot-dashed light green). Our lens bins are marked with vertical lines.}
\label{cfhtls:fig:massSelection}
\end{figure}
\begin{table}
\caption{Details of the stellar mass bins. (1) Stellar mass range $[h_{70}^{-2}\,M_{\odot}]$; (2) Number of lenses; (3) Mean redshift; (4) Fraction of lenses that are blue.}
\label{cfhtls:tab:stelbins1}
\begin{center}
\begin{tabular}{@{}lcccc}
\hline
Sample 	& $\log_{10}M_*$$^{(1)}$ & $n_{\rm lens}$$^{(2)}$ & $\langle z\rangle$$^{(3)}$ & $f_{\rm blue}$$^{(4)}$ \\
\hline
S1 & [9.00,9.50]   & 126406  & 0.33 & 0.981 \\
S2 & [9.50,10.00]  & 78283   & 0.32 & 0.828 \\
S3 & [10.00,10.50] & 48957   & 0.32 & 0.391 \\
S4 & [10.50,11.00] & 37365   & 0.32 & 0.043 \\
S5 & [11.00,11.25] & 7474    & 0.32 & 0.003 \\
S6 & [11.25,11.50] & 2447    & 0.31 & 0.001 \\
S7 & [11.50,11.75] & 396     & 0.30 & 0.000 \\
S8 & [11.75,12.00] & 12      & 0.31 & 0.000 \\
\hline
\end{tabular}
\end{center}
\end{table}
\begin{figure*}
\includegraphics[height=168mm,angle=270]{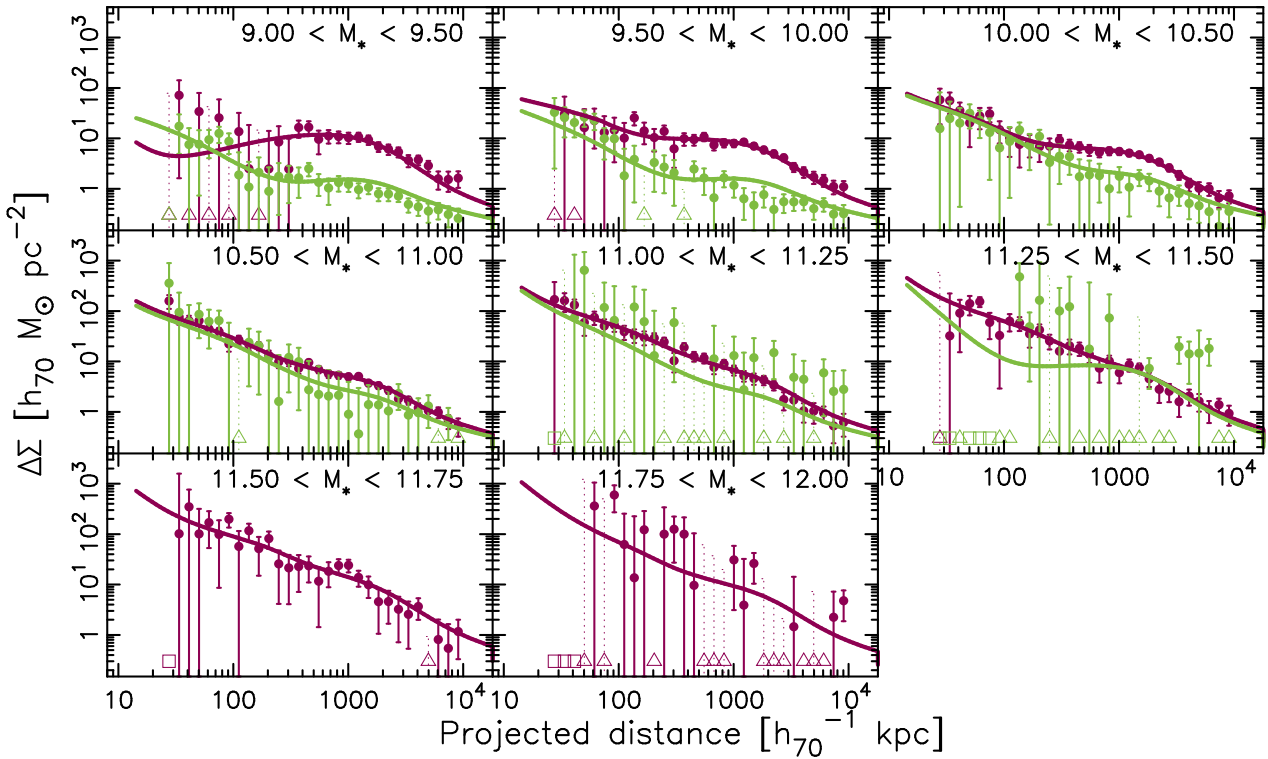}
\caption{Galaxy-galaxy lensing signal around lenses which have been split into stellar mass bins according to Table \ref{cfhtls:tab:stelbins1}, modelled using the halo model described in Section~\ref{cfhtls:sec:halomodel}. The dark purple (light green) dots represent the measured differential surface density, $\Delta\Sigma$, of the red (blue) lenses, and the solid line is the best-fit halo model. Triangles represent negative points that are included unaltered in the model fitting procedure, but that have here been moved up to positive values as a reference. The dotted error bars are the unaltered error bars belonging to the negative points. The squares represent distance bins containing no objects. For a detailed decomposition into the halo model components, we refer to Appendix~\ref{cfhtls:app:stellarmassdetails}.}
\label{cfhtls:fig:stelbins}
\end{figure*}

The galaxy luminosity as a tracer of baryonic content depends both on age and on star formation history. A galaxy's stellar mass does not have such dependence and may therefore be a better indicator of its baryonic content. In this section we study the relation between galaxy stellar mass and the dark matter halo mass, dividing the lenses into 9 stellar mass bins as illustrated in Figure~\ref{cfhtls:fig:massSelection} with details in Table~\ref{cfhtls:tab:stelbins1}. As we did for the luminosity analysis (Section~\ref{cfhtls:sec:luminosity}) we further split each stellar mass bin into a red and a blue sample using their photometric types to approximate early- and late-type galaxies.

We measure the galaxy-galaxy lensing signal for each sample as before, and fit on scales between $50\,h_{70}^{-1}\,\mathrm{kpc}$ and $2\,h_{70}^{-1}\,\mathrm{Mpc}$ using our halo model with the halo mass $M_{200}$ and the satellite fraction $\alpha$ as free parameters. Similarly to the previous section, the results are shown in Figure~\ref{cfhtls:fig:stelbins} for all stellar mass bins and for each red and blue lens sample, with details of the fitted halo model parameters quoted in Table~\ref{cfhtls:tab:stelbins2}. There are no blue lenses available in the two highest stellar mass bins, and in bins S5 and S6 the number of blue lenses is too low to constrain the signal. We therefore remove them from our analysis in the following sections.

\begin{table*}
\caption{Results from the halo model fit for the stellar mass bins. (1) Mean luminosity for red lenses $[10^{10}\,h_{70}^{-2}\,L_{\odot}]$; (2) Mean stellar mass for red lenses $[10^{10}\,h_{70}^{-2}\,M_{\odot}]$; (3) Scatter-corrected best-fit mean halo mass for red lenses $[10^{11}\,h_{70}^{-1}\,M_{\odot}]$; (4) Best-fit satellite fraction for red lenses; (5) Mean luminosity for blue lenses $[10^{10}\,h_{70}^{-2}\,L_{\odot}]$; (6) Mean stellar mass for blue lenses $[10^{10}\,h_{70}^{-2}\,M_{\odot}]$; (7) Scatter-corrected best-fit mean halo mass for blue lenses $[10^{11}\,h_{70}^{-1}\,M_{\odot}]$; (8) Best-fit satellite fraction for blue lenses. The fitted parameters are quoted with their $1\sigma$ errors. Note that the red results from the S1 and S2 bins, and the blue results from the S5 and S6 bins, are not used for fitting the power law relation in Section~\ref{cfhtls:sec:stellarmassScaling}.}
\label{cfhtls:tab:stelbins2}
\begin{center}
\begin{tabular}{@{}lcccccccc}
\hline
Sample 	& $\langle L_r^{\mathrm{red}}\rangle$$^{(1)}$ & $\langle M_*^{\mathrm{red}}\rangle$$^{(2)}$ & $M_h^{\rm red}$$^{(3)}$ & $\alpha^{\rm red}$$^{(4)}$ & $\langle L_r^{\mathrm{blue}}\rangle$$^{(5)}$ & $\langle M_*^{\mathrm{blue}}\rangle$$^{(6)}$ & $M_h^{\rm blue}$$^{(7)}$ & $\alpha^{\rm blue}$$^{(8)}$ \\
\hline
S1 & 0.22 & 0.24 & $0.03^{+1.90}_{-0.02}$ & $0.92^{+0.08}_{-0.28}$ & 0.41 & 0.18 & $1.28^{+0.41}_{-0.33}$ & $0.00^{+0.00}_{-0.00}$\\
S2 & 0.44 & 0.66 & $5.68^{+2.16}_{-1.84}$ & $0.41^{+0.04}_{-0.04}$ & 1.11 & 0.54 & $2.00^{+0.64}_{-0.62}$ & $0.00^{+0.00}_{-0.00}$\\
S3 & 1.06 & 1.97 & $5.81^{+1.67}_{-1.20}$ & $0.23^{+0.02}_{-0.02}$ & 2.87 & 1.59 & $9.14^{+2.37}_{-1.88}$ & $0.00^{+0.01}_{-0.00}$\\
S4 & 2.46 & 5.64 & $26.3^{+3.23}_{-2.88}$ & $0.11^{+0.02}_{-0.02}$ & 7.07 & 4.27 & $26.8^{+11.0}_{-10.3}$ & $0.00^{+0.02}_{-0.00}$\\
S5 & 5.38 & 13.0 & $81.2^{+12.1}_{-8.91}$ & $0.10^{+0.03}_{-0.03}$ & --- & --- & --- & --- \\
S6 & 8.96 & 22.6 & $160^{+28.3}_{-24.2}$ & $0.10^{+0.05}_{-0.05}$ & --- & --- & --- & --- \\
S7 & 14.3 & 38.6 & $388^{+90.7}_{-67.1}$ & $0.20^{+0.00}_{-0.09}$ & --- & --- & --- & --- \\
S8 & 19.1 & 62.7 & $174^{+353}_{-167}$ & $0.20^{+0.00}_{-0.20}$ & --- & --- & --- & --- \\
\hline
\end{tabular}
\end{center}
\end{table*}
The mean mass in each bin increases with increasing stellar mass as expected, resulting in an increased signal amplitude. Similar to the luminosity samples in the previous section, the red lower-mass bins display a bump at scales of \mbox{$\sim0.5\,h^{-1}\,\mathrm{Mpc}$}. Here the lowest bins contain less massive galaxies than the lowest luminosity bins and the bump is more pronounced, indicating that most of the galaxies in these low-mass samples are satellite galaxies. The contribution from nearby haloes is again clearly visible in the best-fit halo model of the lower-mass blue samples, though as noted in Section~\ref{cfhtls:sec:luminosity}, this may be due to an inaccurate galaxy bias description for blue lenses.

\subsection{Stellar mass scaling relations}\label{cfhtls:sec:stellarmassScaling}
\begin{figure}
\includegraphics[height=84mm,angle=270]{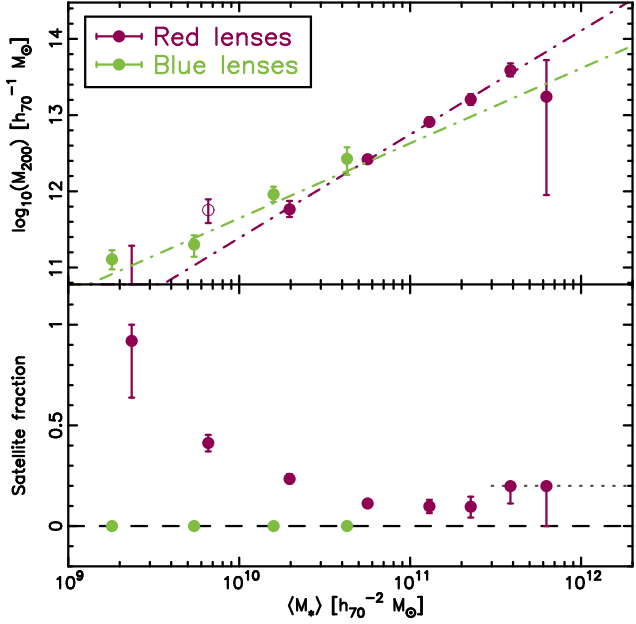}
\caption{Satellite fraction $\alpha$ and halo mass $M_{200}$ as a function of stellar mass. Dark purple (light green) dots represent the results for red (blue) lens galaxies. Open circles show the points that have been excluded from the power law fit because of a high satellite fraction. The dotted line in the lower panel shows the $\alpha$ prior applied to the highest-stellar mass bins.}
\label{cfhtls:fig:stelbinsAlpha}
\end{figure}
\begin{figure}
\includegraphics[height=84mm,angle=270]{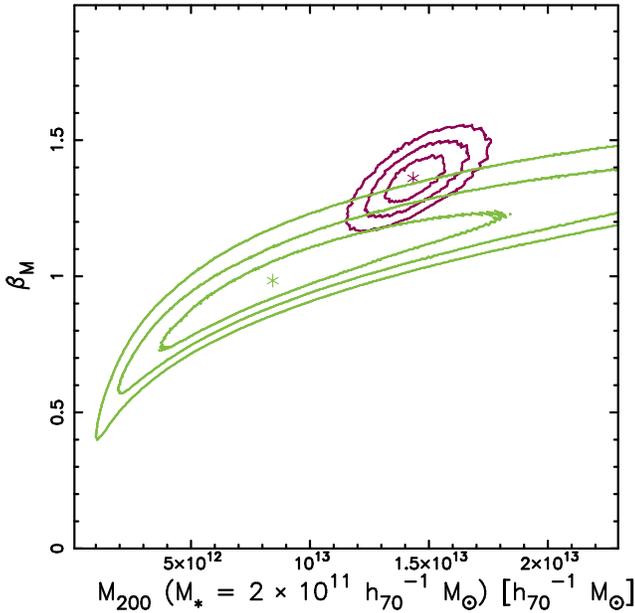}
\caption{Constraints on the power law fits shown in Figure~\ref{cfhtls:fig:stelbinsAlpha}. In dark purple (light green) we show the constraints on the fit for red (blue) lenses, with lines representing the 67.8\%, 95.4\% and 99.7\% confidence limits and stars representing the best-fit value.}
\label{cfhtls:fig:massCon}
\end{figure}

Just as for the luminosity bins, we have to correct the halo mass estimates for two scatter effects: one due to errors in the stellar mass estimates and another due to halo masses not being evenly distributed within a given bin. We describe the correction for these effects in Appendix~\ref{cfhtls:app:binscatter}. The best-fit halo masses, once corrected for these scatter effects, and satellite fractions $\alpha$ for each stellar mass bin are shown in Figure~\ref{cfhtls:fig:stelbinsAlpha}. In the lowest-mass bin, nearly all red lenses are satellites while for higher masses, the majority are located centrally in their halo. As discussed in Section~\ref{cfhtls:sec:satfrac}, this fraction is difficult to constrain for high masses due to the shape of the halo model satellite terms. We therefore apply the same uniform satellite fraction prior to the high-stellar mass bins as we did to the high-luminosity bins, allowing a maximum $\alpha$ of 20\%. The overall low satellite fraction for blue galaxies, suggesting together with low large-scale signal that most blue galaxies are isolated, is consistent with the luminosity results.

To quantify the difference in the relation between dark matter halo and stellar mass between red and blue lenses, we fit a power law to the lensing signals in each bin simultaneously, similarly to our treatment of the luminosity bins in the previous section. The form of the power law is
\begin{equation}
M_{200} = M_{0,M}\left(\frac{M_*}{M_{\mathrm{fid}}}\right)^{\beta_M}
\end{equation}
with $M_{\mathrm{fid}} = 2\times10^{11}\,h_{70}^{-2}\,M_{\odot}$ a scaling factor chosen to be the stellar mass of a fiducial galaxy as in \citetalias{vhv11}. We note that for the lowest red stellar mass bins, though the halo model fits the data very well (see Figure~\ref{cfhtls:fig:stelbins}), the sample consists largely of satellite galaxies as mentioned above. The central halo mass associated with these lenses is therefore effectively inferred from the satellite term, and thus constrained indirectly by the halo model and so we exclude the two lowest stellar mass bins from our analysis.

The resulting best-fit values for red lenses are
\mbox{$M_{0,M}=1.43^{+0.11}_{-0.08}\times10^{13}\,h_{70}^{-1}\,M_{\odot}$}
and
\mbox{$\beta_M=1.36^{+0.06}_{-0.07}$},
and for blue lenses
\mbox{$M_{0,M}=0.84^{+0.20}_{-0.16}\times10^{13}\,h_{70}^{-1}\,M_{\odot}$}
and
\mbox{$\beta_M=0.98^{+0.08}_{-0.07}$}.
We show the constraints and best-fit values in Figure~\ref{cfhtls:fig:massCon}. The red lenses are clearly better constrained than the blue ones due to the stronger signal generated by these generally more massive galaxies. We note here that due to a lack of massive blue lenses in our analysis, the two galaxy type results probe different stellar mass ranges. The blue relation is limited to the low-stellar mass end only, while the red relation is constrained mostly at higher stellar masses.

The baryon fraction, \mbox{$M_*/M_{200}$}, is fairly constant between stellar mass bins though it shows a tendency to decrease for red lenses from 
\mbox{$0.034^{+0.010}_{-0.007}$}
for S3 to
\mbox{$0.010\pm0.002$}
for S7. For blue lenses it conversely shows a slight increase from 
\mbox{$0.014^{+0.892}_{-0.009}$}
for S1 to
\mbox{$0.016\pm0.002$}
for S4. These numbers are indicators of the baryon conversion efficiency, though the particular environment each sample resides in affects the numbers. Since the red and blue samples probe different stellar mass ranges, we cannot directly compare the two.

\section{Comparison with previous results}\label{cfhtls:sec:comparison}

\begin{figure*}
\includegraphics[height=168mm,angle=270]{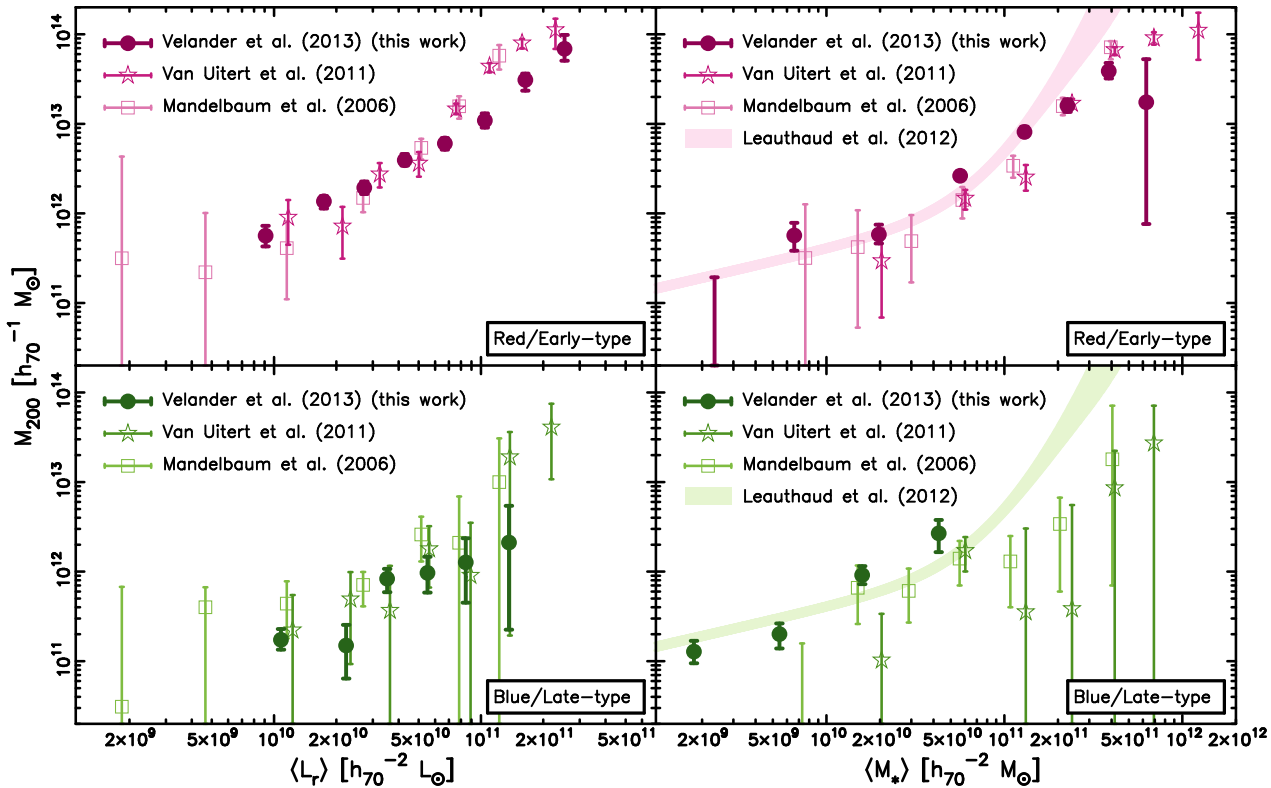}
\caption{Comparison between four different datasets. The left (right) panels show the measured halo mass as a function of luminosity (stellar mass), and the top (bottom) panels show the results for red/early-type (blue/late-type) galaxies. The datasets used are all based on galaxy-galaxy lensing analyses with solid dots showing the CFHTLenS results from this paper. Also shown are halo masses measured using the RCS2 \citepalias[open stars;][]{vhv11}, the SDSS \citep[open squares][]{msk06} and COSMOS \citep[solid band;][]{ltb12}. In the case of COSMOS we use the results from their lowest redshift bin. Also note that no distinction between red and blue lenses was made in the COSMOS analysis, so the same results are shown in both right panels.}
\label{cfhtls:fig:mainComparison1}
\end{figure*}

\begin{figure}
\includegraphics[height=84mm,angle=270]{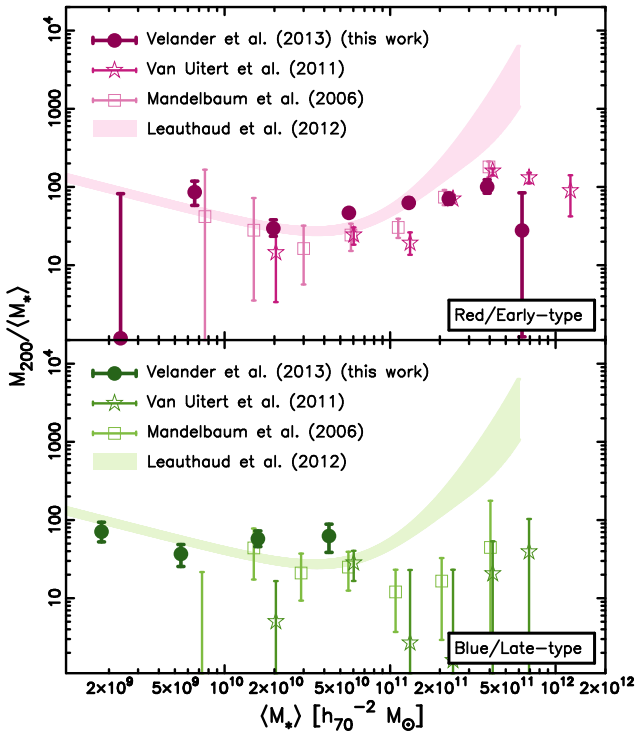}
\caption{Comparison between four different datasets, showing the ratio of measured halo mass to stellar mass as a function of stellar mass. The top (bottom) panels show the results for red/early-type (blue/late-type) galaxies. The datasets used are all based on galaxy-galaxy lensing analyses with solid dots showing the CFHTLenS results from this paper. Also shown are halo masses measured using the RCS2 \citepalias[open stars;][]{vhv11}, the SDSS \citep[open squares][]{msk06} and COSMOS \citep[solid band;][]{ltb12}. In the case of COSMOS we use the results from their lowest redshift bin. Also note that no distinction between red and blue lenses was made in the COSMOS analysis, so the same results are shown in both panels.}
\label{cfhtls:fig:mainComparison2}
\end{figure}

Early galaxy-galaxy lensing based works that have investigated the relation between luminosity and the virial mass of galaxies include \citet{guz02} and \citet{hhy05}. In these works, the mass is found to scale with luminosity as $\propto$$L^{1.4\pm0.2}$ and $\propto$$L^{1.6\pm0.2}$, respectively, in agreement with our findings. We focus, however, on comparing our halo mass results with those from three recent comprehensive galaxy-galaxy lensing halo model analyses which used data from three decidedly different surveys: the very wide but shallow SDSS \citep{msk06}, the moderately deep and wide RCS2 \citepalias{vhv11} and the very deep but narrow COSMOS \citep{ltb12}. All four datasets are shown in Figures~\ref{cfhtls:fig:mainComparison1} and \ref{cfhtls:fig:mainComparison2}, with our results denoted by solid dots.

We begin our comparison by noting that the various works employ different halo models, so we urge the reader to keep the study of the impact of different modelling choices in mind (see Appendix~\ref{cfhtls:app:assumptions}). Furthermore, they use different galaxy type separation criteria. \citet{msk06} and \citetalias{vhv11} base their selection on the brightness profile of the lenses, while we use the SED type. As both selection criteria are found to correlate well with the colours of the lenses, we expect the galaxy samples to be similar --- but not identical --- and the differences between the samples could have some effect. \citet{ltb12} did not split their sample in red and blue, which is why we show the same constraints in both panels of Figures~\ref{cfhtls:fig:mainComparison1} and \ref{cfhtls:fig:mainComparison2}. Further variations between the analyses are discussed in more detail below. With these caveats in mind, we observe that all studies find similar general trends, with a halo mass that increases with increasing luminosity and/or stellar mass.  It is also clear that blue/late-type galaxies tend to reside in haloes of lower mass than red/early-types do. The halo mass estimates of blue galaxies presented in these studies are in excellent agreement. For the red galaxies, our mass estimates are consistent with those from \citetalias{vhv11} and \citet{msk06} except near \mbox{$L_r\sim10^{11}\,h_{70}^{-2}\,L_{\odot}$}, where they are 2-3$\sigma$ lower. However, as a function of stellar mass, our mass estimates of early-types broadly agree with theirs. The halo masses of early-types also agree with the results from \citet{ltb12} at stellar masses below \mbox{$M_*\sim10^{11}\,h_{70}^{-2}\,M_{\odot}$}. At higher stellar masses, the mass estimates are $\sim$2$\sigma$ lower than those from \citet{ltb12}, but we note that this is also the case for the halo masses from \citetalias{vhv11} and \citet{msk06}. We will discuss this in more detail below. In general, a consistent picture of the relation between the baryonic properties of galaxies and their parent haloes is emerging from the four independent studies.

Since our halo model is most closely related to that used by \citetalias{vhv11} (shown as open stars in Figures~\ref{cfhtls:fig:mainComparison1} and \ref{cfhtls:fig:mainComparison2}), a detailed comparison is more straight-forward compared to the other analyses. In \citetalias{vhv11}, $1.7\times10^4$ lens galaxies were studied using the overlap between the SDSS and the RCS2. The combination of the two surveys allowed for accurate baryonic property estimates using the spectroscopic information from the SDSS, and a high source number density of $6.3\,\mathrm{arcmin}^{-2}$ owing to the greater depth and better observing conditions of the RCS2 compared to the SDSS. Because we use photometric redshifts for our analysis our lens sample is more than sixty times that of \citetalias{vhv11}, reflecting the small fraction of galaxies that have spectroscopic redshifts determined by SDSS. The even greater depth of the CFHTLenS compared to the RCS2 means that our source density is a factor of 1.7 higher. Furthermore, in contrast to \citetalias{vhv11} we have individual redshift estimates available for all our sources. The increased number density and redshift resolution in our analysis results in significantly tighter constraints on the relations between halo mass and luminosity, and between halo mass and stellar mass.

As evidenced by Figure~\ref{cfhtls:fig:mainComparison1}, our halo masses agree well with those found by \citetalias{vhv11} in general, though our halo mass relations are shallower; for red lenses we measure a power law slope for the relation between halo mass and luminosity of
\mbox{$1.32\pm0.06$},
and between halo mass and stellar mass of
\mbox{$1.36^{+0.06}_{-0.07}$},
while \citetalias{vhv11} find slopes\footnote{The RCS2 halo masses shown in Figures~\ref{cfhtls:fig:mainComparison1} and~\ref{cfhtls:fig:mainComparison2}, and the power law slopes quoted in the text have been updated since the publication of \citetalias{vhv11} to account for an issue with the halo modelling software. The issue was discovered and resolved during the preparation of this paper. We note that the change to the RCS2 results is within their reported observational uncertainties.} of \mbox{$2.2\pm0.1$}, and \mbox{$1.8\pm0.1$}, respectively, using the same power law definitions. The general trend with stellar mass of a decreasing baryon conversion efficiency for red lenses was observed by \citetalias{vhv11} as well, but they were unable to discern a trend in their late-type sample. There are some differences between the analyses which should be noted, however. As mentioned above, we divide our lens sample in a red and blue one based on the SED type, while \citetalias{vhv11} use the brightness distribution profiles to separate their lenses in a bulge-dominated and a disk-dominated sample. Even though the resulting samples are expected to be fairly similar, they are not identical. As the mass-to-luminosity ratio of galaxies strongly depends on their colour, even small colour differences between the samples could result in different masses. This may explain why our halo mass estimates of the red lenses at the high luminosity end are lower than those of \citetalias{vhv11} and \citet{msk06}, who both use identical galaxy type separation criteria and whose masses agree in this regime. The difference is smaller for the stellar mass results, providing further support for this hypothesis. Furthermore, in our halo model we account for the baryonic mass of each lens, something that was not done in \citetalias{vhv11}. As shown in Appendix~\ref{cfhtls:app:assumptions}, however, the slope and amplitude of our power laws do not change significantly when the baryonic component is removed. Hence this does not explain why \citetalias{vhv11} find a steeper slope than we do.

Another factor to take into account is the fact that we limit our lens samples to redshifts of $0.2\le z_{\mathrm{lens}}\le 0.4$ keeping our mean lens redshift fairly stable at \mbox{$\langle z_{\mathrm{lens}}\rangle\sim0.3$}. This is not done in \citetalias{vhv11}, and as a result the median redshift of our lower luminosity or stellar mass bins is higher than for the same bins in \citetalias{vhv11}, with the opposite being true for the higher bins. Recent numerical simulations indicate that the relation between stellar mass and halo mass will evolve with redshift \citep[see for example][]{con09,msm10}. Lower-mass host galaxies ($M_*<10^{11}\,M_{\odot}$) increase in stellar mass faster than their halo mass increases, i.e.~for higher redshifts the halo mass is lower for the same stellar mass. The opposite trend holds for higher-mass host galaxies ($M_*>10^{11}\,M_{\odot}$). As a result, the relation between halo mass and stellar mass (or an indicator thereof, such as luminosity) steepens with increasing redshift. This means that for the lower-luminosity bins, where our redshifts are higher, we may measure a steeper slope than \citetalias{vhv11} and vice-versa for higher-luminosity bins. The effect is likely small, however, because of the relatively small redshift ranges involved.

Finally we note that the lenses in the sample studied by \citetalias{vhv11} are rather massive and luminous as only galaxies with spectroscopy are used. Our lens sample includes many more low luminosity and low stellar mass objects, however. Hence the difference in slope may be partly due to the fact that we probe different regimes, and that the relation between baryonic observable and halo mass is not simply a power law but turns upward at high luminosities/stellar masses, as the results from \citet{ltb12} suggest.

Having compared our analysis to that of \citetalias{vhv11}, we now turn our attention to the comparison with the \citet{msk06} analysis of $3.5\times10^5$ lenses in the SDSS DR4, shown as open squares in Figures~\ref{cfhtls:fig:mainComparison1} and \ref{cfhtls:fig:mainComparison2}. Their lens sample is, similarly to the \citetalias{vhv11} sample, also divided into early- and late-type galaxies based on their brightness profiles. To allow for a comparison between our results and theirs we first have to consider the differences in the luminosity definition. \citet{msk06} use absolute magnitudes which are based on a K correction to a redshift of $z=0.1$ and a distance modulus calculated using $h=1.0$. Furthermore, their luminosities are corrected for passive evolution by applying a factor $1.6(z-0.1)$. However, \citetalias{vhv11} convert their luminosities, which are similar to ours, using the \citet{msk06} definition and find that for low-luminosity low-redshift samples the difference between the two definitions is negligible. The more luminous lenses reside at higher redshifts and for them the correction is found to be greater, most likely due to the difference in the passive evolution corrections. Since our lenses are confined to relatively low redshifts, and since the main difference between luminosity definitions is the passive evolution factor, we can compare our results to \citet{msk06} without correcting the luminosities. Our halo mass definition is also different to that used by \citet{msk06} though. 
\citet{msk06} define the mass within the radius where the density is 180 times the mean background density while we set it to be 200 times the critical density. The correction factor stemming from the different definitions amounts to \mbox{$\sim30\%$}. Having corrected for this, our results are then very similar to those from \citet{msk06}, but the same concerns of object selection and baryonic contribution discussed above apply here as well. The relation that \citet{msk06} find between halo mass and luminosity for red lenses is shallower than the one found by \citetalias{vhv11}, as discussed therein, and are therefore more in agreement with our results. For the stellar mass relation, however, they find a steeper power law slope, though this result is mostly driven by their highest stellar mass bin as pointed out by \citetalias{vhv11}.

Finally, \citet{ltb12} perform a combined analysis of galaxy-galaxy lensing, galaxy clustering and galaxy number densities using data from the COSMOS survey, shown as a solid band in the right panels of Figure~\ref{cfhtls:fig:mainComparison1} and in Figure~\ref{cfhtls:fig:mainComparison2}. For our comparison we select the results from their lowest redshift bin, since its redshift range of $0.22<z<0.48$ is very similar to the redshift range used here. Contrary to the other datasets, \citet{ltb12} did not separate their lens sample according to galaxy type. The results shown in the top panel of Figures~\ref{cfhtls:fig:mainComparison1} and \ref{cfhtls:fig:mainComparison2} are therefore identical to those shown in the bottom panel. Note that at high stellar masses, their sample is expected to be dominated by red galaxies, and at low stellar masses by blue galaxies, as these are generally more abundant in the respective regimes (see Table \ref{cfhtls:tab:stelbins1}). For stellar masses lower than $10^{11}\,h_{70}^{-1}M_*$, the agreement between \citet{ltb12} and the other galaxy-galaxy lensing results is excellent for both galaxy types. For higher stellar masses, however, \citet{ltb12} find higher halo masses than what has been observed in the lensing only analyses discussed above. This may be explained if a larger fraction of the galaxies used in the \citet{ltb12} analysis reside in dense environments and can be associated with galaxy groups and clusters such that their halo masses correspond to the total mass of these structures. This theory is corroborated by Figure 10 of \citet{ltb12} which shows that for large stellar masses, the ratio of stellar mass to halo mass is very similar to that determined for a set of X-ray luminous clusters in \citet{hoe07}, indicating that we are entering the cluster regime. Furthermore, the sampling variance is not taken into account in the COSMOS error range. This is likely to affect the higher stellar mass bins more because the number of objects there is sparse. Additionally, the results from the COSMOS analysis of X-ray selected groups presented in \citet{lfk10}, which is centred on a redshift similar to ours and also shown in Figure 10 of \citet{ltb12} as grey squares, agree better with our results for higher stellar masses. We note, however, that another possibility is that the high stellar mass end constraints from \citet{ltb12} may be driven mainly by the stellar mass function rather than by the lensing measurements. This, combined with the differences in the two halo model implementations, could also contribute to the observed discrepancy.

A further subtlety discussed in Section \ref{cfhtls:sec:satfrac} is that the satellite fraction of galaxies with high stellar masses is not well constrained by galaxy-galaxy lensing only. Since the satellite fraction and halo mass are weakly anti-correlated \citepalias[see][]{vhv11}, our halo masses may be slightly underestimated if the satellite fractions are too high. Furthermore, the modelling of the shear signal from satellites in this mass range is a bit uncertain as they may have been stripped of more than the 50\% of their dark matter we have assumed so far, and this could also have some effect. However, we estimate that these modelling uncertainties only have a small effect on our best-fit halo masses, and that it is not sufficient to explain the differences between the results.

\section{Conclusion}\label{cfhtls:sec:conclusion}
In this work we have used high-quality weak lensing data produced by the CFHTLenS collaboration to place galaxy-galaxy lensing constraints on the relation between dark matter halo mass and the baryonic content of the lenses, quantified through luminosity and stellar mass estimates. The combination of large area and high source number density in this survey has made it possible to achieve tighter constraints compared to previous lensing surveys such as the SDSS, COSMOS or the RCS2. We also extended our study to lower stellar masses than have been studied before using a halo model such as the one described here.

In this paper we have included a halo model constituent which was neglected by most earlier implementations: the baryonic component. Since the lensing signal is a response to the total mass of a system, it is essential to account for baryons in order to not overestimate the mass contained in the dark matter halo. We have shown, however, that care has to be taken when including a baryonic component since doing so has a greater impact on the fitted halo mass than one might na\"{\i}vely expect due to the complicated interplay between stellar mass, satellite fraction and halo mass.

As luminosity and stellar mass increases, the halo mass increases as well. For red lenses, the halo mass increases with greater baryonic content at a higher rate than for blue galaxies, independent of whether the measure of baryonic content is luminosity or stellar mass. The two measures thus produce comparable results. For each we fit power law relations to quantify the rate of increase in halo mass. We find a best-fit slope of
\mbox{$1.32\pm0.06$}
and a normalisation of
\mbox{$1.19^{+0.06}_{-0.07}\times10^{13}\,h_{70}^{-1}\,M_{\odot}$}
for a fiducial luminosity of \mbox{$L_{\mathrm{fid}}=10^{11}\,h_{70}^{-2}L_{\odot}$} for red galaxies, while for blue galaxies we find a slope of
\mbox{$1.09^{+0.20}_{-0.13}$}
and a normalisation
\mbox{$0.18^{+0.04}_{-0.05}\times10^{13}\,h_{70}^{-1}\,M_{\odot}$}.
The power law relation between stellar mass and halo mass has a slope of 
\mbox{$1.36^{+0.06}_{-0.07}$}
and a normalisation of
\mbox{$1.43^{+0.11}_{-0.08}\times10^{13}\,h_{70}^{-1}\,M_{\odot}$}
for a fiducial mass of \mbox{$M_{\mathrm{fid}}=2\times10^{11}\,h_{70}^{-2}\,M_{\odot}$} for red galaxies, and for blue galaxies we find a slope of
\mbox{$0.98^{+0.08}_{-0.07}$}
and a normalisation of
\mbox{$0.84^{+0.20}_{-0.16}\times10^{13}\,h_{70}^{-1}\,M_{\odot}$}.

For our blue galaxy selection, the satellite fraction is low across all luminosities and stellar masses considered here. The signal at large scales for these samples is also generally low in the lowest luminosity and stellar mass bins, indicating that these galaxies are relatively isolated and reside in less clustered environments than the red galaxies do and that we may be overestimating the galaxy bias for these samples. At low luminosity/stellar mass, a considerable fraction of red galaxies are satellites within a larger dark matter halo. This fraction decreases steadily with increasing luminosity or stellar mass. In general, the satellite fractions show that at these redshifts the galaxies in denser regions are mostly red while for the same luminosity or stellar mass isolated galaxies tend to be bluer and thus star forming. This indicates that the star formation history of galaxies differs depending on the density of the environment they are residing in.

Another finding of this work is that for faint and low stellar mass blue galaxies, the amplitude of the lensing signal at projected separations larger than $\sim$2$h_{70}^{-1}$Mpc is lower than the corresponding best-fit halo model. For the red galaxies, the halo model fits the data well over all scales. This could indicate that while the bias description works well for red galaxies, it is not optimal for blue galaxies. If this is the case, then the environments in which the two samples reside are radically different and the difference will have to be taken into account in the future. Alternatively, the discrepancy could be caused by other choices that affect the 1-halo to 2-halo transition regime in our halo model implementation. Currently, we do not have enough data to favour or rule out either scenario, but we plan to explore this further in upcoming works.

The relations between baryonic content indicators and dark matter halo mass presented in this work, as well as the dependence of the satellite fraction on luminosity and stellar mass, improve our understanding of the mechanisms behind galaxy formation since they provide constraints that can be directly compared to numerical simulations that model different galaxy formation scenarios. With currently ongoing \citep[for instance DES\footnote{www.darkenergysurvey.org} or KiDS;][]{dkk13} and planned \citep[such as LSST\footnote{www.lsst.org}, HSC\footnote{www.subarutelescope.org/Projects/HSC} or Euclid\footnote{www.euclid-ec.org};][]{laa11} surveys, weak lensing analyses will become yet more powerful than the one presented in this paper. In preparation for the future there are therefore several sources of uncertainty that should be investigated. As mentioned above, the galaxy bias description may not be optimal for blue lenses and with future data this bias can likely be constrained directly using galaxy-galaxy lensing observations. Recent simulations have also indicated that there is a redshift evolution of the halo mass relations, and this evolution can be studied with weak lensing \citep[see][]{ctm12,hud12}. Other possible improvements to the halo model used here include studies of the distribution of satellites within a galaxy dark matter halo, a more accurate description of the regime where the 1-halo term and 2-halo term overlap (i.e. halo exclusion), and investigations into the stripping of satellite haloes. The analysis presented in this paper is a significant advance on recent analyses, and with future surveys we will be able to use galaxy-galaxy lensing to study the connection between baryons and dark matter in exquisite detail.

\section*{Acknowledgments}
We thank R. Bielby, O. Ilbert and the TERAPIX team for making the WIRDS stellar mass catalogue available to us, and Peter Schneider for valuable comments on the manuscript. Additionally, we thank the anonymous referee for their insightful comments which helped improve this paper and ensure its robustness.

This work is based on observations obtained with MegaPrime/MegaCam, a joint project of CFHT and CEA/DAPNIA, at the Canada-France-Hawaii Telescope (CFHT) which is operated by the National Research Council (NRC) of Canada, the Institut National des Sciences de l'Univers of the Centre National de la Recherche Scientifique (CNRS) of France, and the University of Hawaii. This research used the facilities of the Canadian Astronomy Data Centre operated by the National Research Council of Canada with the support of the Canadian Space Agency. We thank the CFHT staff for successfully conducting the CFHTLS observations and in particular Jean-Charles Cuillandre and Eugene Magnier for the continuous improvement of the instrument calibration and the Elixir detrended data that we used. We also thank TERAPIX for the quality assessment and validation of individual exposures during the CFHTLS data acquisition period, and Emmanuel Bertin for developing some of the software used in this study. CFHTLenS data processing was made possible thanks to significant computing support from the NSERC Research Tools and Instruments grant program, and to HPC specialist Ovidiu Toader. The early stages of the CFHTLenS project was made possible thanks to the support of the European Commission's Marie Curie Research Training Network DUEL (MRTN-CT-2006-036133) which directly supported six members of the CFHTLenS team (LF, HH, PS, BR, CB, MV) between 2007 and 2011 in addition to providing travel support and expenses for team meetings.

MV acknowledges support from the European DUEL Research-Training Network (MRTN-CT-2006-036133), from the Netherlands Organization for Scientific Research (NWO) and from the Beecroft Institute for Particle Astrophysics and Cosmology.
H.~Hoekstra acknowledges support from  Marie Curie IRG grant 230924, the Netherlands Organisation for Scientific Research (NWO) grant number 639.042.814 and from the European Research Council under the EC FP7 grant number 279396.
TE is supported by the Deutsche Forschungsgemeinschaft through project ER 327/3-1 and the Transregional Collaborative Research Centre TR 33 - "The Dark Universe".
CH acknowledges support from the European Research Council under the EC FP7 grant number 240185.
H.~Hildebrandt is supported by the Marie Curie IOF 252760, a CITA National Fellowship, and the DFG grant Hi 1495/2-1.
TDK acknowledges support from a Royal Society University Research Fellowship.
YM acknowledges support from CNRS/INSU (Institut National des Sciences de l'Univers) and the Programme National Galaxies et Cosmologie (PNCG).
LVW acknowledges support from the Natural Sciences and Engineering Research Council of Canada (NSERC) and the Canadian Institute for Advanced Research (CIfAR, Cosmology and Gravity program).
LF acknowledges support from NSFC grants 11103012 \& 10878003, Innovation Program 12ZZ134 and Chen Guang project 10CG46 of SMEC, and STCSM grant 11290706600 \& Pujiang Program 12PJ1406700. 
SG acknowledges support from the Netherlands Organization for Scientific Research (NWO) through VIDI grant 639.042.814.
MJH acknowledges support from the Natural Sciences and Engineering Research Council of Canada (NSERC).
BR acknowledges support from the European Research Council in the form of a Starting Grant with number 24067.
TS acknowledges support from NSF through grant AST-0444059-001, SAO through grant GO0-11147A, and NWO.
ES acknowledges support from the Netherlands Organisation for Scientific Research (NWO) grant number 639.042.814 and support from the European Research Council under the EC FP7 grant number 279396.

{\small Author Contributions: All authors contributed to the development and writing of this paper. The authorship list reflects the lead authors of this paper (MV, EvU and H.~Hoekstra) followed by two alphabetical groups. The first alphabetical group includes key contributors to the science analysis and interpretation in this paper, the founding core team and those whose long-term significant effort produced the final CFHTLenS data product. The second group covers members of the CFHTLenS team who made a significant contribution to the project and/or this paper. The CFHTLenS collaboration was co-led by CH and LVW and the CFHTLenS Galaxy-Galaxy Lensing Working Group was led by BR and CB.}

\bibliographystyle{mn2e_old}
\bibliography{velander.cfhtlens.2}

\begin{thebibliography}{78}
\expandafter\ifx\csname natexlab\endcsname\relax\def\natexlab#1{#1}\fi

\bibitem[{{Adelman-McCarthy} {et~al.}(2006){Adelman-McCarthy}, {Ag{\"u}eros},
  {Allam}, {Anderson}, {Anderson}, {Annis}, {Bahcall}, {Baldry}, {Barentine},
  {Berlind}, {Bernardi}, {Blanton}, {Boroski}, {Brewington}, {Brinchmann},
  {Brinkmann}, {Brunner}, {Budav{\'a}ri}, {Carey}, {Carr}, {Castander},
  {Connolly}, {Csabai}, {Czarapata}, {Dalcanton}, {Doi}, {Dong}, {Eisenstein},
  {Evans}, {Fan}, {Finkbeiner}, {Friedman}, {Frieman}, {Fukugita}, {Gillespie},
  {Glazebrook}, {Gray}, {Grebel}, {Gunn}, {Gurbani}, {de Haas}, {Hall},
  {Harris}, {Harvanek}, {Hawley}, {Hayes}, {Hendry}, {Hennessy}, {Hindsley},
  {Hirata}, {Hogan}, {Hogg}, {Holmgren}, {Holtzman}, {Ichikawa}, {Ivezi{\'c}},
  {Jester}, {Johnston}, {Jorgensen}, {Juri{\'c}}, {Kent}, {Kleinman}, {Knapp},
  {Kniazev}, {Kron}, {Krzesinski}, {Kuropatkin}, {Lamb}, {Lampeitl}, {Lee},
  {Leger}, {Lin}, {Long}, {Loveday}, {Lupton}, {Margon},
  {Mart{\'{\i}}nez-Delgado}, {Mandelbaum}, {Matsubara}, {McGehee}, {McKay},
  {Meiksin}, {Munn}, {Nakajima}, {Nash}, {Neilsen}, {Newberg}, {Newman},
  {Nichol}, {Nicinski}, {Nieto-Santisteban}, {Nitta}, {O'Mullane}, {Okamura},
  {Owen}, {Padmanabhan}, {Pauls}, {Peoples}, {Pier}, {Pope}, {Pourbaix},
  {Quinn}, {Richards}, {Richmond}, {Rockosi}, {Schlegel}, {Schneider},
  {Schroeder}, {Scranton}, {Seljak}, {Sheldon}, {Shimasaku}, {Smith}, {Smol{\v
  c}i{\'c}}, {Snedden}, {Stoughton}, {Strauss}, {SubbaRao}, {Szalay},
  {Szapudi}, {Szkody}, {Tegmark}, {Thakar}, {Tucker}, {Uomoto}, {Vanden Berk},
  {Vandenberg}, {Vogeley}, {Voges}, {Vogt}, {Walkowicz}, {Weinberg}, {West},
  {White}, {Xu}, {Yanny}, {Yocum}, {York}, {Zehavi}, {Zibetti}, \&
  {Zucker}}]{aaa06}
{Adelman-McCarthy} J.~K., {Ag{\"u}eros} M.~A., {Allam} S.~S., {Anderson}
  K.~S.~J., {Anderson} S.~F., {Annis} J., {Bahcall} N.~A., {Baldry} I.~K.,
  {Barentine} J.~C., {Berlind} A., {Bernardi} M., {Blanton} M.~R., {Boroski}
  W.~N., {Brewington} H.~J., {Brinchmann} J., {Brinkmann} J., {Brunner} R.~J.,
  {Budav{\'a}ri} T., {Carey} L.~N., {Carr} M.~A., {Castander} F.~J., {Connolly}
  A.~J., {Csabai} I., {Czarapata} P.~C., {Dalcanton} J.~J., {Doi} M., {Dong}
  F., {Eisenstein} D.~J., {Evans} M.~L., {Fan} X., {Finkbeiner} D.~P.,
  {Friedman} S.~D., {Frieman} J.~A., {Fukugita} M., {Gillespie} B.,
  {Glazebrook} K., {Gray} J., {Grebel} E.~K., {Gunn} J.~E., {Gurbani} V.~K.,
  {de Haas} E., {Hall} P.~B., {Harris} F.~H., {Harvanek} M., {Hawley} S.~L.,
  {Hayes} J., {Hendry} J.~S., {Hennessy} G.~S., {Hindsley} R.~B., {Hirata}
  C.~M., {Hogan} C.~J., {Hogg} D.~W., {Holmgren} D.~J., {Holtzman} J.~A.,
  {Ichikawa} S.-i., {Ivezi{\'c}} {\v Z}., {Jester} S., {Johnston} D.~E.,
  {Jorgensen} A.~M., {Juri{\'c}} M., {Kent} S.~M., {Kleinman} S.~J., {Knapp}
  G.~R., {Kniazev} A.~Y., {Kron} R.~G., {Krzesinski} J., {Kuropatkin} N.,
  {Lamb} D.~Q., {Lampeitl} H., {Lee} B.~C., {Leger} R.~F., {Lin} H., {Long}
  D.~C., {Loveday} J., {Lupton} R.~H., {Margon} B., {Mart{\'{\i}}nez-Delgado}
  D., {Mandelbaum} R., {Matsubara} T., {McGehee} P.~M., {McKay} T.~A.,
  {Meiksin} A., {Munn} J.~A., {Nakajima} R., {Nash} T., {Neilsen} Jr. E.~H.,
  {Newberg} H.~J., {Newman} P.~R., {Nichol} R.~C., {Nicinski} T.,
  {Nieto-Santisteban} M., {Nitta} A., {O'Mullane} W., {Okamura} S., {Owen} R.,
  {Padmanabhan} N., {Pauls} G., {Peoples} Jr. J., {Pier} J.~R., {Pope} A.~C.,
  {Pourbaix} D., {Quinn} T.~R., {Richards} G.~T., {Richmond} M.~W., {Rockosi}
  C.~M., {Schlegel} D.~J., {Schneider} D.~P., {Schroeder} J., {Scranton} R.,
  {Seljak} U., {Sheldon} E., {Shimasaku} K., {Smith} J.~A., {Smol{\v c}i{\'c}}
  V., {Snedden} S.~A., {Stoughton} C., {Strauss} M.~A., {SubbaRao} M., {Szalay}
  A.~S., {Szapudi} I., {Szkody} P., {Tegmark} M., {Thakar} A.~R., {Tucker}
  D.~L., {Uomoto} A., {Vanden Berk} D.~E., {Vandenberg} J., {Vogeley} M.~S.,
  {Voges} W., {Vogt} N.~P., {Walkowicz} L.~M., {Weinberg} D.~H., {West} A.~A.,
  {White} S.~D.~M., {Xu} Y., {Yanny} B., {Yocum} D.~R., {York} D.~G., {Zehavi}
  I., {Zibetti} S., {Zucker} D.~B., 2006, ApJS, 162, 38

\bibitem[{{Arnouts} {et~al.}(1999){Arnouts}, {Cristiani}, {Moscardini},
  {Matarrese}, {Lucchin}, {Fontana}, \& {Giallongo}}]{acm99}
{Arnouts} S., {Cristiani} S., {Moscardini} L., {Matarrese} S., {Lucchin} F.,
  {Fontana} A., {Giallongo} E., 1999, MNRAS, 310, 540

\bibitem[{{Arnouts} {et~al.}(2007){Arnouts}, {Walcher}, {Le F{\`e}vre},
  {Zamorani}, {Ilbert}, {Le Brun}, {Pozzetti}, {Bardelli}, {Tresse}, {Zucca},
  {Charlot}, {Lamareille}, {McCracken}, {Bolzonella}, {Iovino}, {Lonsdale},
  {Polletta}, {Surace}, {Bottini}, {Garilli}, {Maccagni}, {Picat},
  {Scaramella}, {Scodeggio}, {Vettolani}, {Zanichelli}, {Adami}, {Cappi},
  {Ciliegi}, {Contini}, {de la Torre}, {Foucaud}, {Franzetti}, {Gavignaud},
  {Guzzo}, {Marano}, {Marinoni}, {Mazure}, {Meneux}, {Merighi}, {Paltani},
  {Pell{\`o}}, {Pollo}, {Radovich}, {Temporin}, \& {Vergani}}]{awl07}
{Arnouts} S., {Walcher} C.~J., {Le F{\`e}vre} O., {Zamorani} G., {Ilbert} O.,
  {Le Brun} V., {Pozzetti} L., {Bardelli} S., {Tresse} L., {Zucca} E.,
  {Charlot} S., {Lamareille} F., {McCracken} H.~J., {Bolzonella} M., {Iovino}
  A., {Lonsdale} C., {Polletta} M., {Surace} J., {Bottini} D., {Garilli} B.,
  {Maccagni} D., {Picat} J.~P., {Scaramella} R., {Scodeggio} M., {Vettolani}
  G., {Zanichelli} A., {Adami} C., {Cappi} A., {Ciliegi} P., {Contini} T., {de
  la Torre} S., {Foucaud} S., {Franzetti} P., {Gavignaud} I., {Guzzo} L.,
  {Marano} B., {Marinoni} C., {Mazure} A., {Meneux} B., {Merighi} R., {Paltani}
  S., {Pell{\`o}} R., {Pollo} A., {Radovich} M., {Temporin} S., {Vergani} D.,
  2007, A\&A, 476, 137

\bibitem[{{Bell}(2008)}]{bel08}
{Bell} E.~F., 2008, ApJ, 682, 355

\bibitem[{{Bell} \& {de Jong}(2001)}]{bel01}
{Bell} E.~F., {de Jong} R.~S., 2001, ApJ, 550, 212

\bibitem[{{Ben{\'{\i}}tez}(2000)}]{ben00}
{Ben{\'{\i}}tez} N., 2000, ApJ, 536, 571

\bibitem[{{Bielby} {et~al.}(2012){Bielby}, {Hudelot}, {McCracken}, {Ilbert},
  {Daddi}, {Le F{\`e}vre}, {Gonzalez-Perez}, {Kneib}, {Marmo}, {Mellier},
  {Salvato}, {Sanders}, \& {Willott}}]{bhm12}
{Bielby} R., {Hudelot} P., {McCracken} H.~J., {Ilbert} O., {Daddi} E., {Le
  F{\`e}vre} O., {Gonzalez-Perez} V., {Kneib} J.-P., {Marmo} C., {Mellier} Y.,
  {Salvato} M., {Sanders} D.~B., {Willott} C.~J., 2012, A\&A, 545, A23

\bibitem[{{Borch} {et~al.}(2006){Borch}, {Meisenheimer}, {Bell}, {Rix}, {Wolf},
  {Dye}, {Kleinheinrich}, {Kovacs}, \& {Wisotzki}}]{bmb06}
{Borch} A., {Meisenheimer} K., {Bell} E.~F., {Rix} H.-W., {Wolf} C., {Dye} S.,
  {Kleinheinrich} M., {Kovacs} Z., {Wisotzki} L., 2006, A\&A, 453, 869

\bibitem[{{Brainerd} {et~al.}(1996){Brainerd}, {Blandford}, \& {Smail}}]{bbs96}
{Brainerd} T.~G., {Blandford} R.~D., {Smail} I., 1996, ApJ, 466, 623

\bibitem[{{Bruzual} \& {Charlot}(2003)}]{bru03}
{Bruzual} G., {Charlot} S., 2003, MNRAS, 344, 1000

\bibitem[{{Calzetti} {et~al.}(2000){Calzetti}, {Armus}, {Bohlin}, {Kinney},
  {Koornneef}, \& {Storchi-Bergmann}}]{cab00}
{Calzetti} D., {Armus} L., {Bohlin} R.~C., {Kinney} A.~L., {Koornneef} J.,
  {Storchi-Bergmann} T., 2000, ApJ, 533, 682

\bibitem[{{Chabrier}(2003)}]{cha03}
{Chabrier} G., 2003, PASP, 115, 763

\bibitem[{{Choi} {et~al.}(2012){Choi}, {Tyson}, {Morrison}, {Jee}, {Schmidt},
  {Margoniner}, \& {Wittman}}]{ctm12}
{Choi} A., {Tyson} J.~A., {Morrison} C.~B., {Jee} M.~J., {Schmidt} S.~J.,
  {Margoniner} V.~E., {Wittman} D.~M., 2012, ApJ, 759, 101

\bibitem[{{Coe} {et~al.}(2006){Coe}, {Ben{\'{\i}}tez}, {S{\'a}nchez}, {Jee},
  {Bouwens}, \& {Ford}}]{cbs06}
{Coe} D., {Ben{\'{\i}}tez} N., {S{\'a}nchez} S.~F., {Jee} M., {Bouwens} R.,
  {Ford} H., 2006, AJ, 132, 926

\bibitem[{{Coleman} {et~al.}(1980){Coleman}, {Wu}, \& {Weedman}}]{cww80}
{Coleman} G.~D., {Wu} C.-C., {Weedman} D.~W., 1980, ApJS, 43, 393

\bibitem[{{Conroy} \& {Wechsler}(2009)}]{con09}
{Conroy} C., {Wechsler} R.~H., 2009, ApJ, 696, 620

\bibitem[{{Cooray} \& {Sheth}(2002)}]{coo02}
{Cooray} A., {Sheth} R., 2002, Phys.~Rep., 372, 1

\bibitem[{{Coupon} {et~al.}(2012){Coupon}, {Kilbinger}, {McCracken}, {Ilbert},
  {Arnouts}, {Mellier}, {Abbas}, {de la Torre}, {Goranova}, {Hudelot}, {Kneib},
  \& {Le F{\`e}vre}}]{ckm12}
{Coupon} J., {Kilbinger} M., {McCracken} H.~J., {Ilbert} O., {Arnouts} S.,
  {Mellier} Y., {Abbas} U., {de la Torre} S., {Goranova} Y., {Hudelot} P.,
  {Kneib} J.-P., {Le F{\`e}vre} O., 2012, A\&A, 542, A5

\bibitem[{{Davis} {et~al.}(2003){Davis}, {Faber}, {Newman}, {Phillips},
  {Ellis}, {Steidel}, {Conselice}, {Coil}, {Finkbeiner}, {Koo}, {Guhathakurta},
  {Weiner}, {Schiavon}, {Willmer}, {Kaiser}, {Luppino}, {Wirth}, {Connolly},
  {Eisenhardt}, {Cooper}, \& {Gerke}}]{dfn03}
{Davis} M., {Faber} S.~M., {Newman} J., {Phillips} A.~C., {Ellis} R.~S.,
  {Steidel} C.~C., {Conselice} C., {Coil} A.~L., {Finkbeiner} D.~P., {Koo}
  D.~C., {Guhathakurta} P., {Weiner} B., {Schiavon} R., {Willmer} C., {Kaiser}
  N., {Luppino} G.~A., {Wirth} G., {Connolly} A., {Eisenhardt} P., {Cooper} M.,
  {Gerke} B., 2003, SPIE, 4834, 161

\bibitem[{{Davis} {et~al.}(2007){Davis}, {Guhathakurta}, {Konidaris}, {Newman},
  {Ashby}, {Biggs}, {Barmby}, {Bundy}, {Chapman}, {Coil}, {Conselice},
  {Cooper}, {Croton}, {Eisenhardt}, {Ellis}, {Faber}, {Fang}, {Fazio},
  {Georgakakis}, {Gerke}, {Goss}, {Gwyn}, {Harker}, {Hopkins}, {Huang},
  {Ivison}, {Kassin}, {Kirby}, {Koekemoer}, {Koo}, {Laird}, {Le Floc'h}, {Lin},
  {Lotz}, {Marshall}, {Martin}, {Metevier}, {Moustakas}, {Nandra}, {Noeske},
  {Papovich}, {Phillips}, {Rich}, {Rieke}, {Rigopoulou}, {Salim},
  {Schiminovich}, {Simard}, {Smail}, {Small}, {Weiner}, {Willmer}, {Willner},
  {Wilson}, {Wright}, \& {Yan}}]{dgk07}
{Davis} M., {Guhathakurta} P., {Konidaris} N.~P., {Newman} J.~A., {Ashby}
  M.~L.~N., {Biggs} A.~D., {Barmby} P., {Bundy} K., {Chapman} S.~C., {Coil}
  A.~L., {Conselice} C.~J., {Cooper} M.~C., {Croton} D.~J., {Eisenhardt}
  P.~R.~M., {Ellis} R.~S., {Faber} S.~M., {Fang} T., {Fazio} G.~G.,
  {Georgakakis} A., {Gerke} B.~F., {Goss} W.~M., {Gwyn} S., {Harker} J.,
  {Hopkins} A.~M., {Huang} J.-S., {Ivison} R.~J., {Kassin} S.~A., {Kirby}
  E.~N., {Koekemoer} A.~M., {Koo} D.~C., {Laird} E.~S., {Le Floc'h} E., {Lin}
  L., {Lotz} J.~M., {Marshall} P.~J., {Martin} D.~C., {Metevier} A.~J.,
  {Moustakas} L.~A., {Nandra} K., {Noeske} K.~G., {Papovich} C., {Phillips}
  A.~C., {Rich} R.~M., {Rieke} G.~H., {Rigopoulou} D., {Salim} S.,
  {Schiminovich} D., {Simard} L., {Smail} I., {Small} T.~A., {Weiner} B.~J.,
  {Willmer} C.~N.~A., {Willner} S.~P., {Wilson} G., {Wright} E.~L., {Yan} R.,
  2007, ApJL, 660, L1

\bibitem[{{de Jong} {et~al.}(2013){de Jong}, {Verdoes Kleijn}, {Kuijken}, \&
  {Valentijn}}]{dkk13}
{de Jong} J.~T.~A., {Verdoes Kleijn} G.~A., {Kuijken} K.~H., {Valentijn} E.~A.,
  2013, Experimental Astronomy, 35, 25

\bibitem[{{Duffy} {et~al.}(2008){Duffy}, {Schaye}, {Kay}, \& {Dalla
  Vecchia}}]{dsk08}
{Duffy} A.~R., {Schaye} J., {Kay} S.~T., {Dalla Vecchia} C., 2008, MNRAS, 390,
  L64

\bibitem[{{Eddington}(1913)}]{edd13}
{Eddington} A.~S., 1913, MNRAS, 73, 359

\bibitem[{{Eisenstein} {et~al.}(2001){Eisenstein}, {Annis}, {Gunn}, {Szalay},
  {Connolly}, {Nichol}, {Bahcall}, {Bernardi}, {Burles}, {Castander},
  {Fukugita}, {Hogg}, {Ivezi{\'c}}, {Knapp}, {Lupton}, {Narayanan}, {Postman},
  {Reichart}, {Richmond}, {Schneider}, {Schlegel}, {Strauss}, {SubbaRao},
  {Tucker}, {Vanden Berk}, {Vogeley}, {Weinberg}, \& {Yanny}}]{eag01}
{Eisenstein} D.~J., {Annis} J., {Gunn} J.~E., {Szalay} A.~S., {Connolly} A.~J.,
  {Nichol} R.~C., {Bahcall} N.~A., {Bernardi} M., {Burles} S., {Castander}
  F.~J., {Fukugita} M., {Hogg} D.~W., {Ivezi{\'c}} {\v Z}., {Knapp} G.~R.,
  {Lupton} R.~H., {Narayanan} V., {Postman} M., {Reichart} D.~E., {Richmond}
  M., {Schneider} D.~P., {Schlegel} D.~J., {Strauss} M.~A., {SubbaRao} M.,
  {Tucker} D.~L., {Vanden Berk} D., {Vogeley} M.~S., {Weinberg} D.~H., {Yanny}
  B., 2001, AJ, 122, 2267

\bibitem[{{Erben} {et~al.}(2009){Erben}, {Hildebrandt}, {Lerchster}, {Hudelot},
  {Benjamin}, {van Waerbeke}, {Schrabback}, {Brimioulle}, {Cordes}, {Dietrich},
  {Holhjem}, {Schirmer}, \& {Schneider}}]{ehl09}
{Erben} T., {Hildebrandt} H., {Lerchster} M., {Hudelot} P., {Benjamin} J., {van
  Waerbeke} L., {Schrabback} T., {Brimioulle} F., {Cordes} O., {Dietrich}
  J.~P., {Holhjem} K., {Schirmer} M., {Schneider} P., 2009, A\&A, 493, 1197

\bibitem[{{Erben} {et~al.}(2012){Erben}, {Hildebrandt}, {Miller}, {van
  Waerbeke}, {Heymans}, {Hoekstra}, {Kitching}, {Mellier}, {Benjamin}, {Blake},
  {Bonnett}, {Cordes}, {Coupon}, {Fu}, {Gavazzi}, {Gillis}, {Grocutt}, {Gwyn},
  {Holhjem}, {Hudson}, {Kilbinger}, {Kuijken}, {Milkeraitis}, {Rowe},
  {Schrabback}, {Semboloni}, {Simon}, {Smit}, {Toader}, {Vafaei}, {van Uitert},
  \& {Velander}}]{ehm12}
{Erben} T., {Hildebrandt} H., {Miller} L., {van Waerbeke} L., {Heymans} C.,
  {Hoekstra} H., {Kitching} T.~D., {Mellier} Y., {Benjamin} J., {Blake} C.,
  {Bonnett} C., {Cordes} O., {Coupon} J., {Fu} L., {Gavazzi} R., {Gillis} B.,
  {Grocutt} E., {Gwyn} S.~D.~J., {Holhjem} K., {Hudson} M.~J., {Kilbinger} M.,
  {Kuijken} K., {Milkeraitis} M., {Rowe} B.~T.~P., {Schrabback} T., {Semboloni}
  E., {Simon} P., {Smit} M., {Toader} O., {Vafaei} S., {van Uitert} E.,
  {Velander} M., 2012, preprint (astro-ph/1210.8156)

\bibitem[{{Gallazzi} {et~al.}(2005){Gallazzi}, {Charlot}, {Brinchmann},
  {White}, \& {Tremonti}}]{gcb05}
{Gallazzi} A., {Charlot} S., {Brinchmann} J., {White} S.~D.~M., {Tremonti}
  C.~A., 2005, MNRAS, 362, 41

\bibitem[{{Garilli} {et~al.}(2008){Garilli}, {Le F{\`e}vre}, {Guzzo},
  {Maccagni}, {Le Brun}, {de la Torre}, {Meneux}, {Tresse}, {Franzetti},
  {Zamorani}, {Zanichelli}, {Gregorini}, {Vergani}, {Bottini}, {Scaramella},
  {Scodeggio}, {Vettolani}, {Adami}, {Arnouts}, {Bardelli}, {Bolzonella},
  {Cappi}, {Charlot}, {Ciliegi}, {Contini}, {Foucaud}, {Gavignaud}, {Ilbert},
  {Iovino}, {Lamareille}, {McCracken}, {Marano}, {Marinoni}, {Mazure},
  {Merighi}, {Paltani}, {Pell{\`o}}, {Pollo}, {Pozzetti}, {Radovich}, {Zucca},
  {Blaizot}, {Bongiorno}, {Cucciati}, {Mellier}, {Moreau}, \& {Paioro}}]{glg08}
{Garilli} B., {Le F{\`e}vre} O., {Guzzo} L., {Maccagni} D., {Le Brun} V., {de
  la Torre} S., {Meneux} B., {Tresse} L., {Franzetti} P., {Zamorani} G.,
  {Zanichelli} A., {Gregorini} L., {Vergani} D., {Bottini} D., {Scaramella} R.,
  {Scodeggio} M., {Vettolani} G., {Adami} C., {Arnouts} S., {Bardelli} S.,
  {Bolzonella} M., {Cappi} A., {Charlot} S., {Ciliegi} P., {Contini} T.,
  {Foucaud} S., {Gavignaud} I., {Ilbert} O., {Iovino} A., {Lamareille} F.,
  {McCracken} H.~J., {Marano} B., {Marinoni} C., {Mazure} A., {Merighi} R.,
  {Paltani} S., {Pell{\`o}} R., {Pollo} A., {Pozzetti} L., {Radovich} M.,
  {Zucca} E., {Blaizot} J., {Bongiorno} A., {Cucciati} O., {Mellier} Y.,
  {Moreau} C., {Paioro} L., 2008, A\&A, 486, 683

\bibitem[{{Gilbank} {et~al.}(2011){Gilbank}, {Gladders}, {Yee}, \&
  {Hsieh}}]{ggy11}
{Gilbank} D.~G., {Gladders} M.~D., {Yee} H.~K.~C., {Hsieh} B.~C., 2011, AJ,
  141, 94

\bibitem[{{Gillis} {et~al.}(2013){Gillis}, {Hudson}, {Erben}, {Heymans},
  {Hildebrandt}, {Hoekstra}, {Kitching}, {Mellier}, {Miller}, {van Waerbeke},
  {Bonnett}, {Coupon}, {Fu}, {Hilbert}, {Rowe}, {Schrabback}, {Semboloni}, {van
  Uitert}, \& {Velander}}]{ghe13}
{Gillis} B.~R., {Hudson} M.~J., {Erben} T., {Heymans} C., {Hildebrandt} H.,
  {Hoekstra} H., {Kitching} T.~D., {Mellier} Y., {Miller} L., {van Waerbeke}
  L., {Bonnett} C., {Coupon} J., {Fu} L., {Hilbert} S., {Rowe} B.~T.~P.,
  {Schrabback} T., {Semboloni} E., {van Uitert} E., {Velander} M., 2013, MNRAS,
  431, 1439

\bibitem[{{Guo} {et~al.}(2012){Guo}, {Cole}, {Eke}, \& {Frenk}}]{gce12}
{Guo} Q., {Cole} S., {Eke} V., {Frenk} C., 2012, MNRAS, 427, 428

\bibitem[{{Guzik} \& {Seljak}(2002)}]{guz02}
{Guzik} J., {Seljak} U., 2002, MNRAS, 335, 311

\bibitem[{{Heymans} {et~al.}(2012){Heymans}, {Van Waerbeke}, {Miller}, {Erben},
  {Hildebrandt}, {Hoekstra}, {Kitching}, {Mellier}, {Simon}, {Bonnett},
  {Coupon}, {Fu}, {Harnois D{\'e}raps}, {Hudson}, {Kilbinger}, {Kuijken},
  {Rowe}, {Schrabback}, {Semboloni}, {van Uitert}, {Vafaei}, \&
  {Velander}}]{hvm12}
{Heymans} C., {Van Waerbeke} L., {Miller} L., {Erben} T., {Hildebrandt} H.,
  {Hoekstra} H., {Kitching} T.~D., {Mellier} Y., {Simon} P., {Bonnett} C.,
  {Coupon} J., {Fu} L., {Harnois D{\'e}raps} J., {Hudson} M.~J., {Kilbinger}
  M., {Kuijken} K., {Rowe} B., {Schrabback} T., {Semboloni} E., {van Uitert}
  E., {Vafaei} S., {Velander} M., 2012, MNRAS, 427, 146

\bibitem[{{Hildebrandt} {et~al.}(2010){Hildebrandt}, {Arnouts}, {Capak},
  {Moustakas}, {Wolf}, {Abdalla}, {Assef}, {Banerji}, {Ben{\'{\i}}tez},
  {Brammer}, {Budav{\'a}ri}, {Carliles}, {Coe}, {Dahlen}, {Feldmann}, {Gerdes},
  {Gillis}, {Ilbert}, {Kotulla}, {Lahav}, {Li}, {Miralles}, {Purger},
  {Schmidt}, \& {Singal}}]{hak10}
{Hildebrandt} H., {Arnouts} S., {Capak} P., {Moustakas} L.~A., {Wolf} C.,
  {Abdalla} F.~B., {Assef} R.~J., {Banerji} M., {Ben{\'{\i}}tez} N., {Brammer}
  G.~B., {Budav{\'a}ri} T., {Carliles} S., {Coe} D., {Dahlen} T., {Feldmann}
  R., {Gerdes} D., {Gillis} B., {Ilbert} O., {Kotulla} R., {Lahav} O., {Li}
  I.~H., {Miralles} J.-M., {Purger} N., {Schmidt} S., {Singal} J., 2010, A\&A,
  523, A31

\bibitem[{{Hildebrandt} {et~al.}(2012){Hildebrandt}, {Erben}, {Kuijken}, {van
  Waerbeke}, {Heymans}, {Coupon}, {Benjamin}, {Bonnett}, {Fu}, {Hoekstra},
  {Kitching}, {Mellier}, {Miller}, {Velander}, {Hudson}, {Rowe}, {Schrabback},
  {Semboloni}, \& {Ben{\'{\i}}tez}}]{hek12}
{Hildebrandt} H., {Erben} T., {Kuijken} K., {van Waerbeke} L., {Heymans} C.,
  {Coupon} J., {Benjamin} J., {Bonnett} C., {Fu} L., {Hoekstra} H., {Kitching}
  T.~D., {Mellier} Y., {Miller} L., {Velander} M., {Hudson} M.~J., {Rowe}
  B.~T.~P., {Schrabback} T., {Semboloni} E., {Ben{\'{\i}}tez} N., 2012, MNRAS,
  421, 2355

\bibitem[{{Hoekstra}(2007)}]{hoe07}
{Hoekstra} H., 2007, MNRAS, 379, 317

\bibitem[{{Hoekstra} {et~al.}(1998){Hoekstra}, {Franx}, {Kuijken}, \&
  {Squires}}]{hfk98}
{Hoekstra} H., {Franx} M., {Kuijken} K., {Squires} G., 1998, ApJ, 504, 636

\bibitem[{{Hoekstra} {et~al.}(2005){Hoekstra}, {Hsieh}, {Yee}, {Lin}, \&
  {Gladders}}]{hhy05}
{Hoekstra} H., {Hsieh} B.~C., {Yee} H.~K.~C., {Lin} H., {Gladders} M.~D., 2005,
  ApJ, 635, 73

\bibitem[{{Hoekstra} {et~al.}(2004){Hoekstra}, {Yee}, \& {Gladders}}]{hyg04}
{Hoekstra} H., {Yee} H.~K.~C., {Gladders} M.~D., 2004, ApJ, 606, 67

\bibitem[{{Hudson} {et~al.}(2013){Hudson}, {CFHTLenS}, \& et~al.}]{hud12}
{Hudson} M.~J., {CFHTLenS}, et~al., 2013, {CFHTLenS}, preprint

\bibitem[{{Hudson} {et~al.}(1998){Hudson}, {Gwyn}, {Dahle}, \&
  {Kaiser}}]{hgd98}
{Hudson} M.~J., {Gwyn} S.~D.~J., {Dahle} H., {Kaiser} N., 1998, ApJ, 503, 531

\bibitem[{{Ilbert} {et~al.}(2006){Ilbert}, {Arnouts}, {McCracken},
  {Bolzonella}, {Bertin}, {Le F{\`e}vre}, {Mellier}, {Zamorani}, {Pell{\`o}},
  {Iovino}, {Tresse}, {Le Brun}, {Bottini}, {Garilli}, {Maccagni}, {Picat},
  {Scaramella}, {Scodeggio}, {Vettolani}, {Zanichelli}, {Adami}, {Bardelli},
  {Cappi}, {Charlot}, {Ciliegi}, {Contini}, {Cucciati}, {Foucaud}, {Franzetti},
  {Gavignaud}, {Guzzo}, {Marano}, {Marinoni}, {Mazure}, {Meneux}, {Merighi},
  {Paltani}, {Pollo}, {Pozzetti}, {Radovich}, {Zucca}, {Bondi}, {Bongiorno},
  {Busarello}, {de La Torre}, {Gregorini}, {Lamareille}, {Mathez}, {Merluzzi},
  {Ripepi}, {Rizzo}, \& {Vergani}}]{iam06}
{Ilbert} O., {Arnouts} S., {McCracken} H.~J., {Bolzonella} M., {Bertin} E., {Le
  F{\`e}vre} O., {Mellier} Y., {Zamorani} G., {Pell{\`o}} R., {Iovino} A.,
  {Tresse} L., {Le Brun} V., {Bottini} D., {Garilli} B., {Maccagni} D., {Picat}
  J.~P., {Scaramella} R., {Scodeggio} M., {Vettolani} G., {Zanichelli} A.,
  {Adami} C., {Bardelli} S., {Cappi} A., {Charlot} S., {Ciliegi} P., {Contini}
  T., {Cucciati} O., {Foucaud} S., {Franzetti} P., {Gavignaud} I., {Guzzo} L.,
  {Marano} B., {Marinoni} C., {Mazure} A., {Meneux} B., {Merighi} R., {Paltani}
  S., {Pollo} A., {Pozzetti} L., {Radovich} M., {Zucca} E., {Bondi} M.,
  {Bongiorno} A., {Busarello} G., {de La Torre} S., {Gregorini} L.,
  {Lamareille} F., {Mathez} G., {Merluzzi} P., {Ripepi} V., {Rizzo} D.,
  {Vergani} D., 2006, A\&A, 457, 841

\bibitem[{{Ilbert} {et~al.}(2010){Ilbert}, {Salvato}, {Le Floc'h}, {Aussel},
  {Capak}, {McCracken}, {Mobasher}, {Kartaltepe}, {Scoville}, {Sanders},
  {Arnouts}, {Bundy}, {Cassata}, {Kneib}, {Koekemoer}, {Le F{\`e}vre}, {Lilly},
  {Surace}, {Taniguchi}, {Tasca}, {Thompson}, {Tresse}, {Zamojski}, {Zamorani},
  \& {Zucca}}]{isl10}
{Ilbert} O., {Salvato} M., {Le Floc'h} E., {Aussel} H., {Capak} P., {McCracken}
  H.~J., {Mobasher} B., {Kartaltepe} J., {Scoville} N., {Sanders} D.~B.,
  {Arnouts} S., {Bundy} K., {Cassata} P., {Kneib} J.-P., {Koekemoer} A., {Le
  F{\`e}vre} O., {Lilly} S., {Surace} J., {Taniguchi} Y., {Tasca} L.,
  {Thompson} D., {Tresse} L., {Zamojski} M., {Zamorani} G., {Zucca} E., 2010,
  ApJ, 709, 644

\bibitem[{{Kaiser} {et~al.}(1995){Kaiser}, {Squires}, \& {Broadhurst}}]{ksb95}
{Kaiser} N., {Squires} G., {Broadhurst} T., 1995, ApJ, 449, 460

\bibitem[{{Kauffmann} {et~al.}(2003){Kauffmann}, {Heckman}, {White}, {Charlot},
  {Tremonti}, {Brinchmann}, {Bruzual}, {Peng}, {Seibert}, {Bernardi},
  {Blanton}, {Brinkmann}, {Castander}, {Cs{\'a}bai}, {Fukugita}, {Ivezic},
  {Munn}, {Nichol}, {Padmanabhan}, {Thakar}, {Weinberg}, \& {York}}]{khw03}
{Kauffmann} G., {Heckman} T.~M., {White} S.~D.~M., {Charlot} S., {Tremonti} C.,
  {Brinchmann} J., {Bruzual} G., {Peng} E.~W., {Seibert} M., {Bernardi} M.,
  {Blanton} M., {Brinkmann} J., {Castander} F., {Cs{\'a}bai} I., {Fukugita} M.,
  {Ivezic} Z., {Munn} J.~A., {Nichol} R.~C., {Padmanabhan} N., {Thakar} A.~R.,
  {Weinberg} D.~H., {York} D., 2003, MNRAS, 341, 33

\bibitem[{{Kilbinger} {et~al.}(2013){Kilbinger}, {Fu}, {Heymans}, {Simpson},
  {Benjamin}, {Erben}, {Harnois-D{\'e}raps}, {Hoekstra}, {Hildebrandt},
  {Kitching}, {Mellier}, {Miller}, {Van Waerbeke}, {Benabed}, {Bonnett},
  {Coupon}, {Hudson}, {Kuijken}, {Rowe}, {Schrabback}, {Semboloni}, {Vafaei},
  \& {Velander}}]{kfh13}
{Kilbinger} M., {Fu} L., {Heymans} C., {Simpson} F., {Benjamin} J., {Erben} T.,
  {Harnois-D{\'e}raps} J., {Hoekstra} H., {Hildebrandt} H., {Kitching} T.~D.,
  {Mellier} Y., {Miller} L., {Van Waerbeke} L., {Benabed} K., {Bonnett} C.,
  {Coupon} J., {Hudson} M.~J., {Kuijken} K., {Rowe} B., {Schrabback} T.,
  {Semboloni} E., {Vafaei} S., {Velander} M., 2013, MNRAS, 430, 2200

\bibitem[{{Komatsu} {et~al.}(2011){Komatsu}, {Smith}, {Dunkley}, {Bennett},
  {Gold}, {Hinshaw}, {Jarosik}, {Larson}, {Nolta}, {Page}, {Spergel},
  {Halpern}, {Hill}, {Kogut}, {Limon}, {Meyer}, {Odegard}, {Tucker}, {Weiland},
  {Wollack}, \& {Wright}}]{ksd11}
{Komatsu} E., {Smith} K.~M., {Dunkley} J., {Bennett} C.~L., {Gold} B.,
  {Hinshaw} G., {Jarosik} N., {Larson} D., {Nolta} M.~R., {Page} L., {Spergel}
  D.~N., {Halpern} M., {Hill} R.~S., {Kogut} A., {Limon} M., {Meyer} S.~S.,
  {Odegard} N., {Tucker} G.~S., {Weiland} J.~L., {Wollack} E., {Wright} E.~L.,
  2011, ApJS, 192, 18

\bibitem[{{Laureijs} {et~al.}(2011){Laureijs}, {Amiaux}, {Arduini},
  {Augu{\`e}res}, {Brinchmann}, {Cole}, {Cropper}, {Dabin}, {Duvet}, {Ealet},
  \& et~al.}]{laa11}
{Laureijs} R., {Amiaux} J., {Arduini} S., {Augu{\`e}res} J.-L., {Brinchmann}
  J., {Cole} R., {Cropper} M., {Dabin} C., {Duvet} L., {Ealet} A., et~al.,
  2011, preprint (astro-ph/1110.3193)

\bibitem[{{Le F{\`e}vre} {et~al.}(2005){Le F{\`e}vre}, {Vettolani}, {Garilli},
  {Tresse}, {Bottini}, {Le Brun}, {Maccagni}, {Picat}, {Scaramella},
  {Scodeggio}, {Zanichelli}, {Adami}, {Arnaboldi}, {Arnouts}, {Bardelli},
  {Bolzonella}, {Cappi}, {Charlot}, {Ciliegi}, {Contini}, {Foucaud},
  {Franzetti}, {Gavignaud}, {Guzzo}, {Ilbert}, {Iovino}, {McCracken}, {Marano},
  {Marinoni}, {Mathez}, {Mazure}, {Meneux}, {Merighi}, {Paltani}, {Pell{\`o}},
  {Pollo}, {Pozzetti}, {Radovich}, {Zamorani}, {Zucca}, {Bondi}, {Bongiorno},
  {Busarello}, {Lamareille}, {Mellier}, {Merluzzi}, {Ripepi}, \&
  {Rizzo}}]{lvg05}
{Le F{\`e}vre} O., {Vettolani} G., {Garilli} B., {Tresse} L., {Bottini} D., {Le
  Brun} V., {Maccagni} D., {Picat} J.~P., {Scaramella} R., {Scodeggio} M.,
  {Zanichelli} A., {Adami} C., {Arnaboldi} M., {Arnouts} S., {Bardelli} S.,
  {Bolzonella} M., {Cappi} A., {Charlot} S., {Ciliegi} P., {Contini} T.,
  {Foucaud} S., {Franzetti} P., {Gavignaud} I., {Guzzo} L., {Ilbert} O.,
  {Iovino} A., {McCracken} H.~J., {Marano} B., {Marinoni} C., {Mathez} G.,
  {Mazure} A., {Meneux} B., {Merighi} R., {Paltani} S., {Pell{\`o}} R., {Pollo}
  A., {Pozzetti} L., {Radovich} M., {Zamorani} G., {Zucca} E., {Bondi} M.,
  {Bongiorno} A., {Busarello} G., {Lamareille} F., {Mellier} Y., {Merluzzi} P.,
  {Ripepi} V., {Rizzo} D., 2005, A\&A, 439, 845

\bibitem[{{Leauthaud} {et~al.}(2010){Leauthaud}, {Finoguenov}, {Kneib},
  {Taylor}, {Massey}, {Rhodes}, {Ilbert}, {Bundy}, {Tinker}, {George}, {Capak},
  {Koekemoer}, {Johnston}, {Zhang}, {Cappelluti}, {Ellis}, {Elvis}, {Giodini},
  {Heymans}, {Le F{\`e}vre}, {Lilly}, {McCracken}, {Mellier},
  {R{\'e}fr{\'e}gier}, {Salvato}, {Scoville}, {Smoot}, {Tanaka}, {Van
  Waerbeke}, \& {Wolk}}]{lfk10}
{Leauthaud} A., {Finoguenov} A., {Kneib} J.-P., {Taylor} J.~E., {Massey} R.,
  {Rhodes} J., {Ilbert} O., {Bundy} K., {Tinker} J., {George} M.~R., {Capak}
  P., {Koekemoer} A.~M., {Johnston} D.~E., {Zhang} Y.-Y., {Cappelluti} N.,
  {Ellis} R.~S., {Elvis} M., {Giodini} S., {Heymans} C., {Le F{\`e}vre} O.,
  {Lilly} S., {McCracken} H.~J., {Mellier} Y., {R{\'e}fr{\'e}gier} A.,
  {Salvato} M., {Scoville} N., {Smoot} G., {Tanaka} M., {Van Waerbeke} L.,
  {Wolk} M., 2010, ApJ, 709, 97

\bibitem[{{Leauthaud} {et~al.}(2011){Leauthaud}, {Tinker}, {Behroozi}, {Busha},
  \& {Wechsler}}]{ltb11}
{Leauthaud} A., {Tinker} J., {Behroozi} P.~S., {Busha} M.~T., {Wechsler} R.~H.,
  2011, ApJ, 738, 45

\bibitem[{{Leauthaud} {et~al.}(2012){Leauthaud}, {Tinker}, {Bundy}, {Behroozi},
  {Massey}, {Rhodes}, {George}, {Kneib}, {Benson}, {Wechsler}, {Busha},
  {Capak}, {Cort{\^e}s}, {Ilbert}, {Koekemoer}, {Le F{\`e}vre}, {Lilly},
  {McCracken}, {Salvato}, {Schrabback}, {Scoville}, {Smith}, \&
  {Taylor}}]{ltb12}
{Leauthaud} A., {Tinker} J., {Bundy} K., {Behroozi} P.~S., {Massey} R.,
  {Rhodes} J., {George} M.~R., {Kneib} J.-P., {Benson} A., {Wechsler} R.~H.,
  {Busha} M.~T., {Capak} P., {Cort{\^e}s} M., {Ilbert} O., {Koekemoer} A.~M.,
  {Le F{\`e}vre} O., {Lilly} S., {McCracken} H.~J., {Salvato} M., {Schrabback}
  T., {Scoville} N., {Smith} T., {Taylor} J.~E., 2012, ApJ, 744, 159

\bibitem[{{Mandelbaum} {et~al.}(2005{\natexlab{a}}){Mandelbaum}, {Hirata},
  {Seljak}, {Guzik}, {Padmanabhan}, {Blake}, {Blanton}, {Lupton}, \&
  {Brinkmann}}]{mhs05}
{Mandelbaum} R., {Hirata} C.~M., {Seljak} U., {Guzik} J., {Padmanabhan} N.,
  {Blake} C., {Blanton} M.~R., {Lupton} R., {Brinkmann} J., 2005{\natexlab{a}},
  MNRAS, 361, 1287

\bibitem[{{Mandelbaum} {et~al.}(2006){Mandelbaum}, {Seljak}, {Kauffmann},
  {Hirata}, \& {Brinkmann}}]{msk06}
{Mandelbaum} R., {Seljak} U., {Kauffmann} G., {Hirata} C.~M., {Brinkmann} J.,
  2006, MNRAS, 368, 715

\bibitem[{{Mandelbaum} {et~al.}(2013){Mandelbaum}, {Slosar}, {Baldauf},
  {Seljak}, {Hirata}, {Nakajima}, {Reyes}, \& {Smith}}]{msb13}
{Mandelbaum} R., {Slosar} A., {Baldauf} T., {Seljak} U., {Hirata} C.~M.,
  {Nakajima} R., {Reyes} R., {Smith} R.~E., 2013, MNRAS, 432, 1544

\bibitem[{{Mandelbaum} {et~al.}(2005{\natexlab{b}}){Mandelbaum}, {Tasitsiomi},
  {Seljak}, {Kravtsov}, \& {Wechsler}}]{mts05}
{Mandelbaum} R., {Tasitsiomi} A., {Seljak} U., {Kravtsov} A.~V., {Wechsler}
  R.~H., 2005{\natexlab{b}}, MNRAS, 362, 1451

\bibitem[{{Miller} {et~al.}(2013){Miller}, {Heymans}, {Kitching}, {van
  Waerbeke}, {Erben}, {Hildebrandt}, {Hoekstra}, {Mellier}, {Rowe}, {Coupon},
  {Dietrich}, {Fu}, {Harnois-D{\'e}raps}, {Hudson}, {Kilbinger}, {Kuijken},
  {Schrabback}, {Semboloni}, {Vafaei}, \& {Velander}}]{mhk13}
{Miller} L., {Heymans} C., {Kitching} T.~D., {van Waerbeke} L., {Erben} T.,
  {Hildebrandt} H., {Hoekstra} H., {Mellier} Y., {Rowe} B.~T.~P., {Coupon} J.,
  {Dietrich} J.~P., {Fu} L., {Harnois-D{\'e}raps} J., {Hudson} M.~J.,
  {Kilbinger} M., {Kuijken} K., {Schrabback} T., {Semboloni} E., {Vafaei} S.,
  {Velander} M., 2013, MNRAS, 429, 2858

\bibitem[{{More} {et~al.}(2011){More}, {van den Bosch}, {Cacciato}, {Skibba},
  {Mo}, \& {Yang}}]{mbc11}
{More} S., {van den Bosch} F.~C., {Cacciato} M., {Skibba} R., {Mo} H.~J.,
  {Yang} X., 2011, MNRAS, 410, 210

\bibitem[{{Moster} {et~al.}(2010){Moster}, {Somerville}, {Maulbetsch}, {van den
  Bosch}, {Macci{\`o}}, {Naab}, \& {Oser}}]{msm10}
{Moster} B.~P., {Somerville} R.~S., {Maulbetsch} C., {van den Bosch} F.~C.,
  {Macci{\`o}} A.~V., {Naab} T., {Oser} L., 2010, ApJ, 710, 903

\bibitem[{{Nagai} \& {Kravtsov}(2005)}]{nag05}
{Nagai} D., {Kravtsov} A.~V., 2005, ApJ, 618, 557

\bibitem[{{Navarro} {et~al.}(1996){Navarro}, {Frenk}, \& {White}}]{nfw96}
{Navarro} J.~F., {Frenk} C.~S., {White} S.~D.~M., 1996, ApJ, 462, 563

\bibitem[{{Newman} {et~al.}(2013){Newman}, {Cooper}, {Davis}, {Faber}, {Coil},
  {Guhathakurta}, {Koo}, {Phillips}, {Conroy}, {Dutton}, {Finkbeiner}, {Gerke},
  {Rosario}, {Weiner}, {Willmer}, {Yan}, {Harker}, {Kassin}, {Konidaris},
  {Lai}, {Madgwick}, {Noeske}, {Wirth}, {Connolly}, {Kaiser}, {Kirby},
  {Lemaux}, {Lin}, {Lotz}, {Luppino}, {Marinoni}, {Matthews}, {Metevier}, \&
  {Schiavon}}]{ncd13}
{Newman} J.~A., {Cooper} M.~C., {Davis} M., {Faber} S.~M., {Coil} A.~L.,
  {Guhathakurta} P., {Koo} D.~C., {Phillips} A.~C., {Conroy} C., {Dutton}
  A.~A., {Finkbeiner} D.~P., {Gerke} B.~F., {Rosario} D.~J., {Weiner} B.~J.,
  {Willmer} C.~N.~A., {Yan} R., {Harker} J.~J., {Kassin} S.~A., {Konidaris}
  N.~P., {Lai} K., {Madgwick} D.~S., {Noeske} K.~G., {Wirth} G.~D., {Connolly}
  A.~J., {Kaiser} N., {Kirby} E.~N., {Lemaux} B.~C., {Lin} L., {Lotz} J.~M.,
  {Luppino} G.~A., {Marinoni} C., {Matthews} D.~J., {Metevier} A., {Schiavon}
  R.~P., 2013, ApJS, 208, 5

\bibitem[{{Parker} {et~al.}(2007){Parker}, {Hoekstra}, {Hudson}, {van
  Waerbeke}, \& {Mellier}}]{phh07}
{Parker} L.~C., {Hoekstra} H., {Hudson} M.~J., {van Waerbeke} L., {Mellier} Y.,
  2007, ApJ, 669, 21

\bibitem[{{Paulin-Henriksson} {et~al.}(2008){Paulin-Henriksson}, {Amara},
  {Voigt}, {Refregier}, \& {Bridle}}]{pav08}
{Paulin-Henriksson} S., {Amara} A., {Voigt} L., {Refregier} A., {Bridle} S.~L.,
  2008, A\&A, 484, 67

\bibitem[{{Pozzetti} {et~al.}(2007){Pozzetti}, {Bolzonella}, {Lamareille},
  {Zamorani}, {Franzetti}, {Le F{\`e}vre}, {Iovino}, {Temporin}, {Ilbert},
  {Arnouts}, {Charlot}, {Brinchmann}, {Zucca}, {Tresse}, {Scodeggio}, {Guzzo},
  {Bottini}, {Garilli}, {Le Brun}, {Maccagni}, {Picat}, {Scaramella},
  {Vettolani}, {Zanichelli}, {Adami}, {Bardelli}, {Cappi}, {Ciliegi},
  {Contini}, {Foucaud}, {Gavignaud}, {McCracken}, {Marano}, {Marinoni},
  {Mazure}, {Meneux}, {Merighi}, {Paltani}, {Pell{\`o}}, {Pollo}, {Radovich},
  {Bondi}, {Bongiorno}, {Cucciati}, {de la Torre}, {Gregorini}, {Mellier},
  {Merluzzi}, {Vergani}, \& {Walcher}}]{pbl07}
{Pozzetti} L., {Bolzonella} M., {Lamareille} F., {Zamorani} G., {Franzetti} P.,
  {Le F{\`e}vre} O., {Iovino} A., {Temporin} S., {Ilbert} O., {Arnouts} S.,
  {Charlot} S., {Brinchmann} J., {Zucca} E., {Tresse} L., {Scodeggio} M.,
  {Guzzo} L., {Bottini} D., {Garilli} B., {Le Brun} V., {Maccagni} D., {Picat}
  J.~P., {Scaramella} R., {Vettolani} G., {Zanichelli} A., {Adami} C.,
  {Bardelli} S., {Cappi} A., {Ciliegi} P., {Contini} T., {Foucaud} S.,
  {Gavignaud} I., {McCracken} H.~J., {Marano} B., {Marinoni} C., {Mazure} A.,
  {Meneux} B., {Merighi} R., {Paltani} S., {Pell{\`o}} R., {Pollo} A.,
  {Radovich} M., {Bondi} M., {Bongiorno} A., {Cucciati} O., {de la Torre} S.,
  {Gregorini} L., {Mellier} Y., {Merluzzi} P., {Vergani} D., {Walcher} C.~J.,
  2007, A\&A, 474, 443

\bibitem[{{Salim} {et~al.}(2007){Salim}, {Rich}, {Charlot}, {Brinchmann},
  {Johnson}, {Schiminovich}, {Seibert}, {Mallery}, {Heckman}, {Forster},
  {Friedman}, {Martin}, {Morrissey}, {Neff}, {Small}, {Wyder}, {Bianchi},
  {Donas}, {Lee}, {Madore}, {Milliard}, {Szalay}, {Welsh}, \& {Yi}}]{src07}
{Salim} S., {Rich} R.~M., {Charlot} S., {Brinchmann} J., {Johnson} B.~D.,
  {Schiminovich} D., {Seibert} M., {Mallery} R., {Heckman} T.~M., {Forster} K.,
  {Friedman} P.~G., {Martin} D.~C., {Morrissey} P., {Neff} S.~G., {Small} T.,
  {Wyder} T.~K., {Bianchi} L., {Donas} J., {Lee} Y.-W., {Madore} B.~F.,
  {Milliard} B., {Szalay} A.~S., {Welsh} B.~Y., {Yi} S.~K., 2007, ApJS, 173,
  267

\bibitem[{{Schneider} \& {Rix}(1997)}]{sch97}
{Schneider} P., {Rix} H., 1997, ApJ, 474, 25

\bibitem[{{Scoville} {et~al.}(2007){Scoville}, {Abraham}, {Aussel}, {Barnes},
  {Benson}, {Blain}, {Calzetti}, {Comastri}, {Capak}, {Carilli}, {Carlstrom},
  {Carollo}, {Colbert}, {Daddi}, {Ellis}, {Elvis}, {Ewald}, {Fall},
  {Franceschini}, {Giavalisco}, {Green}, {Griffiths}, {Guzzo}, {Hasinger},
  {Impey}, {Kneib}, {Koda}, {Koekemoer}, {Lefevre}, {Lilly}, {Liu},
  {McCracken}, {Massey}, {Mellier}, {Miyazaki}, {Mobasher}, {Mould}, {Norman},
  {Refregier}, {Renzini}, {Rhodes}, {Rich}, {Sanders}, {Schiminovich},
  {Schinnerer}, {Scodeggio}, {Sheth}, {Shopbell}, {Taniguchi}, {Tyson}, {Urry},
  {Van Waerbeke}, {Vettolani}, {White}, \& {Yan}}]{saa07}
{Scoville} N., {Abraham} R.~G., {Aussel} H., {Barnes} J.~E., {Benson} A.,
  {Blain} A.~W., {Calzetti} D., {Comastri} A., {Capak} P., {Carilli} C.,
  {Carlstrom} J.~E., {Carollo} C.~M., {Colbert} J., {Daddi} E., {Ellis} R.~S.,
  {Elvis} M., {Ewald} S.~P., {Fall} M., {Franceschini} A., {Giavalisco} M.,
  {Green} W., {Griffiths} R.~E., {Guzzo} L., {Hasinger} G., {Impey} C., {Kneib}
  J., {Koda} J., {Koekemoer} A., {Lefevre} O., {Lilly} S., {Liu} C.~T.,
  {McCracken} H.~J., {Massey} R., {Mellier} Y., {Miyazaki} S., {Mobasher} B.,
  {Mould} J., {Norman} C., {Refregier} A., {Renzini} A., {Rhodes} J., {Rich}
  M., {Sanders} D.~B., {Schiminovich} D., {Schinnerer} E., {Scodeggio} M.,
  {Sheth} K., {Shopbell} P.~L., {Taniguchi} Y., {Tyson} N.~D., {Urry} C.~M.,
  {Van Waerbeke} L., {Vettolani} P., {White} S.~D.~M., {Yan} L., 2007, ApJS,
  172, 38

\bibitem[{{Semboloni} {et~al.}(2011){Semboloni}, {Hoekstra}, {Schaye}, {van
  Daalen}, \& {McCarthy}}]{shs11}
{Semboloni} E., {Hoekstra} H., {Schaye} J., {van Daalen} M.~P., {McCarthy}
  I.~G., 2011, MNRAS, 417, 2020

\bibitem[{{Sheldon} {et~al.}(2004){Sheldon}, {Johnston}, {Frieman}, {Scranton},
  {McKay}, {Connolly}, {Budav{\'a}ri}, {Zehavi}, {Bahcall}, {Brinkmann}, \&
  {Fukugita}}]{sjf04}
{Sheldon} E.~S., {Johnston} D.~E., {Frieman} J.~A., {Scranton} R., {McKay}
  T.~A., {Connolly} A.~J., {Budav{\'a}ri} T., {Zehavi} I., {Bahcall} N.~A.,
  {Brinkmann} J., {Fukugita} M., 2004, AJ, 127, 2544

\bibitem[{{Sheth} {et~al.}(2001){Sheth}, {Mo}, \& {Tormen}}]{smt01}
{Sheth} R.~K., {Mo} H.~J., {Tormen} G., 2001, MNRAS, 323, 1

\bibitem[{{Smith} {et~al.}(2003){Smith}, {Peacock}, {Jenkins}, {White},
  {Frenk}, {Pearce}, {Thomas}, {Efstathiou}, \& {Couchman}}]{spj03}
{Smith} R.~E., {Peacock} J.~A., {Jenkins} A., {White} S.~D.~M., {Frenk} C.~S.,
  {Pearce} F.~R., {Thomas} P.~A., {Efstathiou} G., {Couchman} H.~M.~P., 2003,
  MNRAS, 341, 1311

\bibitem[{{Strauss} {et~al.}(2002){Strauss}, {Weinberg}, {Lupton}, {Narayanan},
  {Annis}, {Bernardi}, {Blanton}, {Burles}, {Connolly}, {Dalcanton}, {Doi},
  {Eisenstein}, {Frieman}, {Fukugita}, {Gunn}, {Ivezi{\'c}}, {Kent}, {Kim},
  {Knapp}, {Kron}, {Munn}, {Newberg}, {Nichol}, {Okamura}, {Quinn}, {Richmond},
  {Schlegel}, {Shimasaku}, {SubbaRao}, {Szalay}, {Vanden Berk}, {Vogeley},
  {Yanny}, {Yasuda}, {York}, \& {Zehavi}}]{swl02}
{Strauss} M.~A., {Weinberg} D.~H., {Lupton} R.~H., {Narayanan} V.~K., {Annis}
  J., {Bernardi} M., {Blanton} M., {Burles} S., {Connolly} A.~J., {Dalcanton}
  J., {Doi} M., {Eisenstein} D., {Frieman} J.~A., {Fukugita} M., {Gunn} J.~E.,
  {Ivezi{\'c}} {\v Z}., {Kent} S., {Kim} R.~S.~J., {Knapp} G.~R., {Kron} R.~G.,
  {Munn} J.~A., {Newberg} H.~J., {Nichol} R.~C., {Okamura} S., {Quinn} T.~R.,
  {Richmond} M.~W., {Schlegel} D.~J., {Shimasaku} K., {SubbaRao} M., {Szalay}
  A.~S., {Vanden Berk} D., {Vogeley} M.~S., {Yanny} B., {Yasuda} N., {York}
  D.~G., {Zehavi} I., 2002, AJ, 124, 1810

\bibitem[{{Tinker} {et~al.}(2012){Tinker}, {George}, {Leauthaud}, {Bundy},
  {Finoguenov}, {Massey}, {Rhodes}, \& {Wechsler}}]{tgl12}
{Tinker} J.~L., {George} M.~R., {Leauthaud} A., {Bundy} K., {Finoguenov} A.,
  {Massey} R., {Rhodes} J., {Wechsler} R.~H., 2012, ApJL, 755, L5

\bibitem[{{Tinker} {et~al.}(2005){Tinker}, {Weinberg}, {Zheng}, \&
  {Zehavi}}]{twz05}
{Tinker} J.~L., {Weinberg} D.~H., {Zheng} Z., {Zehavi} I., 2005, ApJ, 631, 41

\bibitem[{{van Daalen} {et~al.}(2011){van Daalen}, {Schaye}, {Booth}, \& {Dalla
  Vecchia}}]{vsb11}
{van Daalen} M.~P., {Schaye} J., {Booth} C.~M., {Dalla Vecchia} C., 2011,
  MNRAS, 415, 3649

\bibitem[{{van Uitert} {et~al.}(2011){van Uitert}, {Hoekstra}, {Velander},
  {Gilbank}, {Gladders}, \& {Yee}}]{vhv11}
{van Uitert} E., {Hoekstra} H., {Velander} M., {Gilbank} D.~G., {Gladders}
  M.~D., {Yee} H.~K.~C., 2011, A\&A, 534, A14+

\bibitem[{{Velander} {et~al.}(2011){Velander}, {Kuijken}, \&
  {Schrabback}}]{vks11}
{Velander} M., {Kuijken} K., {Schrabback} T., 2011, MNRAS, 412, 2665

\end{thebibliography}

\appendix
  
\section{Impact of halo model assumptions}\label{cfhtls:app:assumptions}
In this appendix we discuss the impact of the different assumptions which the halo model is necessarily based on. Some of these may be overly stringent or inaccurate and with the accuracy afforded by the CFHTLenS it is important to provide a quantitative impression of how large a role they actually play in determining the halo mass and satellite fractions. Here, we only quote the results from studying red lenses since they are better constrained than the blue lenses, but the results for the latter are qualitatively equivalent. For comparison, we remind the reader that the observational errors we are comparing to are typically in the range of 15--40\% (excluding the highest mass bin).

\begin{figure}
\includegraphics[height=84mm,angle=270]{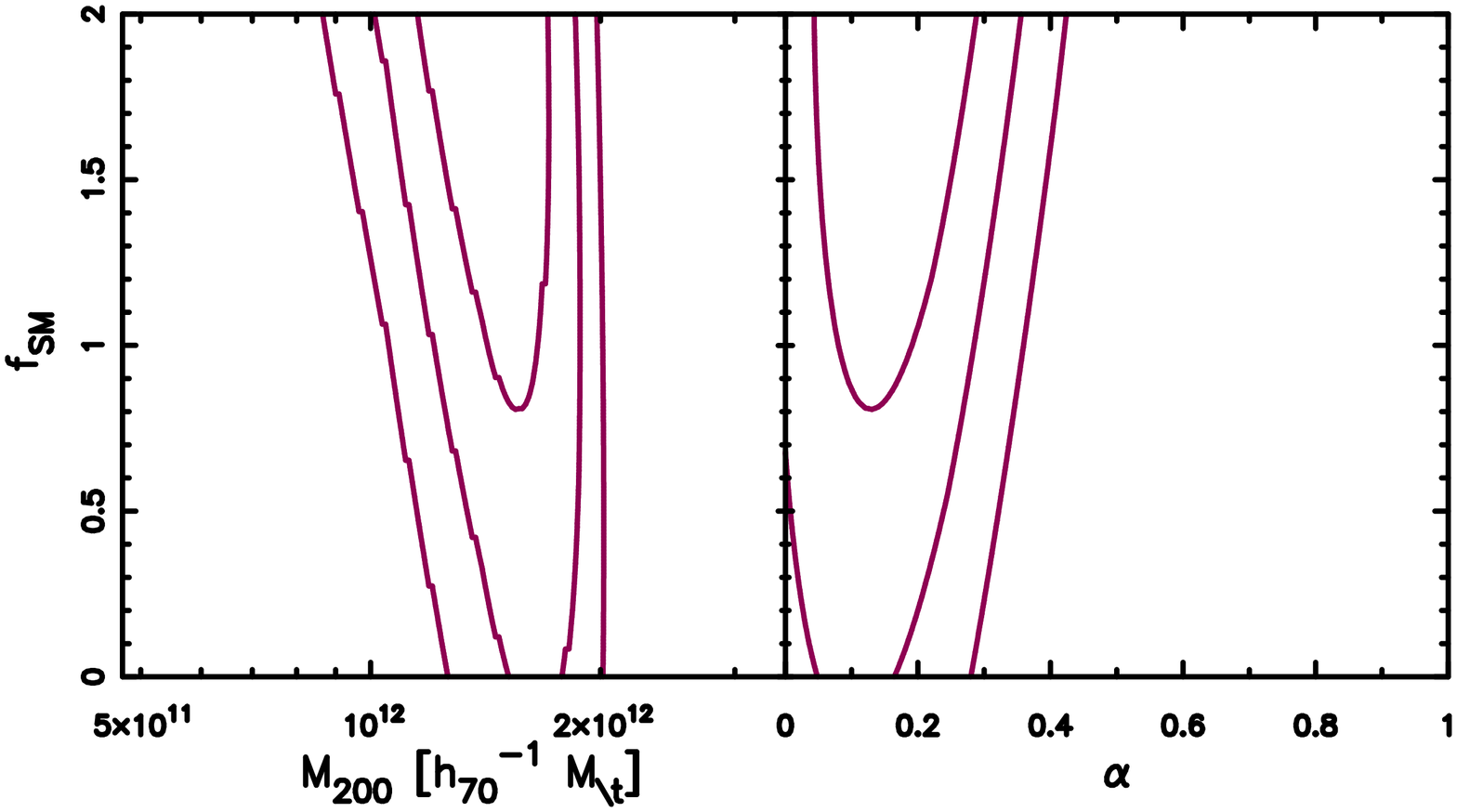}
\caption{Dependence of halo model fitting parameters halo mass $M_{200}$ and satellite fraction $\alpha$ on stellar mass, with $f_{\mathrm{SM}}$ the fraction of true mean stellar mass used in the halo model and contours showing the 67.8\%, 95.4\% and 99.7\% confidence intervals. The left panel shows that including a baryonic component in the model (i.e.~setting $f_{\mathrm{SM}}=1$) will result in a significantly lower best-fit halo mass than not doing so ($f_{\mathrm{SM}}=0$), and the right panel shows that the reason for this is an increased satellite fraction. In our analysis we keep the stellar mass component fixed at $f_{\mathrm{SM}}=1$.}
\label{cfhtls:fig:stellarmassImpact}
\end{figure}

Assumptions that have an effect on small scales where the baryonic, central 1-halo and the stripped satellite terms dominate will translate into an effect on the measured halo mass. To see the impact the inclusion of a baryonic component has, we remove it completely from our model. We find that the masses for some bins then increase by as much as 15\%. It may appear counter-intuitive that including a baryonic component with a mass which is of order 5\% of the total mass should result in a  halo mass estimate that is lowered by a greater amount than that. The explanation lies in the halo model fitting, and specifically in the way the satellite fraction is allowed to vary. Adding a baryonic component on small scales will result in a lowered central halo mass. The central halo profile reaches further than the baryonic component however, and thus power on intermediate scales is also diminished. To compensate for this loss of power, the halo model will increase the satellite 1-halo term by increasing the satellite fraction, which also increases the stripped satellite halo term, lowering the central 1-halo term further until an equilibrium is reached. These mechanisms are illustrated for red galaxies in luminosity bin L4 in Figure~\ref{cfhtls:fig:stellarmassImpact}, where we have allowed halo mass, satellite fraction and stellar mass fraction to vary simultaneously for both panels. This figure also makes clear the degeneracies introduced to the halo model if the stellar mass is left as a free parameter in addition to halo mass and satellite fraction.
Higher luminosity or stellar mass bins are more severely affected by this effect than the lower end due to the lack of a prominent satellite 1-halo feature. To study the effect on the best-fit power law parameters, we re-fit the halo models excluding the baryonic component. The resulting slope and amplitude of the power law do not change significantly. We note, however, that our baryonic component only accounts for the stars in the lens and not for example the hot gas. The influence of feedback on the gas distribution in galaxies is a complicated issue which may also affect our results, as discussed in \citet{vsb11} and \citet{shs11}, but it is an effect which we do not attempt to model here. However, as future lensing surveys grow more powerful and the data allow for greater accuracy this will be an important effect to study.

The two dark matter terms which dominate on these small scales are mainly affected by two implementation choices: the profile types of our dark matter haloes (NFW, possibly stripped, in this case) and the relation between the halo mass and its NFW concentration for which we have selected the relation described by \citet{dsk08}. To estimate the magnitude of the impact we change our central 1-halo term while keeping everything else the same. Because the relative amplitudes of the different terms in our halo model are intimately connected, this will only give an approximate idea of the influence of these choices, since we have not changed the stripped satellite term, or the distribution of satellites which still follows the original NFW. First we change the concentrations of our NFW haloes. The $1\sigma$ error intervals of the three \citet{dsk08} relation parameters results in a variation in concentration of at most 4\% for our halo masses. If we instead were to assume that the haloes in our sample were fully relaxed, the concentration may increase by as much as 25\% for the lowest stellar mass bin. To test the effect of such a change in concentration, we multiply the original \citet{dsk08} concentration of the central NFW halo by a factor of 1.25 which results in the same mass being contained within a smaller radius. In general that means that the satellite 1-halo term has to compensate on intermediate scales, leading to a greater satellite fraction and therefore a lower halo mass. The lower the luminosity or stellar mass, the less affected the estimated halo mass since the satellite 1-halo term feature is clearly visible in the signal and therefore well constrained. For the highest luminosity or stellar mass bins the estimated halo mass is then up to about 10\% less than our original estimate, a variation which is subdominant to the observational errors in all bins. As mentioned, the satellite fraction is also affected by this, increasing by about 30\% for the higher luminosity or stellar mass bins while staying roughly the same for the lower bins.

Moving on to the modelling of the satellite halo, we choose to strip 50\% (corresponding to a truncation radius of $0.4r_{200}$) of the satellite dark matter irrespective of type or distance to the centre of the main halo. Though this is a somewhat simplistic modelling choice, we can test how the measured halo mass is affected by a change in the amount of dark matter that is stripped from the satellite haloes. \citet{ghe13} find that for groups in the CFHTLenS, high density environment galaxies with a stellar mass between $10^9$ and $10^{10.5}$ and located at redshifts between 0.2 and 0.8 have been tidally stripped of 57\% of their mass. This corresponds to a truncation radius of $(0.26\pm0.14)r_{200}$. Furthermore, the two extreme cases where either all or none of the mass is stripped from the satellite haloes have both been ruled out \citep[see][]{msk06}. We therefore test two more sensible truncation radii: $0.2r_{200}$ and $0.6r_{200}$. In the first case, more dark matter is stripped from the average satellite than for our standard choice, while the opposite is true in the second case. For the range in luminosities and stellar masses used in this work, the best-fit satellite fractions do not change much with the different truncation radii (at most it decreases by about 10\% for the case where the truncation radius is smaller). As the truncation radius is reduced, some signal is lost on small scales and the modelling software compensates by increasing the halo mass by about 10--15\% at most. Similarly, the best-fit halo mass is slightly smaller when a greater truncation radius is used, though the effect is less pronounced. The larger the satellite fraction, the more the signal is affected and the greater the effect is on the fitted halo mass. The effect is more pronounced for the reduced truncation radius than for the increased one due to the shape of the halo profile, though it is still smaller than the observational errors. To further investigate what range of truncation radii is reasonable requires the use of high-resolution hydrodynamical simulations, and that is beyond the scope of this work. Since it is unlikely that the majority of satellites are strongly stripped \citep{msk06} we therefore choose to not take this effect into account. With the statistical improvement offered by the next generation of weak lensing surveys, however, a more sophisticated description of the stripping of satellite haloes, possibly as a function of distance from the centre of the main halo, is needed. 

We now turn our focus to the factors that influence the model on intermediate scales, i.e.~where the satellite 1-halo term dominates. The shape of the satellite 1-halo term is determined by the distribution of satellites within the main halo, while the amplitude is affected by the HOD \citep{mts05}. Here we assume that the distribution of satellites follows the distribution of the dark matter exactly. It may very well be, however, that the satellites are less concentrated than the dark matter halo is \citep[see, for example,][]{nag05,gce12}. To assess the impact of using a different concentration parameter for the satellites than for the dark matter, we try two cases: $c_{\mathrm{sat}}=2c_{\mathrm{dm}}$ and $c_{\mathrm{sat}}=0.5c_{\mathrm{dm}}$. This check has already been carried out by \citetalias{vhv11}, and their best-fit parameters do not change significantly, but with the greater signal-to-noise of our signal we consider it appropriate to repeat the test. Doubling (halving) the NFW concentration of the satellite galaxies implies a somewhat reduced (added) satellite 1-halo contribution on small scales. This results in a $<10\%$ decrease (increase) of the satellite fraction and a decrease (increase) in the estimated halo mass ranging from 2--20\% over the luminosity and stellar mass range included in our analysis. This fits within our error bars, but with future signal precision this is another assumption that requires some scrutiny.

Moving on to the choice of HOD, we note that it would be very difficult to determine the number of satellites expected for a given mass, the HOD, from a galaxy-galaxy analysis such as this. The reason is that it is nearly completely degenerate with the satellite fraction. The satellite fraction is mainly determined from these scales where the satellite 1-halo represents the main contribution to the total signal. Changing the amplitude of the satellite 1-halo term by changing the HOD therefore mimics a change in satellite fraction. We note, however, that \citet{mts05} can recover a simulated satellite fraction with an accuracy of 10\% using a HOD identical to the one in this paper. To see the impact such an error may have on our halo masses we take our best-fit satellite fraction in each luminosity or stellar mass bin, increase it by 10\% and fit a new halo mass estimate. The most affected bins are again the ones with the highest satellite fraction, with the new halo estimate being less than 10\% lower than the original one for nearly all bins used in this analysis, reaching 15\% and 20\% for S3 and L1 respectively.

On scales beyond $\sim1\,h_{70}^{-1}\,\mathrm{Mpc}$ the 2-halo terms become important, and the choice of bias influences these terms. The prescription we adopt for the bias in our halo model does not include non-linear effects. Figure~1 from \citet{msb13} shows that non-linear bias affects the galaxy-dark matter cross-correlation coefficient at the 2\% level at a comoving separation of $4\,h_{70}^{-1}\,\mathrm{Mpc}$. The magnitude of the effect diminishes with increasing distance to 1\% at $10\,h_{70}^{-1}\,\mathrm{Mpc}$, and the influence on our 2-halo terms should be comparable. The affected regime, where the 1-halo and 2-halo terms overlap, is notoriously difficult to model however. One major issue is that of halo exclusion which attempts to account for the way neighbouring dark matter haloes overlap. To illustrate the influence of the 2-halo terms on our best-fit parameters we can choose to limit our fit to scales where these terms do not play a major role, i.e.~fit out to $0.5\,h_{70}^{-1}\,\mathrm{Mpc}$ rather than to our default choice of $2\,h_{70}^{-1}\,\mathrm{Mpc}$ (see Section~\ref{cfhtls:sec:luminosity}). The results are then noisier of course, but still well within our error boundaries. For low luminosity or stellar mass, the halo mass is reduced by about 15\%. For the higher luminosity/stellar mass bins, the differences are smaller. The results including or excluding scales where the 2-halo terms are significant are therefore consistent with each other. Thus, since the effect of non-linearity is likely small compared to other modelling uncertainties on these scales, and since the affected range is beyond that used to determine halo masses in this paper, we choose not to include non-linear biasing in our model.

The above study shows that none of the systematic effects considered here will significantly change our best-fit parameters. Re-fitting the power law relations between halo mass and observable (see Sections~\ref{cfhtls:sec:luminosityScaling} and \ref{cfhtls:sec:stellarmassScaling}) in each case confirms that the effect on these relations is subdominant to the observational uncertainties. We note, however, that it is possible for several of these effects to conspire, causing a shift or a tilt in one or more of the power law relations. This should be kept in mind for the next section and for any future comparisons with our results.

\section{Corrections for signal contamination}\label{cfhtls:app:corrections}

\subsection{Photometric redshift bias correction}\label{cfhtls:app:photozbias}
Though the quality of the CFHTLenS photometric redshift estimates is high, there is still a small bias present due to the inherent limitations of template-based Bayesian methods, as discussed in \citet{hek12}. This bias will affect not only the redshift itself, but also the derived quantities such as luminosity and stellar mass. Since our lenses reside at relatively low redshifts we therefore have to correct our lens redshifts and derived quantities for this bias in order to achieve accurate object selection for our dark matter halo relations. Additionally, if this bias is not corrected for, the angular separations between lenses and sources will be altered, causing a coherent shift in the lensing signal radial binning. The resulting halo model fit will then also be affected, further illustrating the importance of this correction.

\begin{table}
\caption{Redshift bias fit parameters for red and blue subsamples. The slope $a$ is kept fixed between magnitude bins while the offset $b$ is allowed to vary.}
\label{cfhtls:tab:zbias}
\begin{center}
\begin{tabular}{@{}lcccc}
\hline
Magnitude bin & $a_{\mathrm{red}}$ & $b_{\mathrm{red}}\,[\times10^{-2}]$ & $a_{\mathrm{blue}}$ & $b_{\mathrm{blue}}\,[\times10^{-2}]$ \\
\hline
(14,19] & 0.99 & -0.62 & 1.08 & -2.52 \\
(19,20] & 0.99 &  0.20 & 1.08 &  7.46 \\
(20,21] & 0.99 &  4.64 & 1.08 &  3.69 \\
(21,22] & 0.99 &  4.64 & 1.08 &  5.07 \\
(22,23] & ---  & ---   & 1.08 &  4.06 \\
\hline
\end{tabular}
\end{center}
\end{table}
Following \citet{hek12}, we perform our correction using spectroscopic redshifts in the overlap with the VIMOS VLT Deep Survey \citep[VVDS;][]{lvg05,glg08}, the DEEP2 galaxy redshift survey \citep{dfn03,dgk07,ncd13} and the SDSS \citep{eag01,swl02}. The completeness of this spectroscopic sample is shown in \citet[][Figure~16]{lvg05} and \citet[][Figure~31]{ncd13}. To ensure a completeness of at least 80\%, we select only lenses with magnitude $i'_{\mathrm{AB}}<23$, as mentioned in Section~\ref{cfhtls:sec:cats}. Since the bias is a function of magnitude and galaxy type, we start by splitting our sample in red and blue subsamples via their photometric type (as described in Section~\ref{cfhtls:sec:stellarmassestimates}) and use several magnitude bins. We then quantify the bias in each bin by fitting a straight line of the form
\begin{equation}
(z_{\mathrm{spec}}-0.3) = a(z_{\mathrm{phot}}-0.3) + b
\end{equation}
where $z_{\mathrm{spec}}$ is the spectroscopic redshift from VVDS/DEEP2/SDSS, $z_{\mathrm{phot}}$ is the CFHTLenS photometric redshift estimate, $a$ is the slope and $b$ is the offset. The pivot point of 0.3 roughly corresponds to the mean redshift of our lens sample, though the correction is insensitive to this number. The slope $a$ is fit simultaneously in all magnitude bins but allowed different values for red and blue samples, while the offset $b$ is allowed to vary between both type and magnitude bins. Keeping the slope fixed allows for a more robust estimate for the bias, though we have verified that allowing it to vary has negligible impact on the results in practice. The resulting fit parameters are shown in Table~\ref{cfhtls:tab:zbias}. Note that there is no correction performed for red galaxies beyond a magnitude of $i'_{\mathrm{AB}}=22$ since we do not use fainter red lenses (see Section~\ref{cfhtls:sec:cats}).

We then use these fit parameters to correct our lens photometric redshift estimates in the range \mbox{$0.2\leq z\leq0.4$}. Calculating the luminosity distances and estimating the k-corrections corresponding to the original and corrected redshifts using the $g'-r'$ colours of the galaxies, we adjust the absolute magnitudes accordingly. We further derive new stellar mass estimates by scaling them to their new luminosities assuming a constant (pre-correction) stellar mass-to-luminosity ratio. The impact on the red galaxy properties is negligible, but for blue galaxies the correction is larger with the average luminosity and stellar mass increasing by \mbox{$\sim12\%$}. We therefore proceed to use the corrected quantities in our luminosity and stellar mass analyses (see Sections~\ref{cfhtls:sec:luminosity} and \ref{cfhtls:sec:stellarmass}).

The sources will also be affected by photometric redshift bias, but its impact on the measured halo masses is expected to be much smaller than the effect of the lens redshift bias. To confirm this hypothesis we shift all sources by a constant bias of 2\% and redo the analysis of Sections~\ref{cfhtls:sec:luminosity} and \ref{cfhtls:sec:stellarmass}. This bias value is most likely slightly larger than necessary \citep[see][Figure 4]{hek12}, but the resulting halo masses agree with the original halo masses within $1\sigma$. We therefore do not need to correct our sources for photometric redshift bias.

\subsection{Photometric redshift scatter correction}\label{cfhtls:app:photozs}
\begin{figure}
\includegraphics[height=84mm,angle=270]{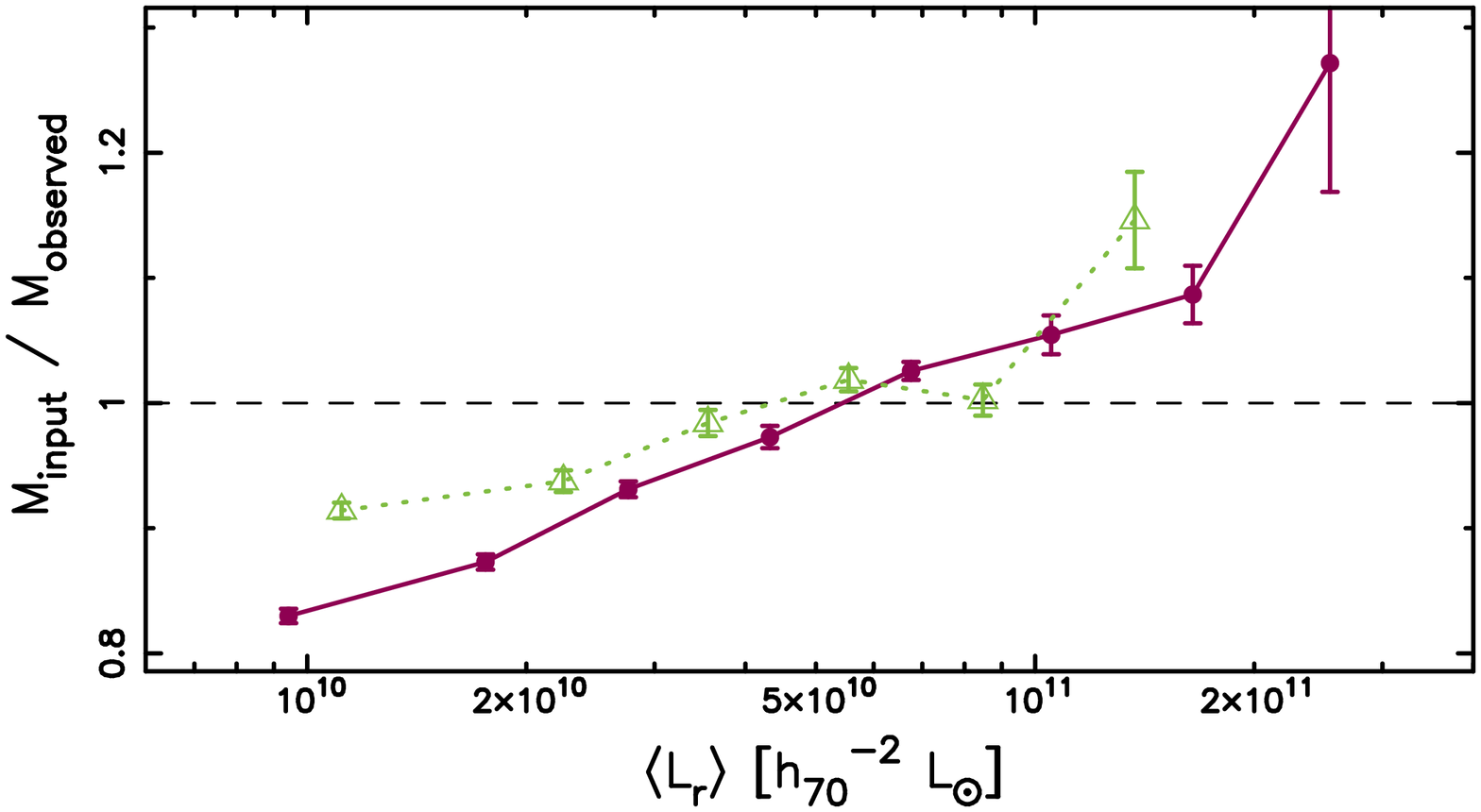}
\caption{Correction factor as a function of luminosity induced through inaccuracies in the photometric redshift estimates. The dark purple solid (light green dotted) line with dots (triangles) shows the scatter correction factor for the red (blue) lens sample. The error bars show the scatter between 10 lens catalogue realisations.}
\label{cfhtls:fig:photozScatter}
\end{figure}
\begin{table}
\caption{Photometric redshift scatter correction factors applied to observed halo masses in each luminosity bin (see Section~\ref{cfhtls:sec:luminosity}) for red and blue lenses. These factors correct both for scatter due to redshift errors, and for the fact that the observed halo mass does not necessarily correspond to the mean halo mass.}
\label{cfhtls:tab:photozScatter}
\begin{center}
\begin{tabular}{@{}lcccc}
\hline
Bin & $f^{\mathrm{lscat}}_{\mathrm{red}}$ & $\sigma_{f,\mathrm{red}}$ & $f^{\mathrm{lscat}}_{\mathrm{blue}}$ & $\sigma_{f,\mathrm{blue}}$ \\
\hline
L1 & 0.86 & 0.01 & 0.94 & 0.01 \\
L2 & 0.89 & 0.01 & 0.96 & 0.01 \\
L3 & 0.96 & 0.01 & 1.01 & 0.01 \\
L4 & 1.02 & 0.01 & 1.05 & 0.01 \\
L5 & 1.09 & 0.01 & 1.04 & 0.01 \\
L6 & 1.13 & 0.02 & 1.20 & 0.04 \\
L7 & 1.16 & 0.02 & --- & --- \\
L8 & 1.36 & 0.10 & --- & --- \\
\hline
\end{tabular}
\end{center}
\end{table}
Before interpreting the luminosity results we have to take into account the effect of Eddington bias \citep{edd13}. The precision of our photometric redshifts is high with a scatter of \mbox{$\sigma_z\sim0.04$} for both lenses and sources \citep{hek12}, but never the less the errors on the redshift estimates have to be taken into account. If the true redshift differs from the estimated one, this will affect all derived quantities. An underestimated redshift, for example, would cause the estimated absolute magnitude to be fainter than the true absolute magnitude and the lens would be placed in the wrong luminosity bin. As can be seen in Figure~\ref{cfhtls:fig:magSelection} there are more faint objects than bright, which means that more objects will scatter from fainter bins into brighter bins than the other way around. This will lower the lensing signal in each bin and bias the observed halo mass low, and the amount of bias will be luminosity dependent. To estimate the impact of redshift scatter we create a simulated version of the CFHTLenS as follows. We fit an initial power law mass-luminosity relation of the form (see Equation~\ref{cfhtls:eq:lumpower}, Section~\ref{cfhtls:sec:luminosityScaling})
\begin{equation}
M_{200} = M_{0,L}\left(\frac{L}{L_{\mathrm{fid}}}\right)^{\beta_L}
\end{equation}
to the raw estimated halo masses, with $L_{\mathrm{fid}}=10^{11}\,h_{70}^{-2}\,L_{r',\odot}$. We then use this relation to assign halo masses to our lenses. Splitting the resulting lens catalogue in the usual magnitude bins for the red and blue sample separately we obtain our `true' halo mass for each bin. Constructing NFW haloes from these halo masses at the photometric redshift of the lenses, we create mock source catalogues with the observed source redshift distribution but with simulated shear estimates with strengths corresponding to those which would be induced by our lens haloes. We then scatter the lenses and sources using the full redshift probability density function, split the lens catalogue according to the scattered magnitudes and measure the resulting signal within $200\,h^{-1}\,\mathrm{kpc}$ of the lenses using our scattered shear catalogue. We only use the small scales for our mass estimate to avoid complications due to insufficient treatment of clustering since on these scales only the central 1-halo signal is relevant, and we force our satellite fraction to zero to obtain a pure NFW fit. This way we obtain the `observed' halo mass for each magnitude bin. The `observed' halo mass is then compared to the `true' value for each bin. To increase the statistical precision of the correction we determine the average of 10 lens catalogue realisations. Since the starting point is a perfect signal, the number of realisations given the area is adequate to retrieve the correction factor. This correction simultaneously accounts for all the effects resulting from any photometric redshift scatter in our analysis, such as the scattering of lenses between luminosity bins, the effect on the lens and source redshift distributions, the smoothing of the signal due to mixing of the projected lens-source separations, and the non-linear dependence of the critical surface density $\Sigma_{\rm crit}$ on the lens and source redshifts. Note that the errors on the correction factors indicate only the propagated photometric redshift uncertainty, and even though they are small compared to the errors on the shear measurements, we have included them in our final error budget. The error on the correction factor does not include the uncertainties of the input parameters. However, we expect these additional uncertainties to be negligible compared to the errors on the halo masses (see the discussion in Appendix \ref{cfhtls:app:binscatter}).

The results from this test are shown in Figure~\ref{cfhtls:fig:photozScatter}. The quality of our photometric redshifts is high which means that the correction factor is small overall, reaching only \mbox{$\sim30\%$} for a luminosity of \mbox{$L_{r'}\sim2.5\times10^{11}\,h_{70}^{-2}\,L_{\odot}$}. Here the contamination is largest due to the shape of the luminosity function causing a larger fraction of low luminosity objects to scatter into the higher-luminosity bin. For our faintest red luminosity bin the correction is \mbox{$\sim20\%$}, in this case caused by larger errors in the photometric redshift estimates. The correction factor is less than unity for lower-luminosity bins due to the turn-over of the distribution of red lenses at \mbox{$M_{r'}\sim-21.2$} (see Figure~\ref{cfhtls:fig:magSelection}). The smaller correction factor for blue lenses is due to their somewhat flatter mass-luminosity relation (see Figure~\ref{cfhtls:fig:magbinsAlpha}). Because of the relative insensitivity of halo mass to changes in luminosity, minor errors in luminosity measurements due to photometric redshift inaccuracies will not strongly affect the halo mass estimate. The process described in this appendix could in principle be iterated over, starting from the fitting of a mass-luminosity relation, until convergence is reached. Since \citet{hhy05} find that different choices for that relation yield similar curves, we choose not to iterate further.

We also have to correct our luminosity bins for a second scatter effect. As discussed in \citetalias{vhv11} (Appendix B), the observed halo mass does not necessarily correspond to the mean halo mass in a given bin since the halo masses in that bin are not evenly distributed and the NFW profiles do not depend linearly on halo mass. The distribution within each bin generally follows a log-normal distribution, and to correct for this effect we follow a similar procedure as the one outlined in Appendix \ref{cfhtls:app:binscatter}, with the difference that we do not scatter the luminosities as we have already corrected for that by accounting for the error in photometric redshift. We stress that this is an intrinsic effect unrelated to any measurement errors. The full correction factor, taking into account both scatter effects discussed here, is shown in Table~\ref{cfhtls:tab:photozScatter}.

The general procedure outlined in this appendix is repeated for the stellar mass bins, though we use the stellar mass-halo mass relation to assign halo masses to the mock lens catalogue, and then bin the lenses according to stellar mass rather than luminosity. In this case we do not use the resulting correction factor, but we do include the errors on said correction factor in our error budget in order to account for the above-mentioned effects in our stellar mass results. The correction factor itself, however, only conveys the impact of photometric redshift uncertainties, and not the additional effects influencing the stellar mass errors. The scatter due to stellar mass errors is accounted for following the method described in the next appendix, and applying this correction factor as well would therefore amount to double-counting.

\subsection{Stellar mass bin scatter correction}\label{cfhtls:app:binscatter}
\begin{figure}
\includegraphics[height=84mm,angle=270]{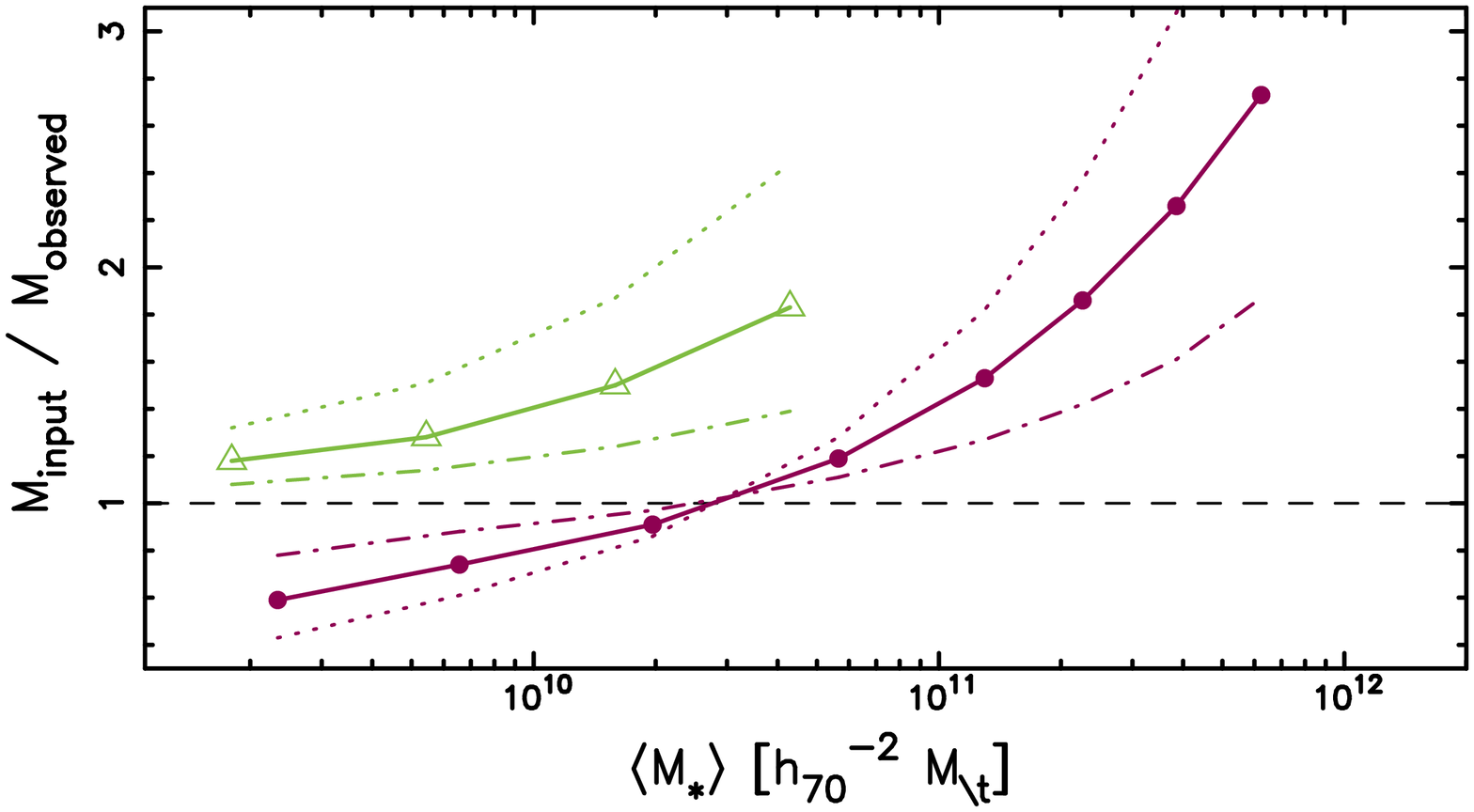}
\caption{Mass correction factor as a function of stellar mass induced through inaccuracies in the stellar mass estimates. The dark purple solid dots (light green open triangles) show the correction factor for the red (blue) lens sample. As discussed in the text, the dot-dashed lines show the correction factors if stellar mass errors of 0.2~dex are assumed, rather than the default 0.3~dex, and the dotted lines show the correction factors derived using stellar mass errors of 0.4~dex.}
\label{cfhtls:fig:massScatter}
\end{figure}
\begin{table}
\caption{Bin scatter correction factors applied to observed halo masses in each stellar mass bin (see Section \ref{cfhtls:sec:stellarmass}) for red and blue lenses. These factors correct both for scatter due to stellar mass errors, and for the fact that the observed halo mass does not necessarily correspond to the mean halo mass. }
\label{cfhtls:tab:binscatter}
\begin{center}
\begin{tabular}{@{}lcc}
\hline
Bin & $f^{\mathrm{mscat}}_{\mathrm{red}}$ & $f^{\mathrm{mscat}}_{\mathrm{blue}}$ \\
\hline
S1 & 0.59 & 1.18  \\
S2 & 0.74 & 1.28  \\
S3 & 0.91 & 1.50  \\
S4 & 1.19 & 1.83  \\
S5 & 1.53 & ---  \\
S6 & 1.86 & ---  \\
S7 & 2.26 & ---  \\
S8 & 2.73 & ---  \\
\hline
\end{tabular}
\end{center}
\end{table}
In a process similar to the scatter in luminosity, objects will scatter between stellar mass bins due to errors on the stellar mass estimates. Though objects scatter randomly according to their individual stellar mass errors, the net effect will be to scatter lenses from greater abundance to lower according to the stellar mass function (SMF). Because the stellar mass function declines steeply at higher stellar mass bins, these will be more severely affected by low-mass object contamination. As a result the observed lensing mass in the highest stellar mass bins will be biased low \citepalias[see Appendix A in][]{vhv11}. Additionally, the lensing halo mass estimates will be affected by the fact that the observed halo mass does not necessarily correspond to the mean halo mass in a given bin, as discussed in Appendix~\ref{cfhtls:app:photozs}.

To assess the impact of both these effects simultaneously we follow a procedure similar to the one used to correct for redshift scatter, as described in the previous appendix. We start by fitting an initial power law halo mass-stellar mass relation using the raw observed lensing halo mass. Drawing a large number of simulated lens galaxies from the stellar mass function, we take these stellar masses to be the true unscattered values and assign a halo mass according to the fitted halo mass-stellar mass relation. As described above, this halo mass will be distributed within the stellar mass bin according to some distribution. Following \citetalias{vhv11} we therefore correct the halo mass for this effect by drawing from a log-normal distribution with a mean given by the original halo mass and a width determined by \citet{mbc11}. We now know the `true' mean halo mass for each bin. Using the resulting simulated lens catalogue we create a source catalogue with shears determined analytically. We then scatter the lenses assuming a Gaussian error distribution with a width of 0.3~dex as appropriate for our stellar mass errors (see Section~\ref{cfhtls:sec:stellarmassestimates}) to create a new simulated lens catalogue, this time containing `observed' stellar masses. Dividing this `observed' lens catalogue according to the usual stellar mass bins for red and blue samples separately, we measure the signal in the simulated shear catalogues and again fit an NFW profile. This way we obtain the `observed' halo mass for each stellar mass bin. By taking the ratio of simulated `observed' to `true' halo mass we arrive at the correction factor for stellar mass scatter as shown in Figure~\ref{cfhtls:fig:massScatter}. We can now apply this factor, as quoted in Table~\ref{cfhtls:tab:binscatter}, to our halo mass estimates to correct for the scatter between stellar mass bins, and for the fact that the observed halo mass does not correspond to the mean halo mass, simultaneously.

The correction factor is relatively sensitive to the adopted value of the stellar mass error, particularly in the regime where the stellar mass function is steep. Therefore, in addition to the correction factor used, we also show in Figure~\ref{cfhtls:fig:massScatter} the correction factors obtained if we adopt a stellar mass error of 0.2~dex or 0.4~dex instead, covering the plausible range of values that the stellar mass error could take. This illustrates how the correction factor coherently shifts if the stellar mass error is different from what we assume. For S8 of the red lenses, the change is largest, with an increase (decrease) of the correction factor by $\sim50\%$ for 0.4~dex (0.2~dex), respectively. We do not use the plausible range of correction factors as the error on the correction, since a different stellar mass error would only lead to a coherent shift of all the correction factors and hence of the corrected halo masses. This property of the correction factors would be lost, and the error bars on the halo masses would be severely overestimated, causing an unjustified loss of information. However, for completeness, we note that the best fit power law normalisation and slope are
\mbox{$1.14^{+0.07}_{-0.08}\times10^{13}\,h_{70}^{-1}\,M_{\odot}$}
(\mbox{$0.84^{+0.20}_{-0.16}\times10^{13}\,h_{70}^{-1}\,M_{\odot}$})
and
\mbox{$1.23^{+0.06}_{-0.07}$}
(\mbox{$0.98^{+0.08}_{-0.07}$})
for red (blue) lenses when we adopt a stellar mass error of 0.2~dex, and 
\mbox{$1.83\pm0.13\times10^{13}\,h_{70}^{-1}\,M_{\odot}$}
(\mbox{$0.84^{+0.20}_{-0.16}\times10^{13}\,h_{70}^{-1}\,M_{\odot}$})
and
\mbox{$1.50^{+0.05}_{-0.07}$}
(\mbox{$0.98^{+0.08}_{-0.07}$})
for red (blue) lenses for a stellar mass error of 0.4~dex.

Additionally, the correction factor has some error due to the uncertainties of the other input parameters, such as in the adopted power law relations, the stellar mass functions, and the scatter in halo mass. \citetalias{vhv11} found that the correction is fairly insensitive to changes in the power law relation; using the power law obtained {\it after} the stellar mass scatter correction only changed the correction factor by at most 4\%. The impact here will be even smaller as the power laws are less steep, and we therefore ignore their effect. Next, the stellar mass function is not the intrinsic stellar mass function as objects have already scattered. However, we cannot reliably obtain the intrinsic stellar mass function where it matters most, i.e. at the high stellar mass range, as the number of galaxies is too low. We therefore do not attempt to obtain the intrinsic stellar mass function, but rather note this as a caveat. Finally, we note that the correction factor is insensitive to the adopted width of the halo mass distribution.

\section{Tests for galaxy-galaxy lensing systematics}\label{cfhtls:app:systematics}

\subsection{Initial consistency analysis of the CFHTLenS catalogue}\label{cfhtls:app:parkerAndRCS2}
\begin{figure}
\includegraphics[height=84mm,angle=270]{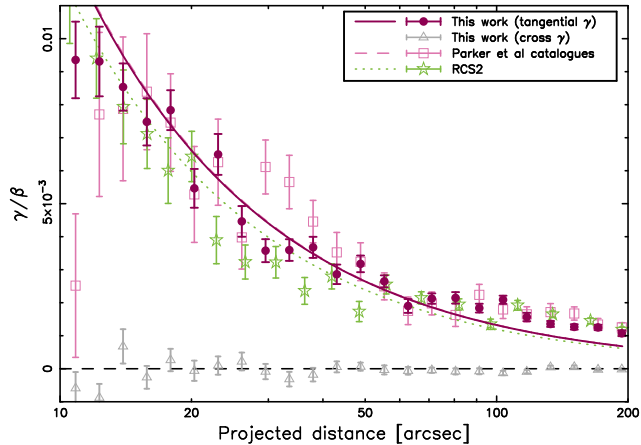}
\caption{Comparison of three data sets: the shear catalogues from \mbox{$\sim22\,\mathrm{deg}^2$} CFHTLS (pink open squares), the results from RCS2 (light green open stars) and our results (dark purple solid dots). The curves show the best-fit singular isothermal sphere for each dataset (with light green and pink nearly identical), and the grey triangles show the cross-shear from our results which should be zero in the absence of systematic errors.}
\label{cfhtls:fig:parkerAndRCS2}
\end{figure}
In this study we use lenses and sources from the full 154~deg$^2$ CFHTLenS catalogue. The accuracy of the CFHTLenS shears has been verified through several rigorous tests aimed at the study of cosmic shear \citep{hvm12,mhk13}, but it is interesting to compare the galaxy-galaxy lensing signal with the results from two previous analyses of a similar nature. The first is the galaxy-galaxy lensing analysis in the CFHTLS-Wide conducted by \citet{phh07}, and the second is based on the shear catalogue from \citetalias{vhv11} (see Section~\ref{cfhtls:sec:comparison}). In \citet{phh07} an area of \mbox{$\sim22\,\mathrm{deg}^2$} in $i'$ was analysed, corresponding to about 14\% of our area. Since they only had data from one band their analysis also lacked redshift estimates for lenses and sources, but they separated lenses from sources using magnitude cuts. The shear estimates for their sources were obtained using a version of the technique introduced by \citet*{ksb95} as outlined in \citet{hfk98}. These shear estimates were measured on a stacked image rather than obtained by fitting all exposures simultaneously \citep[see][for a discussion on this]{mhk13}. To avoid the strong PSF effects at the chip boundaries, \citet{phh07} limited their analysis to the unique chip overlaps. In contrast we are able to use all the data we have at our disposal. The data from \citetalias{vhv11} is the subset of $\sim$400 square degrees of the RCS2 with $i'$-band coverage, which is shallower than the CFHTLS and for which also no redshifts were available for the sources at the time of this analysis.

To compare and contrast our lensing signal with these previous works we mimic the analysis presented in \citet{phh07} as closely as possible and apply the same $i'$-band magnitude cuts as employed in \citet{phh07}, with~$19.0<i'_{\mathrm{AB}}<22.0$ for lenses and $22.5<i'_{\mathrm{AB}}<24.5$ for sources. \citet{phh07} boost their signal to correct for contamination by sources that are physically associated with the lens, and we apply the same correction factor to our values. The resulting galaxy-galaxy signal, scaled with the angular diameter distance ratio $\langle\beta\rangle = \langle D_{ls}/D_s \rangle = 0.49$ from \citet{phh07}, is shown as dark purple solid dots in Figure~\ref{cfhtls:fig:parkerAndRCS2}. We also re-analysed the original shear catalogues used for the \citet{phh07} analysis with the results shown as pink open squares in Figure~\ref{cfhtls:fig:parkerAndRCS2}. The signal from the \citetalias{vhv11} shape measurement catalogues of the RCS2 is obtained using a source selection of $22<r'<24$ instead because the limiting depth in $i'$ is 23.8 for the RCS2. The measurements are also corrected for contamination by physically associated sources, as described in \citetalias{vhv11}, and scaled with $\langle\beta\rangle=0.30$ which is determined by integrating over the lens and source redshift distributions that were obtained from the CFHTLS ``Deep Survey'' fields \citep{iam06}. The measurements are shown as light green open stars. Figure~\ref{cfhtls:fig:parkerAndRCS2} shows that the lensing signals generally agree well. We fit an SIS profile to the shear measurements that have been scaled by $\langle\beta\rangle$  on scales between 7 and 120 arcsec, and find a scaled Einstein radius of $\widetilde{r_{\rm{E}}}=0.277\pm0.006$ arcsec for our results, $\widetilde{r_{\rm{E}}}=0.267\pm0.011$ arcsec for the \citet{phh07} measurements and $\widetilde{r_{\rm{E}}}=0.262\pm0.007$ arcsec for \citetalias{vhv11}, which are broadly consistent.

The best-fit SIS profile corresponds to a velocity dispersion of $\sigma_v=97.9\pm1.0\,\mathrm{km}\,\mathrm{s}^{-1}$, which is lower than the $\sigma_v=132\pm10\,\mathrm{km}\,\mathrm{s}^{-1}$ quoted in \citet{phh07}. However, using the re-analysed \citet{phh07} shear catalogue we find a velocity dispersion of $\sigma_v=96.6\pm2.0\,\mathrm{km}\,\mathrm{s}^{-1}$. For the \citetalias{vhv11} results we find a velocity dispersion of $\sigma_v=95.4\pm1.3\,\mathrm{km}\,\mathrm{s}^{-1}$, slightly lower but in reasonable agreement with our results. Note that there are various small differences between the analyses, such as different effective source redshift distributions and different weights applied to the shears. Additionally we use the multiplicative bias correction factor for our measurements, while the other works did not have such a correction. All these differences could have small but non-negligible effects on the results. The discrepancy with the velocity dispersion quoted in \citet{phh07} remains unexplained, but we conclude that the shear estimates are in fact fully consistent. 
\subsection{Seeing test}\label{cfhtls:app:seeing}
\citet{mhk13} isolated a general multiplicative calibration factor as a function of the signal-to-noise ratio and size of the source galaxy, $m(\nu_{SN},r)$, using simulations. To confirm the successful calibration of the CFHTLenS shears in the context of galaxy-galaxy lensing, we study how a shear bias relates to image quality. In general a round PSF causes circularisation of source images which in turn can cause a multiplicative bias of the measured shapes if it is not properly corrected for. Such a systematic would depend on the size of the PSF. Assuming that the systematic offset due to PSF anisotropy is negligible (a fair assumption given our correction for spurious signal around random lenses; see Section~\ref{cfhtls:sec:wgglensing} and the detailed analysis of PSF residual errors in \citet{hvm12}), and assuming that the shapes of very well resolved galaxies can be accurately recovered, the observed average shear in a galaxy-galaxy lensing azimuthal distance bin is related to the true average shear via
\begin{equation}
\langle\gamma^{\mathrm{obs}}\rangle = \langle\gamma^{\mathrm{true}}\rangle\left[1 + \mathcal{M}\left\langle\left(\frac{r_*}{r_0}\right)^2\right\rangle\right]\;.
\end{equation}
where $\gamma^{\mathrm{obs}}$ is the observed shear, $\gamma^{\mathrm{true}}$ is the true shear, $r_*$ is the PSF size, $r_0$ is the intrinsic (Gaussian) size of the galaxy and $\mathcal{M}$ is a value close to zero representing the multiplicative bias. The particular dependence on PSF size is the result of a full moments analysis \citep[see for example][]{pav08}.

Since the bias depends on the size of the PSF relative to the size of galaxies, data with a spread in seeing should enable us to determine the bias $\mathcal{M}$ directly from the data, thus allowing us to deduce the true performance of the shape measurement pipeline. The CFHTLS images have such a spread, with the best seeing being 0.44~arcsec and the worst being 0.94~arcsec, and therefore provides us with a neat way of determining this bias. Since at small projected separations from the lens, the tangential shear signal is generally well described by an SIS profile: 

\begin{figure}
\includegraphics[height=84mm,angle=270]{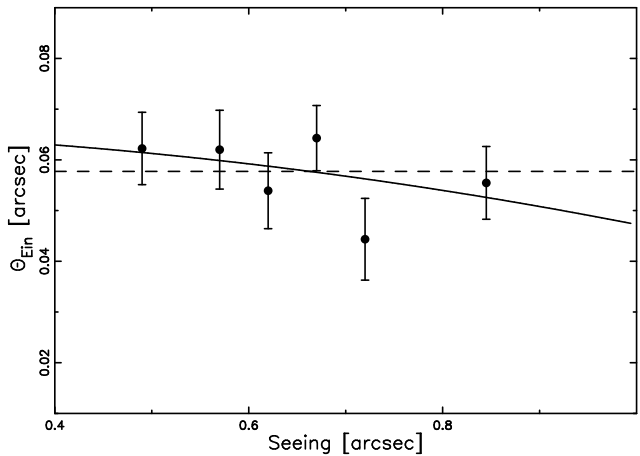}
\caption{Galaxy-galaxy lensing signal quantified through the best-fit Einstein radius (see Equation~\ref{cfhtls:eq:einstein}) as measured in each of 6 seeing bins, according to Table~\ref{cfhtls:tab:seeing}. The solid line shows the best-fit model using Equation~\ref{cfhtls:eq:ThetaEinObs} while the dashed line shows the average Einstein radius assuming no bias.}
\label{cfhtls:fig:seeingbins}
\end{figure}
\begin{table}
\caption{Details of the seeing bins.}\label{cfhtls:tab:seeing}
\begin{center}
\begin{tabular}{@{}lcccc}
\hline
Sample 	& $N_{\mathrm{fields}}$ 	& $\langle r_*\rangle\,[\mathrm{arcsec}]$ 	& $\theta_E\,[\mathrm{arcsec}]$ & $\sigma_{\theta_E}$ \\
\hline
   P1 &    27 &     0.50 &     0.053 &     0.005 \\
   P2 &    23 &     0.57 &     0.044 &     0.006 \\
   P3 &    33 &     0.62 &     0.050 &     0.005 \\
   P4 &    38 &     0.67 &     0.047 &     0.005 \\
   P5 &    28 &     0.72 &     0.040 &     0.006 \\
   P6 &    36 &     0.80 &     0.049 &     0.005 \\
\hline
\end{tabular}
\end{center}
\end{table}

\begin{equation}
\gamma(\theta) = \frac{\theta_E}{2\theta}
\label{cfhtls:eq:einstein}
\end{equation}
where $\theta$ is the distance to the lens and $\theta_E$ is the Einstein radius, we therefore have a simple relationship between the observed Einstein radius and the true one:
\begin{equation}
\theta_E^{\mathrm{obs}} = \theta_E^{\mathrm{true}}\left[1 + \mathcal{M}\left\langle\left(\frac{r_*}{r_0}\right)^2\right\rangle\right]\;.
\label{cfhtls:eq:ThetaEinObs}
\end{equation}
By measuring the Einstein radius of the average lens as a function of seeing we can therefore determine both the true Einstein radius and the performance of the shape measurement pipeline.

We select our lenses in magnitude and redshift as described in the main paper (Section~\ref{cfhtls:sec:cats}), though we do not distinguish between red and blue galaxies, and we also split our data according to Table~\ref{cfhtls:tab:seeing}. Dividing the data according to image quality in this way may imply some minor selection effects, such as redshift and magnitude estimates being less accurate for worse seeing and thus PSF. Since great care has been taken to correct for such effects \citep[see][]{hek12} we will assume here that the lens samples are comparable between seeing bins. Having selected our lenses, we measure the galaxy-galaxy lensing signal in each seeing bin and fit an SIS to the innermost $200\,h_{70}^{-1}\,\mathrm{kpc}$. By fitting only small scales we avoid the influence of neighbouring haloes. The results are shown in Figure~\ref{cfhtls:fig:seeingbins} and quoted in Table~\ref{cfhtls:tab:seeing}. We then fit the relation described by Equation~\ref{cfhtls:eq:ThetaEinObs} to the resulting Einstein radii and find a value of $\mathcal{M} = -0.071 \pm  0.075$. This is consistent with no bias, a fact which is further illustrated in Figure~\ref{cfhtls:fig:seeingbins}; the data points agree with an average Einstein radius of $0.058 \pm 0.003$, shown as a dashed line.

\section{Detailed luminosity bins}\label{cfhtls:app:luminositydetails}
\begin{figure*}
\includegraphics[height=168mm,angle=270]{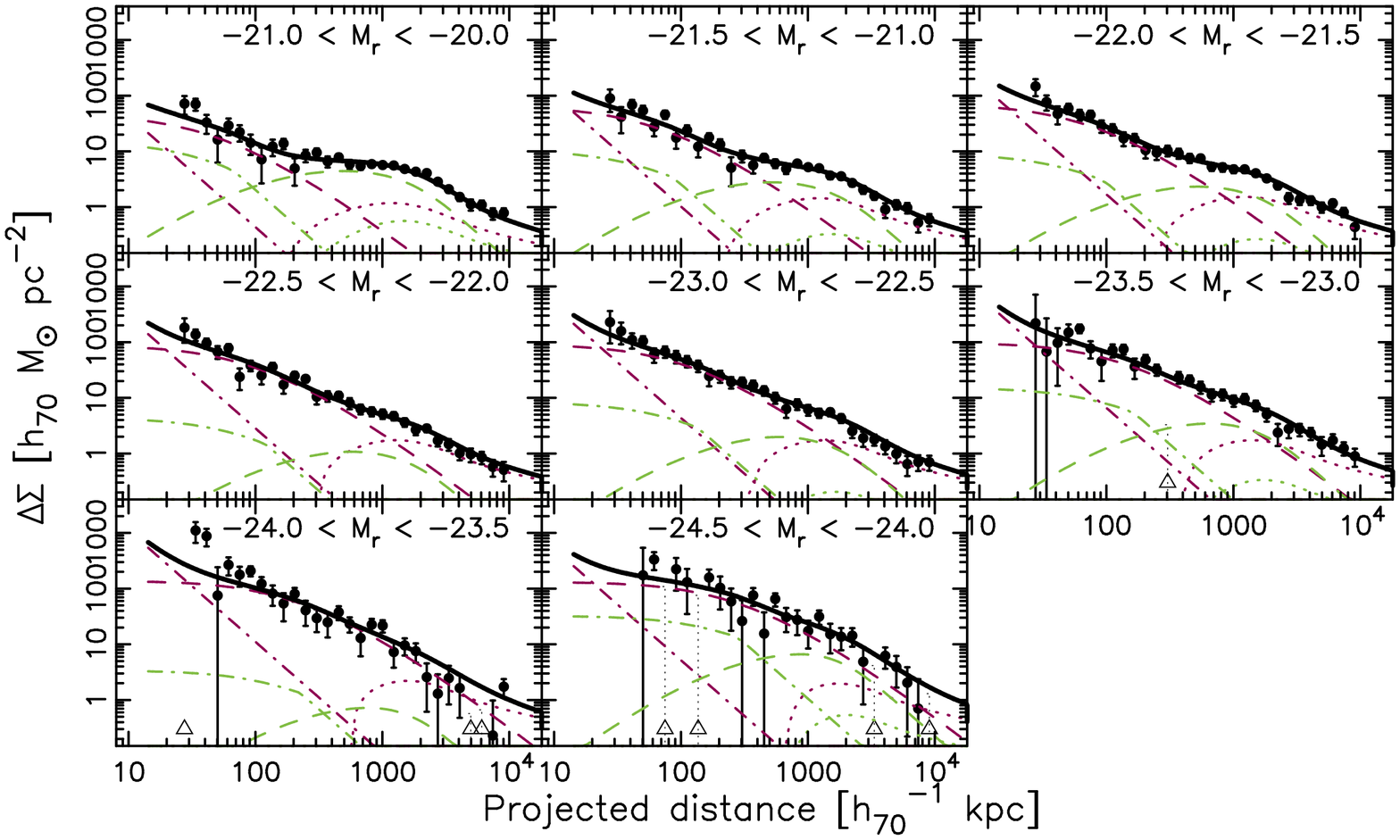}
\caption{Galaxy-galaxy lensing signal around {\em red} lenses which have been split into luminosity bins according to Table \ref{cfhtls:tab:magbins1}, and modelled using the halo model described in Section~\ref{cfhtls:sec:halomodel}. The black dots are the measured differential surface density, $\Delta\Sigma$, and the black line is the best-fit halo model with the separate components displayed using the same convention as in Figure~\ref{cfhtls:fig:haloModelExample}. Grey triangles represent negative points that are included unaltered in the model fitting procedure, but that have here been moved up to positive values as a reference. The dotted error bars are the unaltered error bars belonging to the negative points. The grey squares represent distance bins containing no objects.}
\label{cfhtls:fig:magbinsRed}
\end{figure*}
\begin{figure*}
\includegraphics[height=168mm,angle=270]{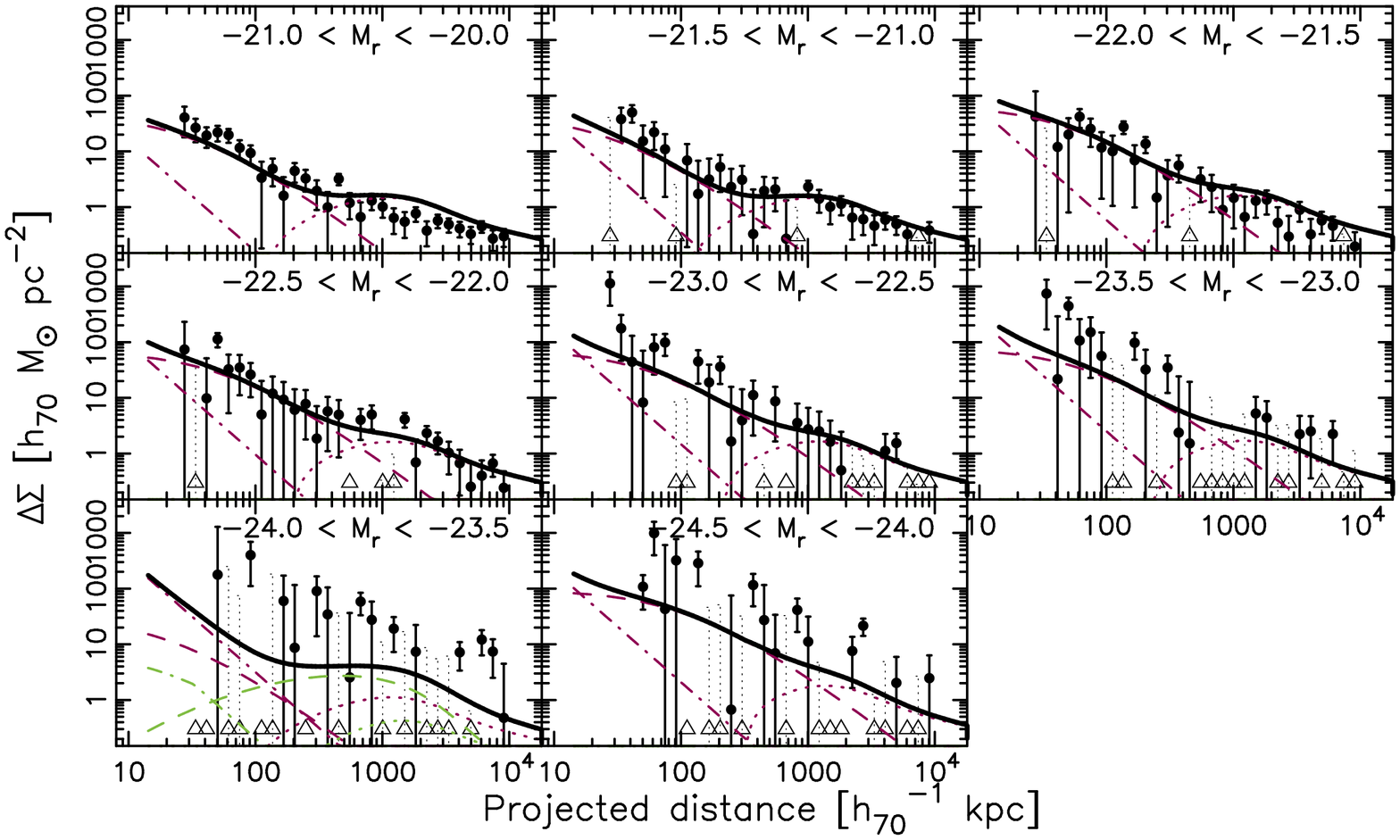}
\caption{Galaxy-galaxy lensing signal around {\em blue} lenses which have been split into luminosity bins according to Table \ref{cfhtls:tab:magbins1}, and modelled using the halo model described in Section~\ref{cfhtls:sec:halomodel}. The black dots are the measured differential surface density, $\Delta\Sigma$, and the black line is the best-fit halo model with the separate components displayed using the same convention as in Figure~\ref{cfhtls:fig:haloModelExample}. Grey triangles represent negative points that are included unaltered in the model fitting procedure, but that have here been moved up to positive values as a reference. The dotted error bars are the unaltered error bars belonging to the negative points. The grey squares represent distance bins containing no objects.}
\label{cfhtls:fig:magbinsBlue}
\end{figure*}

In this Appendix we show the decomposition of the best-fit halo model for red (Figure~\ref{cfhtls:fig:magbinsRed}) and blue (Figure~\ref{cfhtls:fig:magbinsBlue}) lenses, split in luminosity according to Table~\ref{cfhtls:tab:magbins1}. Showing the full decomposition is highly informative because it highlights some of the major trends and clarifies which effects dominate in each case.

The baryonic component based on the mean stellar mass in each bin (dark purple dot-dashed line) becomes more dominant for higher luminosities, but the luminous size of the lenses also increases, making measurement of background source shapes in the innermost distance bins difficult. Thus it is not possible to reliably constrain the baryonic component with our data. Yet the effect of including the baryons in our model is an overall lowering of the dark matter halo profile (dark purple dashed) compared to the model without baryons. For the red lenses we see that a considerable fraction of the sample at lower luminosities necessarily consists of satellite galaxies, since there is a clear bump in the signal at intermediate scales which has to be accounted for. This satellite fraction continuously drops as luminosity increases, and simultaneously becomes more difficult to constrain since the combination of the stripped satellite profile (light green dash-dotted) and satellite 1-halo terms (light green dashed) becomes almost indistinguishable from a single NFW profile for high halo masses. This effect was discussed in more detail in \citetalias{vhv11}, Appendix C.

For the blue lenses, the signal becomes very noisy for the two highest-luminosity bins due to a lack of lenses. These two bins are therefore discarded from the full analysis in Section~\ref{cfhtls:sec:luminosity}. In general, blue galaxies produce a noisier signal than red galaxies for the same luminosity cuts. This could be because blue lenses are in general less massive, and there are fewer of them which results in a weaker signal and a lower signal-to-noise for most bins. It could also be an indicator that the physical correlation between stellar mass and halo mass is noisier for these lenses. We also notice that nearly all blue lenses are galaxies located at the centre of their halo, rather than being satellites. This is consistent with previous findings. It is possible that satellite galaxies in general are redder because they have been stripped of their gas and thus have had their star formation quenched. It could also mean that most blue galaxies in our analysis are isolated; we have made no distinction between field galaxies and galaxies in a more clustered environment. If blue galaxies are more isolated than red ones then the contribution from nearby haloes (dotted lines) would also be less. It is clear from Figure~\ref{cfhtls:fig:magbinsBlue} that the large scales are not optimally fit by our model, and isolation may be one of the reasons since we assume the same mass-bias relation for blue galaxies as for red. With current data it is not possible to constrain the bias as a free parameter, but with future wider surveys this could be done.

\newpage
\clearpage
\newpage

\section{Detailed stellar mass bins}\label{cfhtls:app:stellarmassdetails}
\begin{figure*}
\includegraphics[height=168mm,angle=270]{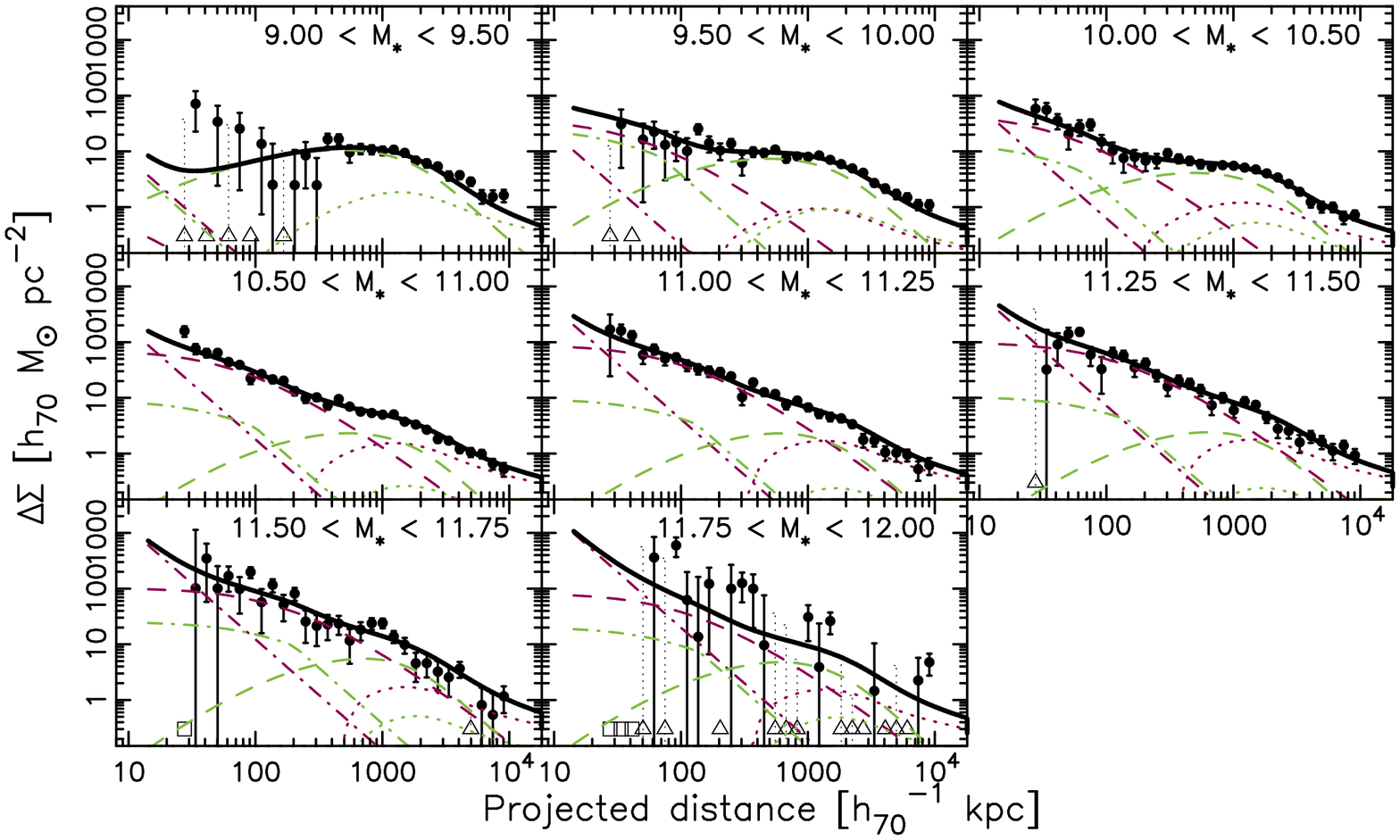}
\caption{Galaxy-galaxy lensing signal around {\em red} lenses which have been split into stellar mass bins according to Table \ref{cfhtls:tab:stelbins1}, and modelled using the halo model described in Section~\ref{cfhtls:sec:halomodel}. The black dots are the measured differential surface density, $\Delta\Sigma$, and the black line is the best-fit halo model with the separate components displayed using the same convention as in Figure~\ref{cfhtls:fig:haloModelExample}. Grey triangles represent negative points that are included unaltered in the model fitting procedure, but that have here been moved up to positive values as a reference. The dotted error bars are the unaltered error bars belonging to the negative points. The grey squares represent distance bins containing no objects.}
\label{cfhtls:fig:stelbinsRed}
\end{figure*}
\begin{figure*}
\includegraphics[height=168mm,angle=270]{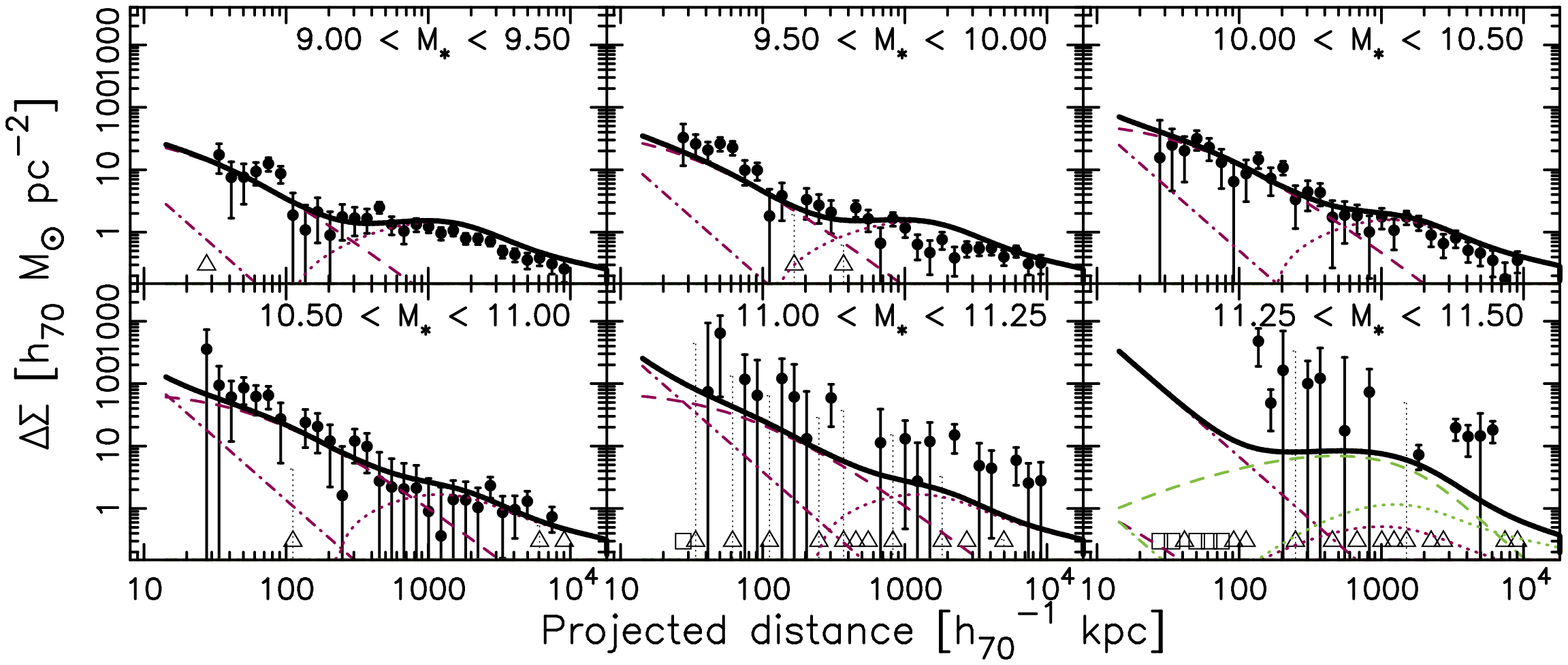}
\caption{Galaxy-galaxy lensing signal around {\em blue} lenses which have been split into stellar mass bins according to Table \ref{cfhtls:tab:stelbins1}, and modelled using the halo model described in Section~\ref{cfhtls:sec:halomodel}. The black dots are the measured differential surface density, $\Delta\Sigma$, and the black line is the best-fit halo model with the separate components displayed using the same convention as in Figure~\ref{cfhtls:fig:haloModelExample}. Grey triangles represent negative points that are included unaltered in the model fitting procedure, but that have here been moved up to positive values as a reference. The dotted error bars are the unaltered error bars belonging to the negative points. The grey squares represent distance bins containing no objects.}
\label{cfhtls:fig:stelbinsBlue}
\end{figure*}

The decomposition of the best-fit halo model for red and blue lenses, divided using stellar mass as detailed in Table~\ref{cfhtls:tab:stelbins1}, is shown in Figures~\ref{cfhtls:fig:stelbinsRed} and \ref{cfhtls:fig:stelbinsBlue} respectively.

By construction the baryonic component amplitude (dark purple dash-dotted line) increases with increasing bin number, and so does the dark matter halo mass (dashed lines). Note that with our stellar mass selections we push to smaller and fainter objects, so the objects in the three lowest-mass bins are on average less massive and less luminous than the galaxies in the faintest luminosity bin. In these bins, nearly all red galaxies are satellites, while for higher stellar mass bins the satellite fraction diminishes, a behaviour which is consistent with the trends we saw for luminosity (Appendix~\ref{cfhtls:app:luminositydetails}). For the higher stellar mass bins, as for the higher luminosity bins, the sum of the satellite stripped and 1-halo terms result in a profile which resembles a single NFW profile, making the satellite fraction more difficult to determine. For the blue lenses we run into the same issues for the highest mass bin as for the highest luminosity bins; the number of lenses is too small to constrain the halo model and so the bin has to be discarded. Furthermore, the satellite fraction is low across all blue lens bins indicating that these lenses are most likely isolated, which is consistent with the low large-scale signal and with our findings for luminosity.

\label{lastpage}
\end{document}